\documentclass[a4paper,onecolumn,10pt, accepted=2024-05-27]{quantumarticle}
\pdfoutput=1

\usepackage{amsthm}
\usepackage{amssymb}
\usepackage{amsmath}
\usepackage{bbm}
\usepackage{mathrsfs} 

\usepackage{multicol}
\usepackage{float}  

\usepackage{graphicx}
\graphicspath{ {./Figures/} }
\usepackage{tikz-cd}  
\usepackage{pgf,tikz}
\usetikzlibrary{arrows}

\definecolor{my-green}{RGB}{0, 96, 81} 

\usepackage{nicefrac} 

\usepackage{titlesec}

\renewcommand{\paragraph}[1]{\vspace{0.4cm}\noindent{\normalfont\sffamily\it#1}}
	
\newcommand{\refprop}[2][]{{\normalfont\hyperref[#2]{\sffamily\if #1\empty \else#1~\fi\ref{#2}\!\!}}
}
\newcommand{\eqrefprop}[2][]{{\normalfont\hyperref[#2]{\sffamily\if #1\empty \else#1~\fi(\ref{#2})\!\!}}
}

\newcommand{\pierre}[1]{#1}

\usepackage{xcolor}
\definecolor{wiring_color}{rgb}{0.65,0.2,0.1}

\usepackage[ruled]{algorithm2e}

\usepackage{enumitem}  

\usepackage{diagbox} 

\usepackage{hyperref} 
\hypersetup{
    colorlinks = true,
    linkcolor = quantumviolet, 
    urlcolor = quantumviolet,
    citecolor = quantumviolet,
}

\newtheoremstyle{mystyle}
  {15pt}
  {15pt}
  {\it}
  {0.cm}
  {\bf\sffamily}
  {.}
  {0.3cm}
  {}

\newtheoremstyle{mystyle-not-italic}
  {15pt}
  {15pt}
  {}
  {0.cm}
  {\it}
  {.}
  {0.3cm}
  {}

\theoremstyle{mystyle}

\newtheorem{theorem}{Theorem}
\newtheorem{proposition}[theorem]{Proposition}
\newtheorem*{statement}{Statement}
\newtheorem{conjecture}[theorem]{Conjecture}
\newtheorem{corollary}[theorem]{Corollary}
\newtheorem{lemma}[theorem]{Lemma}
\newtheorem{fact}[theorem]{Fact}

\newtheorem{definition}[theorem]{Definition}

\theoremstyle{mystyle-not-italic}

\newtheorem{remark}[theorem]{Remark}
\newtheorem{example}{Example}

\DeclareMathOperator{\Aff}{Aff}
\DeclareMathOperator{\Conv}{Conv}
\DeclareMathOperator{\Int}{Int}

\DeclareMathOperator{\Orbit}{Orbit}
\newcommand{\Orbittilde}{\widetilde{\Orbit}}
\DeclareMathOperator{\argmax}{argmax}

\newcommand{\1}{\mathbbm 1}
\newcommand{\A}{\mathcal{A}}
\newcommand{\Atilde}{\widetilde{\mathcal{A}}}
\newcommand{\Abox}{\mathtt{A}}
\newcommand{\B}{\mathcal{B}}
\newcommand{\Bbox}{\mathtt{B}}
\newcommand{\C}{\mathcal{C}}

\newcommand{\CHSH}{{\normalfont\texttt{CHSH}}}

\newcommand{\eg}{\emph{e.g.} }
\newcommand{\etal}{\emph{et al.} }
\newcommand{\G}{{\mathtt{G}}}
\renewcommand{\H}{{\mathcal{H}}}
\newcommand{\ie}{\emph{i.e.}~}
\renewcommand{\iff}{if and only if }
\newcommand{\I}{\mathtt{I}}
\newcommand{\K}{\mathcal{K}}
\renewcommand{\line}{\mathfrak{L}}
\renewcommand{\L}{\mathtt{L}}
\newcommand{\LL}{\mathcal{L}}
\newcommand{\N}{\mathbb{N}}
\newcommand{\NS}{\mathcal{N\!S}}

\renewcommand{\P}{\mathtt{P}}
\newcommand{\proj}{\mathtt{proj}}
\newcommand{\PR}{\mathtt{PR}}
\newcommand{\PL}{\mathtt{P}_{\text{\normalfont L}}}
\newcommand{\PNL}{\mathtt{P}_{\text{\normalfont NL}}}

\newcommand{\Q}{\mathtt{Q}}
\newcommand{\QQ}{\mathcal{Q}}
\newcommand{\R}{\mathbb{R}}
\newcommand{\Rbox}{\mathtt{R}}
\newcommand{\SR}{\mathtt{SR}}
\newcommand{\T}{\mathtt{T}}
\newcommand{\W}{\mathsf{W}}

\newcommand{\WW}{\mathcal{W}}
\newcommand{\Wtriv}{\W_{\text{\normalfont triv}}}
\newcommand{\Wlin}{\W_{\text{\normalfont lin}}}
\newcommand{\Woplus}{\W_{\oplus}}
\newcommand{\Wbs}{\W_{\text{\normalfont BS}}}
\newcommand{\Wdist}{\W_{\text{\normalfont dist}}}
\newcommand{\Wdepth}{\W_{\text{\normalfont depth$3$}}}
\newcommand{\Wand}{\W_{\wedge}}
\newcommand{\Worand}{\W_{\vee\!\wedge}}
\newcommand{\Wold}{\W_{\text{\normalfont old}}}
\newcommand{\Wout}{\W_{\normalfont\text{out}}}
\newcommand{\boxtimesW}{\underset{\W}{\boxtimes}}

\usepackage{subfiles} 

\usepackage[backend=bibtex, style=numeric, url=false, maxbibnames = 10, sortcites=true]{biblatex} 
\AtBeginBibliography{\footnotesize}
\renewbibmacro*{doi+eprint+url}{%
    \printfield{doi}%
    \newunit\newblock%
    \iftoggle{bbx:eprint}{%
        \usebibmacro{eprint}%
    }{}%
    \newunit\newblock%
    \iffieldundef{doi}{%
        \usebibmacro{url+urldate}}%
        {}%
    }
\bibliography{Bibliography}
\newboolean{printBibInSubfiles}
\setboolean{printBibInSubfiles}{true}
\def\bib{\ifthenelse{\boolean{printBibInSubfiles}}
        {\newpage
\begin{multicols}{2}
	\printbibliography
\end{multicols}
        }
        {}
    }

\definecolor{lime}{HTML}{A6CE39}

\newcommand{\myorcid}[1]{\href{https://orcid.org/#1}{\includegraphics[height=0.25cm]{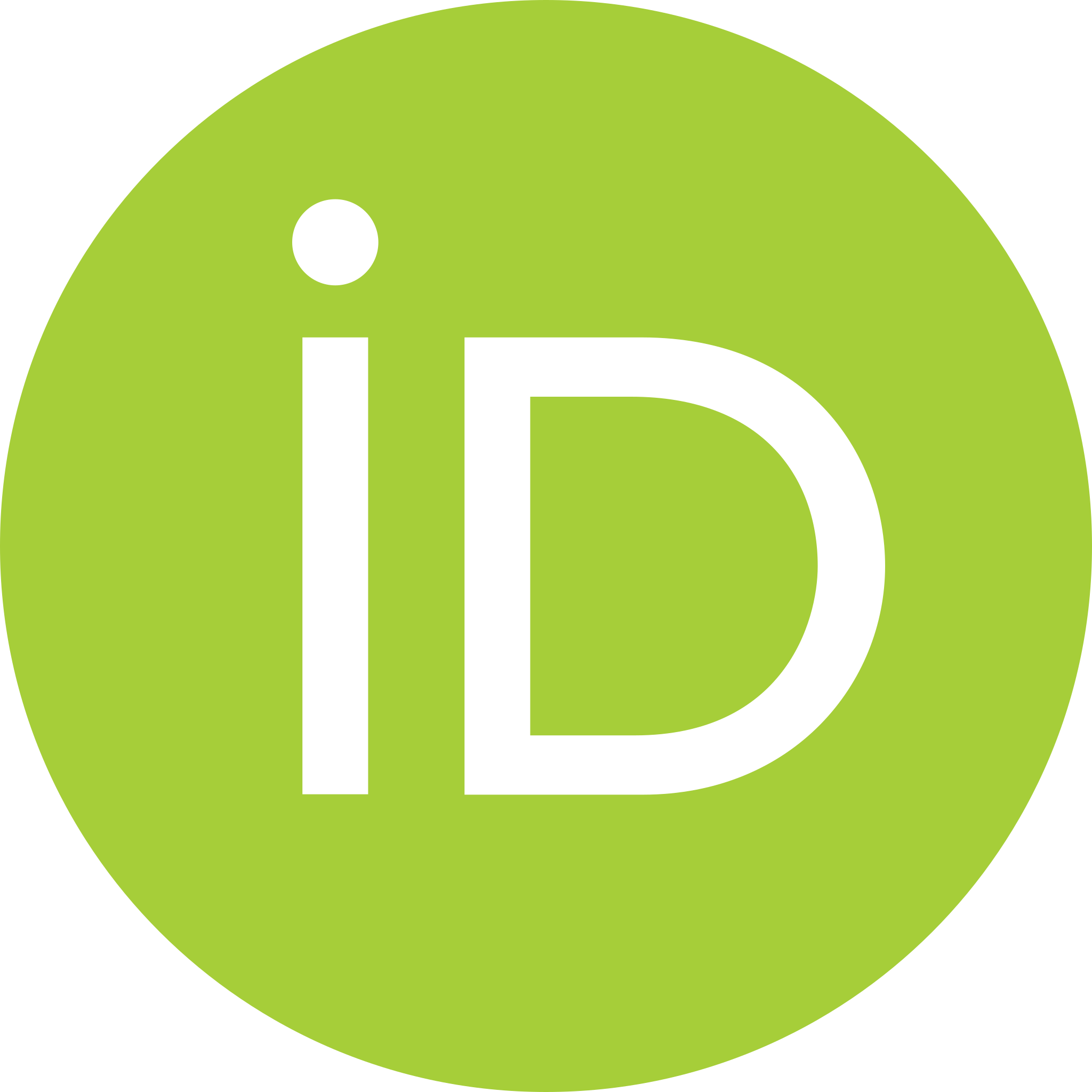}}}

\begin{document}
\setboolean{printBibInSubfiles}{false}

\title{Algebra of Nonlocal Boxes and the \newline Collapse of Communication Complexity}
\date{}

\author{Pierre Botteron~\myorcid{0000-0002-3861-0934}}
\email{pierre.botteron@math.univ-toulouse.fr}
\affiliation{Institut de Mathématiques de Toulouse, Université de Toulouse, France.}
\orcid{0000-0002-3861-0934}

\author{Anne Broadbent~\myorcid{0000-0003-1911-0093}}
\email{abroadbe@uottawa.ca}
\affiliation{Nexus for Quantum Technologies, University of Ottawa, Canada.}
\affiliation{Department of Mathematics and Statistics, University of Ottawa, Canada.}
\orcid{0000-0003-1911-0093}

\author{Reda Chhaibi~\myorcid{0000-0002-0085-0086}}
\email{reda.chhaibi@math.univ-toulouse.fr}
\affiliation{Institut de Mathématiques de Toulouse, Université de Toulouse, France.}
\affiliation{Institut Universitaire de France (IUF), France.}
\orcid{0000-0002-0085-0086}

\author{Ion Nechita~\myorcid{0000-0003-3016-7795}}
\email{ion.nechita@univ-tlse3.fr}
\affiliation{Laboratoire de Physique Théorique, Université de Toulouse, CNRS, France.}
\orcid{0000-0003-3016-7795}

\author{Clément Pellegrini}
\email{clement.pellegrini@math.univ-toulouse.fr}
\affiliation{Institut de Mathématiques de Toulouse, Université de Toulouse, France.}

\maketitle

\begin{abstract}
Communication complexity quantifies how difficult it is for two distant computers to evaluate a function $f(X,Y)$, where the strings $X$ and $Y$ are distributed to the first and second computer respectively,  under the constraint of exchanging as few bits as possible.
Surprisingly, some nonlocal boxes, which are resources shared by the two computers, are so powerful that they allow to \emph{collapse} communication complexity, in the sense that
any Boolean function $f$ can be correctly estimated with the exchange of only one bit of communication.
The Popescu-Rohrlich ($\PR$) box is an example of such a collapsing resource,
but a comprehensive description of the set of collapsing nonlocal boxes remains elusive.

In this work, we carry out an algebraic study of the structure of wirings connecting nonlocal boxes,
thus defining the notion of the ``product of boxes" $\P\boxtimes\Q$, and we show related associativity and commutativity results.
This gives rise to the notion of the ``orbit of a box", unveiling surprising geometrical properties about the alignment and parallelism of distilled boxes.
The power of this new framework is that it allows us to prove previously-reported numerical observations concerning the best way to wire consecutive boxes, and to numerically and analytically recover recently-identified noisy $\PR$ boxes that collapse communication complexity for different types of noise models.
\end{abstract}

\noindent Nonlocal boxes (NLBs) were introduced by Popescu and Rohrlich in 1994 as a theoretical generalization of quantum correlations~\cite{PR94}.
When Alice and Bob share a pair of entangled states $|\Psi\rangle$, each of them can choose to measure their state in a certain basis depending on some ``instructions" $x,y\in\{0,\dots, p\}$, and then each of them can encode their outcomes in respectively $a,b\in\{0,\dots, q\}$.
Similarly, a two-party NLB is a ``black box" shared between Alice and Bob, with some inputs $x,y$ and some outputs $a,b$, and with the rule that Alice has access only to the left part and Bob only to the right part, see \refprop[Figure]{fig:NLB}. This way, we only study the statistics produced by the ``hidden state" inside of the box and not the physical theory describing that state, which is the reason why nonlocal boxes are said to be \emph{device-independent}.
In this work, we consider one of the simplest scenarios, the $\CHSH$ scenario named after Clauser, Horne, Shimony, and Holt~\cite{CHSH69}, where $p=q=1$ and with two parties Alice and Bob (for more general scenarios, see~\cite{BM05b, BP05, DGHMP07, BS13, RBBGMRW20}).

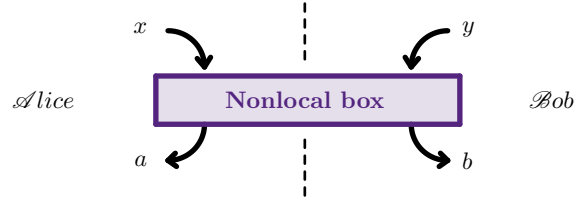
\begin{figure}[h]
	\centering
	\definecolor{quantumviolet}{HTML}{53257F}

\definecolor{qqttzz}{rgb}{0.,0.2,0.6}
\definecolor{rxsfyq}{rgb}{0.09019607843137255,0.1843137254901961,0.5019607843137255}
\definecolor{qqwuqq}{rgb}{0.,0.39215686274509803,0.}
\definecolor{wwqqzz}{rgb}{0.4,0.,0.6}
\definecolor{zzttqq}{rgb}{0.6,0.2,0.}

\begin{tikzpicture}[line cap=round,line join=round,>=triangle 45,x=1.0cm,y=1.0cm,every node/.style={scale=0.9}, scale=0.8]

\newcommand{\mynumber}{0.15}

\draw [line width=2.pt] (-0.2,-1.2 + \mynumber)-- (-0.1,-1.04 + \mynumber);
\draw [line width=2.pt] (-0.32,-1.06 + \mynumber)-- (-0.2,-1.2 + \mynumber);
\draw [shift={(-0.8,-1.2)},line width=2.pt]  plot[domain=0.:1.5707963267948966,variable=\t]({1.*0.6*cos(\t r)+0.*0.6*sin(\t r)},{0.*0.6*cos(\t r)+1.*0.6*sin(\t r) + \mynumber});
\draw [line width=2.pt] (3.2,-1.2 + \mynumber)-- (3.1,-1.04 + \mynumber);
\draw [line width=2.pt] (3.2,-1.2 + \mynumber)-- (3.32,-1.06 + \mynumber);
\draw [shift={(3.8,-1.2)},line width=2.pt]  plot[domain=1.5707963267948966:3.141592653589793,variable=\t]({1.*0.6*cos(\t r)+0.*0.6*sin(\t r)},{0.*0.6*cos(\t r)+1.*0.6*sin(\t r) +  \mynumber});
\draw [line width=2.pt] (-0.8,-2.6)-- (-0.68,-2.48);
\draw [line width=2.pt] (-0.8,-2.6)-- (-0.66,-2.7);
\draw [shift={(3.8,-2.)},line width=2.pt]  plot[domain=3.141592653589793:4.71238898038469,variable=\t]({1.*0.6*cos(\t r)+0.*0.6*sin(\t r)},{0.*0.6*cos(\t r)+1.*0.6*sin(\t r)});
\draw [line width=2.pt] (3.8,-2.6)-- (3.68,-2.48);
\draw [line width=2.pt] (3.8,-2.6)-- (3.66,-2.7);
\draw [shift={(-0.8,-2.)},line width=2.pt]  plot[domain=-1.5707963267948966:0.,variable=\t]({1.*0.6*cos(\t r)+0.*0.6*sin(\t r)},{0.*0.6*cos(\t r)+1.*0.6*sin(\t r)});

\fill[line width=2.pt, color=quantumviolet, fill=quantumviolet, fill opacity=0.15000000596046448] (-1.,-2.) -- (4.,-2.) -- (4.,-1.2) -- (-1.,-1.2) -- cycle;
\draw [line width=2.pt,color=quantumviolet] (-1.,-2.)-- (4.,-2.);
\draw [line width=2.pt,color=quantumviolet] (4.,-2.)-- (4.,-1.2);
\draw [line width=2.pt,color=quantumviolet] (4.,-1.2)-- (-1.,-1.2);
\draw [line width=2.pt,color=quantumviolet] (-1.,-1.2)-- (-1.,-2.);

\draw [line width=1.pt, dashed, color=black] (1.45, 0.)-- (1.45, -1.);
\renewcommand{\mynumber}{-0.18}
\draw [line width=1.pt, dashed, color=black] (1.45, -3.+ \mynumber)-- (1.45, -2.+ \mynumber);

\draw (-1.0, -0.4) node[anchor=east] {$x$};
\draw (3.9, -0.43) node[anchor=west] {$y$};
\draw (-1.0, -2.6) node[anchor=east] {$a$};
\draw (3.9, -2.6) node[anchor=west] {$b$};

\draw [color=quantumviolet](1.45, -1.6) node {\textbf{Nonlocal box}};
\draw[color=black] (-2.2, -1.6) node[anchor=east] {{$\mathscr Alice$}};
\draw[color=black] (5., -1.6) node[anchor=west] {{$\mathscr Bob$}};

\end{tikzpicture}
	\caption{Representation of a nonlocal box.}
	\label{fig:NLB}
\end{figure}

Generally, nonlocal boxes are non-signalling, meaning that they respect the relativistic constraint of no faster-than-light communication between parties (although there is a recent interest for partially-signalling scenarios~\cite{RBBGMRW20}).
These \emph{non-signalling boxes} form a set $\NS$ defined by the following non-negativity and normalization equations, which ensure $\P$ is a well-defined conditional probability distribution:
\begin{align}
	&\forall a,b,x,y\in\{0,1\}, \quad\hspace{0.3cm} \label{eq: non-signalling is a joint proba}
	\P(a,b\,|\,x,y)\geq0 \quad\quad\text{and}\quad\quad\sum_{a,b}\P(a,b\,|\,x,y)=1\,,\quad\quad\\
\intertext{together with the conditions that the marginals of each party are independent of the other party's question:}
	&\forall a,x\in\{0,1\}, \quad\quad\quad \label{eq: non-signalling 1}
	\sum_{b\in\{0,1\}} \P(a, b\,|\,x,0) = \sum_{b\in\{0,1\}} \P(a, b\,|\,x, 1) =: \P(a\,|\,x)\,,\\
	&\forall b, y\in\{0,1\}, \quad\quad\quad \label{eq: non-signalling 2}
	\sum_{a\in\{0,1\}} \P(a, b\,|\,0,y) = \sum_{a\in\{0,1\}} \P(a, b\,|\,1,y) =: \P(b\,|\,y)\,.
\end{align}
The physical interpretation of \eqrefprop[Equations]{eq: non-signalling 1} and~\eqrefprop{eq: non-signalling 2} is that Alice and Bob are space-like separated, so it would take more time for a light ray to move from Bob to Alice than the time needed for Alice to do her protocol and to receive her output of the box, and therefore Alice's marginal does not depend on Bob's input.

The best-known example of a non-signalling box is the $\PR$ box, named after Popescu and Rohrlich~\cite{PR94}. This box is designed to perfectly win at the $\CHSH$-game~\cite{CHSH69}, \ie the box produces outputs $a,b$ such that $a\oplus b=xy$ with probability $1$, where the symbol ``$\oplus$" stands for the sum modulo $2$.\footnote{From a computing perspective, the $\PR$ box turns a distributed AND ($xy$) into a distributed XOR ($a\oplus b$), where ``distributed" means that the computation requires both parties (Alice knows $x$ and $a$, and Bob knows $y$ and $b$).} More precisely, given an input pair $(x,y)$, there are two possibilities for the outputs: either $(a=0, b=xy)$ or $(a=1, b=xy\oplus1)$, each being output with probability $1/2$ by the $\PR$ box.
In fact, it is possible to show that the $\PR$ box thus defined is the only box of $\NS$ that perfectly wins at the $\CHSH$-game.
Let us give two other examples of boxes:
(i) the \emph{fully mixed} box~$\I$, which outputs purely random bits $a$ and $b$; and
(ii) the deterministic boxes $\P_0,\P_1$ that always output $(0,0),(1,1)$ independently of the inputs.
All these boxes are in $\NS$ and can be written as conditional probability distributions:
\begin{equation} \label{eq:definition-of-PR-P0-P1-I}
\begin{array}{ll}
	\PR\big(a,b\,|\,x,y\big)
	\,:=\,
	{\textstyle\frac12}\1_{a\oplus b=x y}\,,\quad\quad\quad
	&
	\I\big(a,b\,|\,x,y\big)
	\,:=\,
	{\textstyle\frac14}\,,
	\\
	\\
	\P_0\big(a,b\,|\,x,y\big)
	\,:=\,
	\1_{a=b=0}\,,
	&
	\P_1\big(a,b\,|\,x,y\big)
	\,:=\,
	\1_{a=b=1}\,,
\end{array}
\end{equation}
where the symbol ``$\1$" stands for the indicator function.
Two other important classes of correlations are the \emph{local} set $\LL$ and the \emph{quantum} (tensor) set $\QQ$, with strict inclusions $\LL\subsetneq \QQ \subsetneq \NS$, defined as:
\begin{align*}
	& \LL \,=\, \Bigg\{ \P(a,b\, |\, x,y)=\int_\lambda \mathbb P_A(a\,|\, x, \lambda)\, \mathbb P_B(b\,|\,y, \lambda)\, \mu(\lambda) \,:\, \mathbb P_A, \mathbb P_B, \mu \text{ are probability distributions} \Bigg\}\,,\\
	& \QQ \,=\Bigg\{ \P(a,b\, |\, x,y)= \langle \Psi| \Big(E^x_a \otimes F^y_b\Big) |\Psi\rangle \,:\!
	\begin{array}{l}
		\text{\small$\{E^x_a\}_a,\{F^y_b\}_b \text{ are POVMs over Hilbert spaces } \H_A, \H_B$}\,,\\
		\text{\small$|\Psi\rangle \text{ is a vector of } \H_A\otimes\H_B \text{ of norm $1$}$}
	\end{array}
	\Bigg\}\,.
\end{align*}
For more details on correlation sets and their separation, see~\cite{JNVWY20}.

Interestingly, some nonlocal boxes collapse what is called \emph{communication complexity} (CC), a notion introduced by Yao in~\cite{Yao79}
and reviewed in~\cite{KN96, RY20}
that quantifies the difficulty of performing a distributed computation. Say we want to evaluate a Boolean function $f:\{0,1\}^n\times\{0,1\}^m\to\{0,1\}$ using two distant computers, where the first computer receives as input $X\in\{0,1\}^n$, and the other computer receives as input $Y\in\{0,1\}^m$.
The CC of $f$ is then defined as the minimal number of bits that the computers need to communicate in order for the first computer to output the value $f(X,Y)$.
For instance, when $n=m=2$, $X=(x_1, x_2)$, $Y=(y_1, y_2)$, the CC of $f_1:= x_1\cdot(y_1 \oplus y_2)$ equals~$1$, using the communication bit $y_1 \oplus y_2$, whereas it is possible to show that the CC of $f_2:= x_1\cdot y_1 \oplus x_2\cdot y_2$ equals~$2$, using communication bits $y_1$ and $y_2$; therefore $f_2$ is more complex than $f_1$ in the sense of CC.
Yao also introduced in~\cite{Yao79} a probabilistic version of CC, in which the computers can access shared randomness, and where for all $X,Y$ the first computer has to output the correct value $f(X,Y)$ with probability at least $p>1/2$ ($p$ being independent of $X$ and $Y$).
Now, if a nonlocal box $\P$ is used in the protocol to compute the value $f(X,Y)$, we say that the box $\P$  \emph{collapses} communication complexity if there exists a fixed $p>1/2$ for which \emph{any} Boolean function $f$, with arbitrary input size, can be correctly computed with only \emph{one} bit of communication and probability $p$.
In this definition, an arbitrary number of copies of the box $\P$ can be used in the protocol.
Such a collapse is strongly believed to be unachievable in Nature since it would imply the absurdity that a single bit of communication is sufficient to distantly estimate any value of any Boolean function $f$~\cite{vD99, BBLMTU06, BS09, BG15}.
For more details on the link between nonlocal boxes and communication complexity, see~\cite{Botteron22}.

\vspace{0.4cm}
\textbf{Open question.}~Among the four examples of boxes listed in \eqrefprop[Equation]{eq:definition-of-PR-P0-P1-I}, only the $\PR$ box collapses communication complexity~\cite{vD99}, meaning that this box is very ``powerful".
We also know that some noisy versions of the $\PR$ box collapse CC for different types of noise~\cite{BBLMTU06, BS09, BMRC19, BBP23, EWC22PRL}.
On the other hand, we know that quantum correlations do \emph{not} collapse communication complexity~\cite{CvDNT99}, and neither does a slightly wider set named ``almost quantum correlations"~\cite{NGHA15}.
To this day, the question is still open whether the remaining non-signalling boxes are collapsing, meaning that there is still a gap to be filled.
We refer to \cite[Fig. 2]{BBP23} for a figure that summarizes the situation.

\vspace{0.4cm}
{\textbf{Recent results.}~Two recent papers make progress on the question above.
In~\cite{BMRC19} Brito, Moreno, Rai, and Chaves study correlation distillation in \emph{quantum voids}~\cite{RDBC19}, which are subsets of a face of $\NS$ where all nonlocal points are non-quantum. They prove strong distillation properties in $1$- and $2$- dimensional quantum voids and deduce that these regions collapse communication complexity, which partially answers the open question.
More recently, in~\cite{EWC22PRL}, Eftaxias, Weilenmann, and Colbeck propose a \emph{sequential algorithm} to find a suitable sequence of wirings to collapse communication complexity, where a \emph{wiring} is defined as a connection between boxes that allows the creation of a new box out of copies of a box (see \refprop[Section]{section:Algebra-of-boxes}). This allows us to numerically determine a collapsing region of nonlocal boxes, which is again a partial answer to the open question.
}

\vspace{0.4cm}
\textbf{Our Results.}~{We provide a new mathematical framework and algorithms in working towards addressing the open question. The ideas are based in part on the M.Sc.~thesis of one of the authors~\cite{Botteron22}.
\begin{enumerate}[label=(\roman*)]

	\item \label{our-result-1}We introduce a new framework that we call the \emph{algebra of boxes}. After recalling the definition of a wiring~$\W$, we use it to introduce a \emph{product of boxes} $\P\boxtimes_\W\Q$. This leads to a natural embedding of the non-signalling set $\NS$ in an algebra, which we call an \emph{algebra of boxes} and for which we characterize associativity and commutativity (see \refprop[Proposition]{prop:Characterization-of-commutativity-and-associativity}). This gives an algebraic perspective on protocols for correlation distillation---for instance, the non-associativity of the algebra of boxes tells us that the order in which the boxes are wired matters.
	
	\item \label{our-result-2} This framework gives rise to the fascinating notion of what we call the \emph{orbit of a box}. The orbit of~$\P\in\NS$ is roughly the set of all possible boxes that can be produced by wiring arbitrarily many copies of $\P$. This allows interesting visualizations of the hidden structure of boxes (see \refprop[Figure]{figure: orbit of a box}), and surprisingly we observe that these orbits satisfy strong alignment and parallelism properties as shown in \refprop[Theorem]{theo: alignment and parallelism} and \refprop[Corollary]{coro:parallelism}. Moreover, we derive the expression of the highest $\CHSH$-valued box of the (tilted) orbit in \refprop[Theorem]{theo: highest box}, which explains the numerical observation reported in~\cite[Supplementary Material, II]{EWC22PRL}, and for which we derive an insightful linear-time algorithm that is exponentially more efficient compared to the naive exponential-time computation of the entire orbit. In addition, we recover in \refprop[Theorem]{theo: new known collapsing boxes} a similar result as in~\cite{EWC22PRL} stating that those methods lead to finding collapsing boxes via the recursive application of the multiplication $\cdot\boxtimes\P$ on the right.
	
	\item \label{our-result-3}We provide algorithms in our GitHub page~\cite{GitHub-algebra-of-boxes} for the following task: given a box $\P$ that we want to show is collapsing, find an appropriate wiring $\W$ such that the orbit contains a collapsing box. The idea is to repeat several times in parallel a variant of the Gradient Descent Algorithm in order to find the most appropriate wiring $\W$.
	These algorithms allow us to recover in \refprop[Figure]{fig:numerical-new-collapsing-regions} similar new collapsing areas as in~\cite[Figure~3]{EWC22PRL}.
	
	\item \label{our-result-4}In \refprop[Theorems]{thm: the triangle PR-P0-P1 is collapsing} and \ref{thm:new-collapsing-triangles}, we show that our framework also allows us to recover some analytical results with a new proof based on the algebra of boxes: some triangles in the boundary of $\NS$ are collapsing~\cite{BMRC19}. To that end, algorithms of~\ref{our-result-3} above were performed in order to identify a convenient wiring.
	Moreover, in \refprop[Corollary]{coro:quantum-boxes-in-the-boundary-of-NS}, we recover a result from~\cite{RDBC19} with a new proof, based on communication complexity, showing that the triangle joining $\PR,\P_0,\P_1$ is a ``quantum void".

\end{enumerate}

}

\vspace{0.4cm}
{
\textbf{Further Comparison with Recent Results.} A contribution of our work is in providing a new algebraic framework for a unified perspective on three recent results~\cite{EWC22PRL, BMRC19, RDBC19}.
 Compared to~\cite{EWC22PRL}, our work concurrently and independently\footnote{Our results were first reported in the M.Sc.~thesis of one of the authors~\cite{Botteron22}, which appeared in the same month (June 2022) as the arXiv version of  \cite{EWC22PRL}.} derives the existence of new collapsing boxes using right multiplication (\refprop[Theorem]{theo: new known collapsing boxes})
 and we report a similar numerical result (\refprop[Figure]{fig:numerical-new-collapsing-regions}). However, as detailed in \refprop[Remark]{rem:two-other-existing-methods}, we note that the methods complement each other: as the authors of~\cite{EWC22PRL}, we implement an algorithm finding a non-constant sequence of wirings, but we also implement an algorithm finding a constant sequence of wirings for analytical purposes.
Moreover, as mentioned above, our \refprop[Theorem]{theo: highest box}\,\,analytically proves a numerical observation reported in~\cite[Supplementary Material, II]{EWC22PRL}.
As for~\cite{BMRC19}, we recover some of their results in \refprop[Theorems]{thm: the triangle PR-P0-P1 is collapsing} and~\ref{thm:new-collapsing-triangles}, and although our proof reproduces previously-known analytical collapsing areas, we view our contribution as a new approach based on the algebra of boxes.  Regarding~\cite{RDBC19}, we recover one of their result in \href{coro:quantum-boxes-in-the-boundary-of-NS}{Corollary~\ref{coro:quantum-boxes-in-the-boundary-of-NS}} with a new proof, based on communication complexity, showing that the triangle joining $\PR,\P_0,\P_1$ is a ``quantum void".
In summary, our new algebraic viewpoint unifies the results of~\cite{EWC22PRL} (see \refprop[Subsection]{subsec:numerical-results}),~\cite{BMRC19} (see \refprop[Subsection]{subsection:analytical-results}), and~\cite{RDBC19} (see \refprop[Subsection]{subsection:analytical-results}).
}

\vspace{0.4cm}
\textbf{Structure.}~This work is divided into four sections, a conclusion, and some appendices:
\begin{itemize}
	\setlength\itemsep{0.2em}
	\item \refprop[Section]{section:Algebra-of-boxes}: given a wiring $\W$, we introduce the product of nonlocal boxes $\P\boxtimes_\W\Q$ and we study the new framework of \emph{algebra of boxes}.
	\item \refprop[Section]{section:Orbit-of-a-box}: we define the notion of the \emph{orbit} of a box $\P$, which consists of all boxes produced using copies of $\P$ and the product $\boxtimes$  previously defined, and we investigate its surprising geometric structure.
	\item \refprop[Section]{sec:numerical-approach}: we present algorithms for the following task: given a box $\P$ that we want to show is collapsing, find an appropriate wiring $\W$ such that the orbit contains a collapsing box. These algorithms are based on Gradient Descents methods and are entirely accessible via our GitHub page~\cite{GitHub-algebra-of-boxes}.
	\item \refprop[Section]{section: Analytical Proofs of the New Collapsing Regions}: we find collapsing boxes in two different ways: (i)~numerically, using the algorithms of the previous section,
	and (ii)~analytically, using the algebra of boxes and the orbit of a~box.
	\item {\sffamily\hyperref[conclusion]{Conclusion}}: we conclude with some discussion, open problems, and avenues for future work.
	\item {\sffamily\hyperref[sec:drawing-of-some-orbits]{Appendices}}: we complete this work with supplementary figures and proofs.
\end{itemize}

\bib

	\section{Algebra of Boxes}
	\label{section:Algebra-of-boxes}
	
The set of non-signalling boxes $\NS$ is the compact convex subset of the vector space $\B=\big\{\P:\{0,1\}^4\to\R \big\}$ satisfying \eqrefprop[Equations]{eq: non-signalling is a joint proba}, \eqref{eq: non-signalling 1},~\eqref{eq: non-signalling 2}. In this section, we propose to endow the vector space $\B$ with a multiplication $\boxtimes_\W$, so that $\B$ becomes an algebra that we call \emph{algebra of boxes} and that we denote $\B_\W$.
To that end, we recall the notion of \emph{wiring} $\W$ (deterministic then mixed), which, for the sake of simplicity, we define for only \emph{two} boxes being connected --- see \refprop[Remark]{rem: not the most general framework, but it is enough} for more generality. Then we provide some typical examples of wirings from the literature, and we finally introduce the algebra of boxes and characterize its associativity and commutativity.

		\subsection{Intuition Behind Wirings}

Given two non-signalling boxes $\P$ and $\Q$, it is possible to build a new box by \emph{wiring} them together.
This notion of wiring has found a great interest in the last two decades, especially with the following two goals:
(i)~attempting nonlocality distillation, \ie we want to build a box that is ``strongly nonlocal" starting from some boxes that are ``weakly nonlocal"~\cite{DW08, FWW09, BS09, HR10, EW14, BG15, BMRC19, EWC22PRL, NSSRRB22PRL, EWC22b};
(ii)~finding sets that are closed under wirings, because it is argued that a consistent physical theory should, in principle, be closed under natural simple operations as wirings~\cite{ABLPSV09, NW09, LVN14, BG15, NGHA15}.

As one might guess, a wiring simply connects some outputs to some inputs under some rules, and it applies some pre- and post- processing operations to the carried bits. An example of wiring is presented in  \refprop[Figure]{figure: example of wiring}~(a), where the wiring indeed connects some outputs to some inputs, but is counter-intuitive at first, since Alice and Bob do not use their share of the boxes in the same order: while Alice uses $\P$ then $\Q$, Bob uses $\Q$ then $\P$.
This independence on the choice of the box order for each player generalizes quantum mechanics, in the sense that if Alice and Bob were sharing two entangled pairs instead of two nonlocal boxes, Alice would be able to measure her first particle and then the second one, while Bob would be able to do the converse, and they would still receive the outputs ``instantaneously".
Now, as in the quantum case, Alice receives an answer from the box $\P$ instantaneously even if Bob has not yet inputted a bit in his side of $\P$, and she can use the output $a_1$ as a parametrization for the input $x_2$ of the box $\Q$; similarly for Bob. This ``instantaneous-answer" property of a box is typical of non-signalling correlations, as modelled by \eqrefprop[Equations]{eq: non-signalling 1} and~\eqref{eq: non-signalling 2} saying that Alice's marginal is independent of Bob's input, and vice-versa.
Note that a wiring cannot link Alice's side to Bob's side, nor the opposite, since otherwise it could create a signalling box: there would be communication between parties.

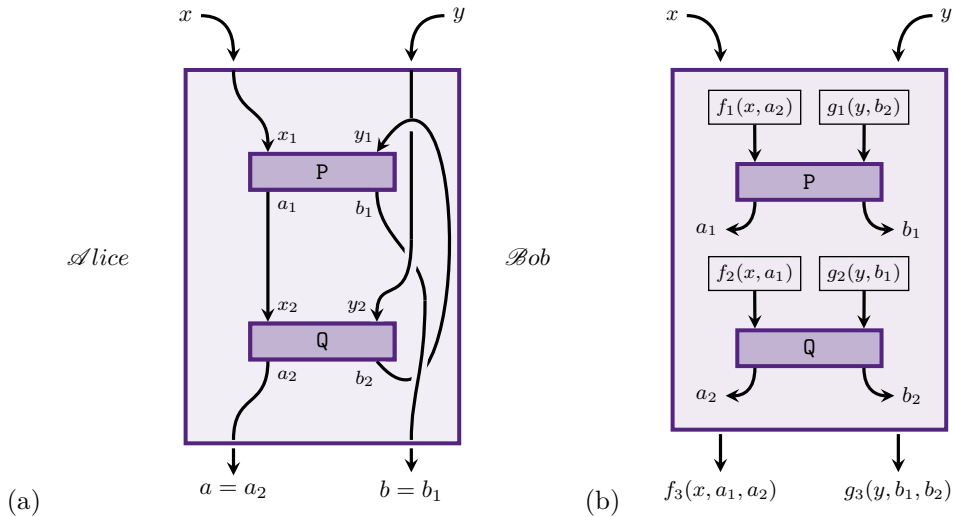
\begin{figure}[h]
	\hspace{-0.8cm}
	(a)
	{
\begin{tikzpicture}[scale=0.9, every node/.style={scale=0.9},
small-NLB/.style={rectangle, draw=quantumviolet, fill=quantumviolet!35, line width=1.5pt, minimum width=60, minimum height=15, anchor=center},
big-NLB/.style={rectangle, draw=quantumviolet, fill=quantumviolet!8, line width=1.5pt, minimum width=60, minimum height=15},
operation/.style={rectangle, draw=black, line width=0.5pt},
my-arrow/.style={->, very thick, >=stealth},
my-line-violet/.style={line width=0.5em, color=quantumviolet!8},
my-line/.style={very thick},
]

	\draw[big-NLB] (0., -1) rectangle (4., 4.5);
		
	\node[small-NLB] at (2, 3) (P) {\normalsize$\P$};
	\node[small-NLB] at (2, 0.5) (Q) {\normalsize$\Q$};
	\node at (0., 5.3) (Alice1) {$x$};
	\node at (0.7, 4.5) (Alice2) {};
	\node at (0.7, -0.95) (Alice7) {};
	\node at (0.7, -1.7) (Alice8) {$a=a_2$};
	\node at (4., 5.3) (Bob1) {$y$};
	\node at (3.3, 4.5) (Bob2) {};
	\node at (3.3, 2.) (Bob3) {};
	\node at (3.5, 1.1) (Bob4) {};
	\node at (4.2, 5.4) (Bob4bis) {};
	\node at (4.2, -1.5) (Bob6) {};
	\node at (4-0.7, -0.95) (Bob7) {};
	\node at (4-0.7, -1.7) (Bob8) {$b=b_1$};
	
	\draw[my-arrow] (Alice1) .. controls +(right:2em) and +(up:1em).. (Alice2);
	\draw[my-arrow] (Alice2.center) .. controls  +(down:2em) and +(up:2em) ..  (P.160) node[above right]{\footnotesize$x_1$};
	\draw[my-arrow] (P.200) node[below right]{\footnotesize$a_1$} to (Q.160) node[above right]{\footnotesize$x_2$};
	\draw[my-line] (Q.200) node[below right]{\footnotesize$a_2$} ..controls +(down:2em) and + (up:2em) .. (Alice7.center);
	\draw[my-arrow] (Alice7) to (Alice8);
	
	\draw[my-arrow] (Bob1) .. controls +(left:2em) and +(up:1em).. (Bob2);
	\draw[my-line] (Bob2.center) to (Bob3.center);
	\draw[my-line-violet] (Q.340)  .. controls (Bob6) and (Bob4bis) .. (P.20);
	\draw[my-arrow] (Q.340) node[below left]{\footnotesize$b_2$\!\!}  .. controls (Bob6) and (Bob4bis) .. (P.20) node[above left]{\footnotesize$y_1$\!\!};
	\draw[my-line] (P.340) node[below left]{\footnotesize$b_1$\!\!} .. controls +(down:2em) and +(up:2em).. (Bob4.center);
	\draw[my-line-violet] (Bob4.center) .. controls +(down:3em) and +(up:2em).. (Bob7.center);
	\draw[my-line] (Bob4.center) .. controls +(down:3em) and +(up:2em).. (Bob7.center);
	\draw[my-line-violet] (Bob3) .. controls +(down:3.em) and +(up:2em) .. (Q.20);
	\draw[my-arrow] (Bob3.center) .. controls +(down:3.em) and +(up:2em) .. (Q.20) node[above left]{\footnotesize$y_2$};
	\draw[my-arrow] (Bob7) to (Bob8);
	
	\node[small-NLB] at (2, 3) (P) {\normalsize$\P$};
	\node[small-NLB] at (2, 0.5) (Q) {\normalsize$\Q$};
	
	\draw[color=black] (-0.7, 1.75) node[anchor=east] {{$\mathscr Alice$}};
	\draw[color=black] (4.55, 1.75) node[anchor=west] {{$\mathscr Bob$}};

\end{tikzpicture}
}
	(b)
	{
\newcommand{\shiftP}{-0.65}
\newcommand{\shiftQ}{-0.8}
\small
\begin{tikzpicture}[scale=0.9, every node/.style={scale=0.9},
small-NLB/.style={rectangle, draw=quantumviolet, fill=quantumviolet!35, line width=1.5pt, minimum width=60, minimum height=15, anchor=center},
big-NLB/.style={rectangle, draw=quantumviolet, fill=quantumviolet!9, line width=1.5pt, minimum width=60, minimum height=15},
operation/.style={rectangle, draw=black, line width=0.5pt},
my-arrow/.style={->, very thick, >=stealth},
my-arrow-violet/.style={->, line width=0.5em, color=quantumviolet!8},
my-line/.style={very thick},
]

	\draw[big-NLB] (0., -1+0.2) rectangle (4., 4.5);
		
	\node[small-NLB] at (2, 3.5+\shiftP) (P) {\normalsize$\P$};
	\node[small-NLB] at (2, 1.2+\shiftQ) (Q) {\normalsize$\Q$};
	\node at (0., 5.3) (Alice1) {$x$};
	\node at (0.7, 4.5) (Alice2) {};
	\node[operation] at (1.2, 4.6+\shiftP) (Alice3) {\footnotesize$f_1(x,a_2)$};
	\node at (0.5, 2.8+\shiftP) (Alice4) {$a_1$};
	\node[operation] at (1.2, 2.3+\shiftQ) (Alice5) {\footnotesize$f_2(x,a_1)$};
	\node at (0.5, 0.5+\shiftQ) (Alice6) {$a_2$};
	\node at (0.7, -0.95+0.2) (Alice7) {};
	\node at (0.7, -1.7) (Alice8) {$f_3(x, a_1, a_2)$};
	\node at (4., 5.3) (Bob1) {$y$};
	\node at (3.3, 4.5) (Bob2) {};
	\node[operation] at (4-1.2, 4.6+\shiftP) (Bob3) {\footnotesize$g_1(y,b_2)$};
	\node at (4-0.5, 2.8+\shiftP) (Bob4) {$b_1$};
	\node[operation] at (4-1.2, 2.3+\shiftQ) (Bob5) {\footnotesize$g_2(y,b_1)$};
	\node at (4-0.5, 0.5+\shiftQ) (Bob6) {$b_2$};
	\node at (4-0.7, -0.95+0.2) (Bob7) {};
	\node at (4-0.7, -1.7) (Bob8) {$g_3(y,b_1,b_2)$};
	
	\draw[my-arrow] (Alice1) .. controls +(right:2em) and +(up:1em).. (Alice2);
	\draw[my-arrow] (Alice3.south)  .. controls  +(down:1em) and +(up:0.8em) ..  (P.160);
	\draw[my-arrow] (P.200)  .. controls  +(down:1em) and +(right:0.8em) ..  (Alice4.east);
	\draw[my-arrow] (Alice5.south)  .. controls  +(south:1em) and +(up:0.8em) ..  (Q.160);
	\draw[my-arrow] (Q.200)  .. controls  +(down:1em) and +(right:0.8em) ..  (Alice6.east);
	\draw[my-arrow] (Alice7) to (Alice8);
	
	\draw[my-arrow] (Bob1) .. controls +(left:2em) and +(up:1em).. (Bob2);
	\draw[my-arrow] (Bob3.south)  .. controls  +(down:1em) and +(up:0.8em) .. (P.20);
	\draw[my-arrow] (P.340)  .. controls  +(down:1em) and +(left:0.8em) ..  (Bob4.west);
	\draw[my-arrow] (Bob5.south)  .. controls  +(down:1em) and +(up:0.8em).. (Q.20);
	\draw[my-arrow] (Q.340)  .. controls  +(down:1em) and +(left:0.8em) ..  (Bob6.west);
	\draw[my-arrow] (Bob7) to (Bob8);

\end{tikzpicture}
}
	\centering
	\caption{(a) Example of wiring between two boxes $\P$ and $\Q$.
	(b) General wiring between two boxes $\P$ and $\Q$.
	}
	\label{figure: example of wiring}
	\label{figure: general wiring}
\end{figure}

	\subsection{Deterministic Wirings}

Two boxes $\P$ and $\Q$ can be wired as in \refprop[Figure]{figure: general wiring}~(b),
using functions $f_i$ and $g_j$ depending on the global entries $x$ and $y$ and on the outputs $a_k$ and $b_\ell$ of the boxes.
Nevertheless, to be a valid wiring, the inputs on Alice's side must be in a valid order: the input $x_2$ of $\Q$ can depend on the output $a_1$ of $\P$ only if the input $x_1$ of $\P$ does not depend on the output $a_2$ of $\Q$; the same should also hold on Bob's side.
In other words, the functions $f_1(x,a_2)$ and $f_2(x,a_1)$ cannot both depend on $a_2$ and $a_1$ respectively for the same value of $x$, and similarly for $g_1(y, b_2)$ and $g_2(y, b_1)$.
These conditions are formalized in \eqref{eq: condition 1 for valid wiring} and \eqref{eq: condition 2 for valid wiring} of the following definition:

\begin{definition}[Deterministic wiring]
	A \emph{deterministic wiring} $\W$ between two boxes $\P,\Q\in\NS$ consists in six Boolean functions $f_1, f_2, g_1, g_2:\{0,1\}^2\to\{0,1\}$ and $f_3, g_3:\{0,1\}^3\to\{0,1\}$ satisfying the \emph{non-cyclicity conditions}:
	\begin{align}
		\forall x, \quad\quad     \label{eq: condition 1 for valid wiring}
		&\big( f_1(x,0) - f_1(x,1) \big) \big( f_2(x,0) - f_2(x,1) \big) \,=\,0\,,\\
		\forall y, \quad\quad	\label{eq: condition 2 for valid wiring}
		&\big( g_1(y,0) - g_1(y,1) \big) \big( g_2(y,0) - g_2(y,1) \big) \,=\,0\,.
	\end{align}
\end{definition}

Given a wiring $\W$ and two boxes $\P,\Q\in\NS$, we obtain a new box that we denote $\P\boxtimes_\W\Q$.
Formally, this new box is defined as the following conditional probability distribution:
\begin{multline}   \label{eq: definition of the box product}
		\small
	  	\P\boxtimesW \Q(a,b\,|\,x,y)
		:=\,
		\hspace{-0.4cm}
		\sum_{a_1, a_2, b_1, b_2}
		\P\Big(a_1,\,b_1\,|\, {f_1(x,a_2)},\,{g_1(y, b_2)}\Big) \,
		\hspace{-0.2cm}\times\,\Q\Big(a_2,\,b_2\,|\, {f_2(x,a_1)},\,{g_2(y, b_1)}\Big) \\
		\small
		\hspace{-0.2cm}\times\,\1_{a = f_3(x,a_1,a_2)} \,\times\,\1_{b = g_3(y,b_1,b_2)}\,,
	\end{multline}
where the symbol $\1$ stands for the indicator function, taking value $1$ if the indexed condition is satisfied, and $0$ otherwise.
It is important to specify the condition $\P,\Q\in\NS$, since in that case $\P\boxtimes_\W\Q$ is indeed a conditional probability distribution as shown in the proof of \refprop[Fact]{fact: NS closed under deterministic wirings}; otherwise, if one requires only the condition on $\P,\Q$ to be conditional probability distributions (not necessarily lying in $\NS$), then it might happen that the product $\P\boxtimes_\W\Q$ is not a well-defined probability distribution: consider for example $\P=\Q=\1_{a=y}\1_{b=x}$ and the deterministic wiring $\W=(f_1=x, f_2=a_1, g_1=b_2, g_2=y, f_3=0, g_3=0)$.

\begin{definition}[Closed under wirings]
A set $X\subseteq\NS$ is said to be \emph{closed under wirings} if for all boxes $\P, \Q$ in $X$ and all wirings~$\W$, the new box $\P\boxtimes_\W\Q$ is in $X$ as well.\footnote{Notice that there exists as well a more general definition involving $k$ boxes and $m$ parties, but our manuscript is restricted to the simpler case of $k=m=2$, which is the reason why we give a simpler definition here. In the general framework, many sets are known to be closed under wirings~\cite{ABLPSV09, NW09, LVN14, BG15, NGHA15}.}
\end{definition}

For the sake of completeness, we recall the fact that the non-signalling polytope $\NS$ is an example of a set that is closed under wirings.

\begin{fact}\hspace{-0.2cm}{\normalfont\cite{ABLPSV09}}   \label{fact: NS closed under deterministic wirings}
	$\NS$ is closed under deterministic wirings.
\end{fact}

\begin{proof}
	We need to show that the box $\P\boxtimes_\W\Q$ given in \eqref{eq: definition of the box product} is a well-defined conditional probability distribution that satisfies the non-signalling conditions \eqref{eq: non-signalling 1} and \eqref{eq: non-signalling 2}.
	First, by non-negativity of $\P$ and $\Q$, the new box $\P\boxtimes_\W\Q$ is non-negative as well.
	Now, fix $x,y\in\{0,1\}$.
	The non-cyclicity conditions \eqref{eq: condition 1 for valid wiring} and \eqref{eq: condition 2 for valid wiring} tell us that $f_1$ or $f_2$ is constant in the second variable, and similarly for $g_1$ and $g_2$.
	Without loss of generality, up to changing the roles of both $f_1$, $g_1$ with respectively $f_2,g_2$, we only need to consider the following two non-exclusive cases:
	\begin{itemize}
		\item Case~1: the functions $f_1(x, a_2)$ and $g_1(y, b_2)$ are constant in the second variable, and we denote them $f_1(x)$ and $g_1(y)$;
		\item Case~2: the functions $f_2(x, a_1)$ and $g_1(y, b_2)$ are constant in the second variable, and we denote them $f_2(x)$ and $g_1(y)$.
	\end{itemize}
	In Case~1, we see that coefficients sum to one by normalization of $\P$ and $\Q$:
	\[
		\sum_{a,b}  \P\boxtimesW \Q(a,b\,|\,x,y)
		\,=\,
		\sum_{a_1, b_1}
		\P\big(a_1,\,b_1\,|\, {f_1(x)},\,{g_1(y)}\big)
		\times\Bigg( \small
		\underbrace{
		\sum_{a_2, b_2}
		\Q\Big(a_2,\,b_2\,|\, {f_2(x,a_1)},\,{g_2(y, b_1)}\Big)
		}_{=1}
		\Bigg)
		\,=\,
		1\,.\hspace{-0.4cm}
	\]
	In Case~2, using the non-signalling conditions on $\P$ and $\Q$, we see that coefficients sum to one again:
	\begin{align*}
		\sum_{a,b}  \P\boxtimesW \Q(a,b\,|\,x,y)
		\,=\,
		&\small
		\sum_{a_2, b_1}\!
		\Bigg(\!
		\underbrace{
		\sum_{a_1}
		\P\big(a_1,\,b_1\!\,|\,\! {f_1(x,\! a_2\!)},\,{g_1(y)}\big)
		}_{=\, \P(b_1\,|g_1(y))}\!
		\Bigg)\!
		\Bigg(\!
		\underbrace{
		\sum_{b_2}
		\Q\Big(a_2,\,b_2\!\,|\, \!{f_2(x)},\,{g_2(y,\! b_1\!)}\Big)
		}_{=\, \Q(a_2\,|f_2(x))}\!
		\Bigg)
		\\
		\,=\,
		&
		\Bigg( \sum_{b_1} \P\big(b_1\,|g_1(y)\big) \Bigg)
		\,
		\Bigg( \sum_{a_2} \Q\big(a_2\,|f_2(x)\big) \Bigg)
		\,=\,
		1\,.
	\end{align*}
	Hence $\P\boxtimes_\W\Q$ is a conditional probability distribution.
	It only remains to check that $\P\boxtimes_\W\Q$ satisfies the non-signalling conditions \eqref{eq: non-signalling 1} and \eqref{eq: non-signalling 2}. Fix $x,a\in\{0,1\}$. In Case~1, we have for all $y\in\{0,1\}$:
	\begin{align*}
		\hspace{0.2cm}\sum_{b}  \P\boxtimesW \Q(a,b\,|\,x,y)
		\,=\, &
		\sum_{a_1,a_2} \P\big(a_1\,|\, f_1(x)\big)\,
		\Q\big(a_2\,|\, f_2(x, a_1)\big)\,
		\1_{a=f_3(x,a_1,a_2)}
		\,=:\,
		\P\boxtimes_W\Q(a\,|\,x)\,,
	\end{align*}
	\ie the result does not depend on $y$,
	which means that the marginal in $b$ is well-defined.
	This is similar in Case~2, changing $f_1(x)$ into $f_1(x, a_2)$ and $f_2(x, a_1)$ into $f_2(x)$.
	Hence the first non-signalling condition \eqref{eq: non-signalling 1} is satisfied, and the other one \eqref{eq: non-signalling 2} follows in a similar way.
\end{proof}

		\subsection{Mixed Wirings}

Using local randomness, one can generalize the class of deterministic wirings to the one of \emph{mixed wirings}. The difference is that the functions $f_i$ and $g_j$ take values in $[0,1]$ instead of $\{0,1\}$.
For instance, if $f_1(x,a_1)=p\in[0,1]$ for some fixed bits $x$ and $a_1$, it means that Alice uses a Bernoulli distribution $\mathcal B(p)$ to input the bit $1$ with probability $p$, or the bit $0$ with probability $1-p$.
In other words, we have $32$ Bernoulli variables $(B_1, \dots, B_{32})$, whose parameters are stored in $\W=\big(\text{\small $f_1(0,0), f_1(0,1),  \dots, g_3(1,1,1)$}\big)\in\R^{32}$, and the box product $\P\boxtimes \Q$ becomes the expected value of the deterministic wirings:
\begin{equation}  \label{eq:wiring-as-an-expectation}
	\P\boxtimesW\Q \,=\, \mathbb E\Big[ \P\underset{\{\!B_i\!\}}\boxtimes\Q \Big]\,.
\end{equation}
Note that this generalization of wirings does not change the definition of nonlocal boxes: the inputs and outputs of a box are still classical bits, not any real number between $0$ and $1$.
In order to ensure a well-defined local order for both Alice and Bob, we will need to add a dependence relation between the variables $B_i$, namely the non-cyclicity condition, as for the deterministic wirings:

\begin{definition}[Mixed wiring]     \label{def: mixed wiring}
	A \emph{mixed wiring} $\W$ between two boxes $\P,\Q\in\NS$ consists in six functions $f_1, f_2, g_1, g_2:\{0,1\}^2\to[0,1]$ and $f_3, g_3:\{0,1\}^3\to[0,1]$ satisfying the \emph{non-cyclicity conditions} \eqref{eq: condition 1 for valid wiring} and \eqref{eq: condition 2 for valid wiring}.
	Mixed wirings form a set that we denote $\WW$.
\end{definition}

The set of mixed wirings $\W$ is not convex because non-cyclicity conditions \eqref{eq: condition 1 for valid wiring} and \eqref{eq: condition 2 for valid wiring} are non-affine equalities.
For instance, consider the wirings $\W$, $\W'$ with all coefficients $0$ except the one corresponding to respectively $f_1(0,0)=1$, $f'_2(0,0)=1$; each of these wirings satisfies the non-cyclicity conditions \eqref{eq: condition 1 for valid wiring} and \eqref{eq: condition 2 for valid wiring}, but the mean $\W''=(\W+\W')/2$ does not:
\[
	\big( f''_1(0,0) - f''_1(0,1) \big) \big( f''_2(0,0) - f''_2(0,1) \big)
	\,=\,
	\big( 1/2 - 0 \big) \big( 1/2 - 0 \big)
	\, \neq\,
	0\,,
\]
hence the non-convexity of $\WW$.
The expression of $\P\boxtimes_\W\Q$ is the same as before, with the convention that $\P(a,b\,|\,\alpha,\beta)$ with $\alpha,\beta\in[0,1]$ means $(1-\alpha)(1-\beta)\P(ab\,|\,00) + (1-\alpha)\beta\P(ab\,|\,01) + \alpha(1-\beta)\P(ab\,|\,10) + \alpha\beta\P(ab\,|\,11)$, which gives:

\definecolor{quantumviolet}{HTML}{53257F}
\begin{multline}    \label{eq: P times Q with mixed wiring}
		\P\boxtimes_{\color{quantumviolet!95}\W} \Q(a,b\,|\,x,y)
		\\
		=\,
		\hspace{-0.5cm}
		\sum_{a_1, a_2, b_1, b_2\in\{0,1\}}
		\hspace{-0.3cm}
		\Big[
		\P\big(a_1,\,b_1\,|\, 0,\,0\big) (1-\text{\footnotesize\color{quantumviolet!95} $f_1(x,a_2)$})\,(1-\text{\footnotesize\color{quantumviolet!95} $g_1(y,b_2)$})
		+
		\P\big(a_1,\,b_1\,|\, 0,\,1\big) (1-\text{\footnotesize\color{quantumviolet!95} $f_1(x,a_2)$}) \,\text{\footnotesize\color{quantumviolet!95} $g_1(y,b_2)$}
		 \\
		+\, \P\big(a_1,\,b_1\,|\, 1,\,0\big) \, \text{\footnotesize\color{quantumviolet!95} $f_1(x,a_2)$}\,(1-\text{\footnotesize\color{quantumviolet!95} $g_1(y,b_2)$})
		+
		\P\big(a_1,\,b_1\,|\, 1,\,1\big) \, \text{\footnotesize\color{quantumviolet!95} $f_1(x,a_2)$}\, \text{\footnotesize\color{quantumviolet!95} $g_1(y,b_2)$} \Big]\\
		\times
		\Big[
		\Q\big(a_2,\,b_2\,|\, 0,\,0\big) (1-\text{\footnotesize\color{quantumviolet!95} $f_2(x,a_1)$})\,(1-\text{\footnotesize\color{quantumviolet!95} $g_2(y,b_1)$})
		+
		\Q\big(a_2,\,b_2\,|\, 0,\,1\big) (1-\text{\footnotesize\color{quantumviolet!95} $f_2(x,a_1)$})\, \text{\footnotesize\color{quantumviolet!95} $g_2(y,b_1)$}
		 \\
		+\, \Q\big(a_2,\,b_2\,|\, 1,\,0\big) \, \text{\footnotesize\color{quantumviolet!95} $f_2(x,a_1)$}\,(1-\text{\footnotesize\color{quantumviolet!95} $g_2(y,b_1)$})
		+
		\Q\big(a_2,\,b_2\,|\, 1,\,1\big) \, \text{\footnotesize\color{quantumviolet!95} $f_2(x,a_1)$}\, \text{\footnotesize\color{quantumviolet!95} $g_2(y,b_1)$} \Big]\\
		\times\,\Big[(1-\text{\footnotesize\color{quantumviolet!95} $f_3(x,a_1, a_2)$})\,\1_{a=0} + \text{\footnotesize\color{quantumviolet!95} $f_3(x,a_1, a_2)$}\,\1_{a=1} \Big]
		\times\,\Big[(1-\text{\footnotesize\color{quantumviolet!95} $g_3(y,b_1,b_2)$})\,\1_{b=0} + \text{\footnotesize\color{quantumviolet!95} $g_3(y,b_1,b_2)$}\,\1_{b=1} \Big]  \,.
	\end{multline}

Using the probabilistic point of view of \eqrefprop[Equation]{eq:wiring-as-an-expectation}, we see that a mixed wiring is a convex combination of deterministic wirings (the realization of Bernoulli variables being either $0$ or $1$). Hence, by linearity of the expectation $\mathbb E$, we deduce from \refprop[Fact]{fact: NS closed under deterministic wirings} that the set $\NS$ is closed under mixed wirings:

\begin{fact}   \label{lemma: NS closed under mixed wirings}
	$\NS$ is closed under mixed wirings.\qed
\end{fact}

\begin{remark}		\label{rem: not the most general framework, but it is enough}
Note that the formalism presented here is deliberately not the most general one, since this simpler version is enough to state our results in the next sections.
For a more general framework, see~\cite{BG15}.
For instance, here we require a deterministic local box order: knowing $x$, Alice perfectly knows which box she will use first, and similarly for Bob knowing $y$, but a more general mixed wiring would consist in setting a probability distribution on the different permutations of Alice's boxes and another one on Bob's boxes.
In addition, here we only defined \emph{wirings of depth~$2$}, but it is possible to have more complex wirings using $k$ nonlocal boxes, thus obtaining a \emph{wirings of depth~$k$}, \pierre{which play a significant role in the context of trivial communication complexity~\cite{EWC22PRL}.}
\end{remark}

		\subsection{Typical Examples of Wirings}
		\label{subsec:typical-examples-of-wirings}
		
We now review typical wirings that are studied in the literature. See \refprop[Figure]{fig: Drawings of the wirings} for an illustration of these wirings.
Note that all of these wirings are \emph{deterministic} wirings.

\newcommand{\scalewirings}{0.79}
\newlength{\hsep}
\setlength{\hsep}{-0.5cm}

\definecolor{color-box1}{RGB}{96, 0, 81}
\definecolor{color-box2}{RGB}{96, 0, 81}
\definecolor{color-box3}{RGB}{30, 70, 120}
\definecolor{color-box4}{RGB}{0, 96, 81}
\definecolor{color-box5}{RGB}{0, 96, 81}
\definecolor{color-box6}{RGB}{30, 70, 120}
\definecolor{color-box7}{RGB}{150, 30, 30}
\definecolor{color-box8}{RGB}{30, 70, 120}

\newcommand{\x}{4}
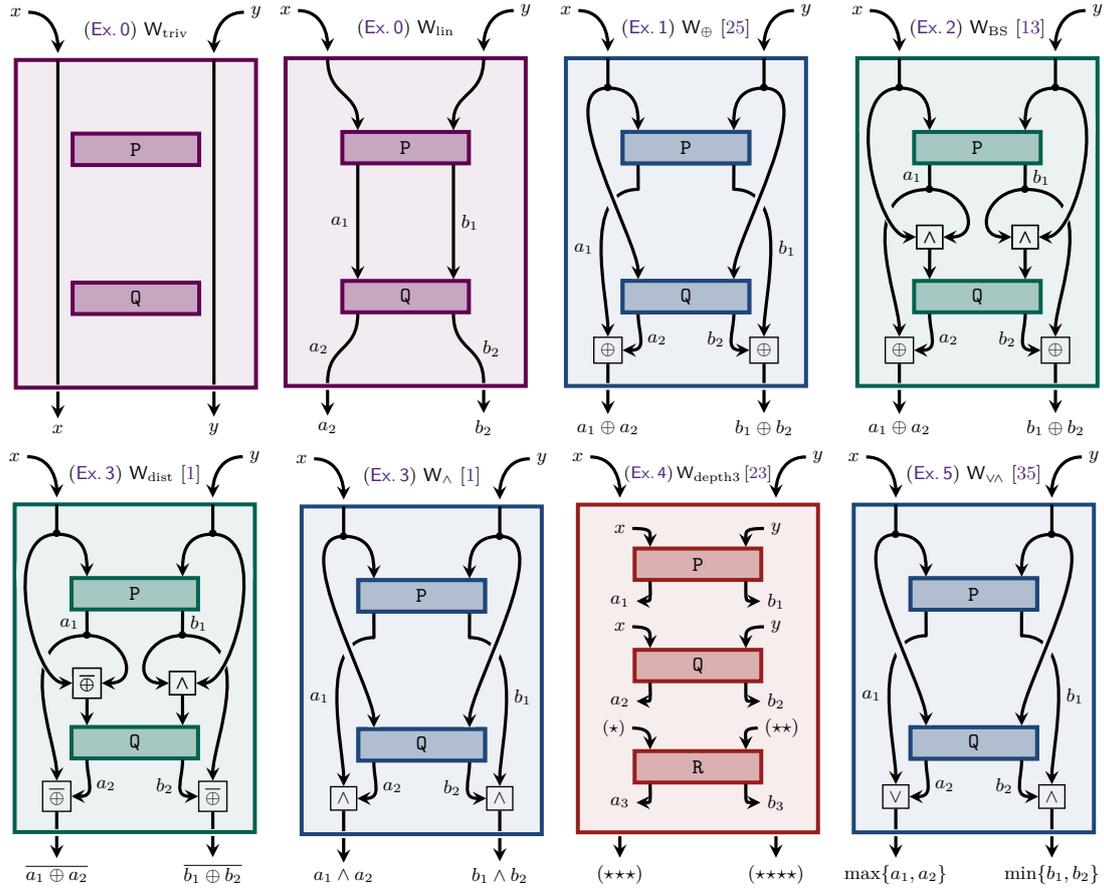
\begin{figure}[h]
	\centering
	{
\small
\begin{tikzpicture}[scale=\scalewirings, every node/.style={scale=\scalewirings},
small-NLB/.style={rectangle, draw=color-box1, fill=color-box1!35, line width=1.5pt, minimum width=60, minimum height=15, anchor=center},
big-NLB/.style={rectangle, draw=color-box1, fill=color-box1!8, line width=1.5pt, minimum width=60, minimum height=15},
operation/.style={rectangle, draw=black, line width=0.5pt},
my-arrow/.style={->, very thick, >=stealth},
my-arrow-violet/.style={->, line width=0.5em, color=color-box1!8},
my-line/.style={very thick},
]

	\draw[big-NLB] (0., -1) rectangle (4., 4.5);
	\node at (2, 5) {(\refprop[Ex.\!]{paragraph:example-trivial})~$\Wtriv$};
		
	\node[small-NLB] at (2, 3) (P) {\normalsize$\P$};
	\node[small-NLB] at (2, 0.5) (Q) {\normalsize$\Q$};
	\node at (0., 5.3) (Alice1) {$x$};
	\node at (0.7, 4.5) (Alice2) {};
	\node at (0.7, -0.95) (Alice7) {};
	\node at (0.7, -1.7) (Alice8) {$x$};
	\node at (4., 5.3) (Bob1) {$y$};
	\node at (3.3, 4.5) (Bob2) {};
	\node at (4-0.7, -0.95) (Bob7) {};
	\node at (4-0.7, -1.7) (Bob8) {$y$};
	
	\draw[my-arrow] (Alice1) .. controls +(right:2em) and +(up:1em).. (Alice2);
	\draw[my-line] (Alice2.center) to (Alice7.center);
	\draw[my-arrow] (Alice7) to (Alice8);
	
	\draw[my-arrow] (Bob1) .. controls +(left:2em) and +(up:1em).. (Bob2);
	\draw[my-line] (Bob2.center) to (Bob7.center);
	\draw[my-arrow] (Bob7) to (Bob8);

\end{tikzpicture}
}
	\hspace{\hsep}
	{
\small
\begin{tikzpicture}[scale=\scalewirings, every node/.style={scale=\scalewirings},
small-NLB/.style={rectangle, draw=color-box2, fill=color-box2!35, line width=1.5pt, minimum width=60, minimum height=15, anchor=center},
big-NLB/.style={rectangle, draw=color-box2, fill=color-box2!8, line width=1.5pt, minimum width=60, minimum height=15},
operation/.style={rectangle, draw=black, line width=0.5pt},
my-arrow/.style={->, very thick, >=stealth},
my-arrow-violet/.style={->, line width=0.5em, color=color-box2!8},
my-line/.style={very thick},
]

	\draw[big-NLB] (0., -1) rectangle (4., 4.5);
	\node at (2, 5) {(\refprop[Ex.\!]{paragraph:example-trivial})~$\Wlin$};
		
	\node[small-NLB] at (2, 3) (P) {\normalsize$\P$};
	\node[small-NLB] at (2, 0.5) (Q) {\normalsize$\Q$};
	\node at (0., 5.3) (Alice1) {$x$};
	\node at (0.7, 4.5) (Alice2) {};
	\node at (0.7, -0.95) (Alice7) {};
	\node at (0.7, -1.7) (Alice8) {$a_2$};
	\node at (4., 5.3) (Bob1) {$y$};
	\node at (3.3, 4.5) (Bob2) {};
	\node at (4-0.7, -0.95) (Bob7) {};
	\node at (4-0.7, -1.7) (Bob8) {$b_2$};
	
	\draw[my-arrow] (Alice1) .. controls +(right:2em) and +(up:1em).. (Alice2);
	\draw[my-arrow] (Alice2.center) .. controls  +(down:2em) and +(up:2em) ..  (P.160);
	\draw[my-arrow] (P.200) to node[left]{$a_1$} (Q.160);
	\draw[my-line] (Q.200) ..controls +(down:2em) and + (up:2em) .. node[left]{$a_2$\,\,} (Alice7.center);
	\draw[my-arrow] (Alice7) to (Alice8);
	
	\draw[my-arrow] (Bob1) .. controls +(left:2em) and +(up:1em).. (Bob2);
	\draw[my-arrow] (Bob2.center) .. controls  +(down:2em) and +(up:2em) .. (P.20);
	\draw[my-arrow] (P.340) to node[right]{$b_1$} (Q.20);
	\draw[my-line] (Q.340) ..controls +(down:2em) and + (up:2em) .. node[right]{\,\,$b_2$} (Bob7.center);
	\draw[my-arrow] (Bob7) to (Bob8);

\end{tikzpicture}
}
	\hspace{\hsep}
	{
\small
\begin{tikzpicture}[scale=\scalewirings, every node/.style={scale=\scalewirings},
small-NLB/.style={rectangle, draw=color-box3, fill=color-box3!35, line width=1.5pt, minimum width=60, minimum height=15, anchor=center},
big-NLB/.style={rectangle, draw=color-box3, fill=color-box3!8, line width=1.5pt, minimum width=60, minimum height=15},
operation/.style={rectangle, draw=black, line width=0.5pt},
my-arrow/.style={->, very thick, >=stealth},
my-line-violet/.style={line width=0.5em, color=color-box3!8},
my-line/.style={very thick},
]

	\draw[big-NLB] (0., -1) rectangle (4., 4.5);
	\node at (2, 5) {(\refprop[Ex.\!]{paragraph:example-FWW09})~$\Woplus$~\cite{FWW09}};
		
	\node[small-NLB] at (2, 3) (P) {\normalsize$\P$};
	\node[small-NLB] at (2, 0.5) (Q) {\normalsize$\Q$};
	\node at (0., 5.3) (Alice1) {$x$};
	\node at (0.7, 4.5) (Alice2) {};
	\node at (0.7, 4.) (Alice3) {$\bullet$};
	\node at (1.205, 2.3) (Alice4) {};
	\node[operation] at (0.7, -0.4) (Alice6) {$\oplus$};
	\node at (0.7, -0.95) (Alice7) {};
	\node at (0.7, -1.7) (Alice8) {$a_1\oplus a_2$};
	\node at (4., 5.3) (Bob1) {$y$};
	\node at (3.3, 4.5) (Bob2) {};
	\node at (3.3, 4.) (Bob3) {$\bullet$};
	\node at (4-1.205, 2.3) (Bob4) {};
	\node[operation] at (4-0.7, -0.4) (Bob6) {$\oplus$};
	\node at (4-0.7, -0.95) (Bob7) {};
	\node at (4-0.7, -1.7) (Bob8) {$b_1\oplus b_2$};
	
	\draw[my-arrow] (Alice1) .. controls +(right:2em) and +(up:1em).. (Alice2);
	\draw[my-line] (Alice2.center) to (Alice3.center);
	\draw[my-arrow] (Alice3.center)  .. controls  +(right:1.5em) and +(up:1em) ..  (P.160);
	\draw[my-line, line cap=round] (P.200) to (Alice4.center);
	\draw[my-arrow] (Alice4.center) .. controls +(left:3.em) and +(up:2em) .. node[left]{$a_1$} (Alice6.north);
	\draw[my-arrow] (Q.200) ..controls +(down:1em) and + (right:1.3em) .. node[right]{$a_2$} (Alice6.east);
	\draw[my-line-violet] (Alice3) .. controls +(left:3.em) and +(up:2em) .. (Q.160);
	\draw[my-arrow] (Alice3.center) .. controls +(left:3.em) and +(up:2em) .. (Q.160);
	\draw[my-line] (Alice6.south) to (Alice7.center);
	\draw[my-arrow] (Alice7) to (Alice8);
	
	\draw[my-arrow] (Bob1) .. controls +(left:2em) and +(up:1em).. (Bob2);
	\draw[my-line] (Bob2.center) to (Bob3.center);
	\draw[my-arrow] (Bob3.center)  .. controls  +(left:1.5em) and +(up:1em) .. (P.20);
	\draw[my-line, line cap=round] (P.340) to (Bob4.center);
	\draw[my-arrow] (Q.340) ..controls +(down:1em) and + (left:1.3em) .. node[left]{$b_2$} (Bob6.west);
	\draw[my-arrow] (Bob4.center) .. controls +(right:3em) and +(up:2em).. node[right]{$b_1$} (Bob6.north);
	\draw[my-line-violet] (Bob3) .. controls +(right:3.em) and +(up:2em) .. (Q.20);
	\draw[my-arrow] (Bob3.center) .. controls +(right:3.em) and +(up:2em) .. (Q.20);
	\draw[my-line] (Bob6.south) to (Bob7.center);
	\draw[my-arrow] (Bob7) to (Bob8);
	
	\node[small-NLB] at (2, 3) (P) {\normalsize$\P$};
	\node[small-NLB] at (2, 0.5) (Q) {\normalsize$\Q$};

\end{tikzpicture}
}
	\hspace{\hsep}
	{
\small
\begin{tikzpicture}[scale=\scalewirings, every node/.style={scale=\scalewirings},
small-NLB/.style={rectangle, draw=color-box4, fill=color-box4!35, line width=1.5pt, minimum width=60, minimum height=15, anchor=center},
big-NLB/.style={rectangle, draw=color-box4, fill=color-box4!8, line width=1.5pt, minimum width=60, minimum height=15},
operation/.style={rectangle, draw=black, line width=0.5pt},
my-arrow/.style={->, very thick, >=stealth},
my-line-violet/.style={line width=0.5em, color=color-box4!8},
my-line/.style={very thick},
]

	\draw[big-NLB] (0., -1) rectangle (4., 4.5);
	\node at (2, 5) {(\refprop[Ex.\!]{paragraph:example-BS09})~$\Wbs$~\cite{BS09}};
		
	\node[small-NLB] at (2, 3) (P) {\normalsize$\P$};
	\node[small-NLB] at (2, 0.5) (Q) {\normalsize$\Q$};
	\node at (0., 5.3) (Alice1) {$x$};
	\node at (0.7, 4.5) (Alice2) {};
	\node at (0.7, 4.) (Alice3) {$\bullet$};
	\node at (1.205, 2.3) (Alice4) {$\bullet$};
	\node[operation] at (1.205, 1.5) (Alice5) {$\wedge$};
	\node[operation] at (0.7, -0.4) (Alice6) {$\oplus$};
	\node at (0.7, -0.95) (Alice7) {};
	\node at (0.7, -1.7) (Alice8) {$a_1\oplus a_2$};
	\node at (4., 5.3) (Bob1) {$y$};
	\node at (3.3, 4.5) (Bob2) {};
	\node at (3.3, 4.) (Bob3) {$\bullet$};
	\node at (4-1.205, 2.3) (Bob4) {$\bullet$};
	\node[operation] at (4-1.205, 1.5) (Bob5) {$\wedge$};
	\node[operation] at (4-0.7, -0.4) (Bob6) {$\oplus$};
	\node at (4-0.7, -0.95) (Bob7) {};
	\node at (4-0.7, -1.7) (Bob8) {$b_1\oplus b_2$};
	
	\draw[my-arrow] (Alice1) .. controls +(right:2em) and +(up:1em).. (Alice2);
	\draw[my-line] (Alice2.center) to (Alice3.center);
	\draw[my-arrow] (Alice3.center)  .. controls  +(right:1.5em) and +(up:1em) ..  (P.160);
	\draw[my-line] (P.200) to node[left]{$a_1$} (Alice4.center);
	\draw[my-arrow] (Alice4.center) .. controls +(right:2.em) and +(right:2em) ..  (Alice5.east);
	\draw[my-arrow] (Alice5.south) to (Q.160);
	\draw[my-arrow] (Q.200) ..controls +(down:1em) and + (right:1.3em) .. node[right]{$a_2$} (Alice6.east);
	\draw[my-arrow] (Alice4.center) .. controls +(left:4em) and +(up:2em).. (Alice6.north);
	\draw[my-line-violet] (Alice3) .. controls +(left:3.em) and +(left:2em) .. (Alice5.west);
	\draw[my-arrow] (Alice3.center) .. controls +(left:3.em) and +(left:2em) .. (Alice5.west);
	\draw[my-line] (Alice6.south) to (Alice7.center);
	\draw[my-arrow] (Alice7) to (Alice8);
	
	\draw[my-arrow] (Bob1) .. controls +(left:2em) and +(up:1em).. (Bob2);
	\draw[my-line] (Bob2.center) to (Bob3.center);
	\draw[my-arrow] (Bob3.center)  .. controls  +(left:1.5em) and +(up:1em) .. (P.20);
	\draw[my-line] (P.340) to node[right]{$b_1$} (Bob4.center);
	\draw[my-arrow] (Bob4.center) .. controls +(left:2.em) and +(left:2em) ..  (Bob5.west);
	\draw[my-arrow] (Bob5.south) to (Q.20);
	\draw[my-arrow] (Q.340) ..controls +(down:1em) and + (left:1.3em) .. node[left]{$b_2$} (Bob6.west);
	\draw[my-arrow] (Bob4.center) .. controls +(right:4em) and +(up:2em)..  (Bob6.north);
	\draw[my-line-violet] (Bob3) .. controls +(right:3.em) and +(right:2em) .. (Bob5.east);
	\draw[my-arrow] (Bob3.center) .. controls +(right:3.em) and +(right:2em) .. (Bob5.east);
	\draw[my-line] (Bob6.south) to (Bob7.center);
	\draw[my-arrow] (Bob7) to (Bob8);
	
	\node[operation] at (1.205, 1.5) (Alice5) {$\wedge$};
	\node[operation] at (4-1.205, 1.5) (Bob5) {$\wedge$};

\end{tikzpicture}
}
	{
\small
\begin{tikzpicture}[scale=\scalewirings, every node/.style={scale=\scalewirings},
small-NLB/.style={rectangle, draw=color-box5, fill=color-box5!35, line width=1.5pt, minimum width=60, minimum height=15, anchor=center},
big-NLB/.style={rectangle, draw=color-box5, fill=color-box5!8, line width=1.5pt, minimum width=60, minimum height=15},
operation/.style={rectangle, draw=black, line width=0.5pt},
my-arrow/.style={->, very thick, >=stealth},
my-line-violet/.style={line width=0.5em, color=color-box5!8},
my-line/.style={very thick},
]

	\draw[big-NLB] (0., -1) rectangle (4., 4.5);
	\node at (2, 5) {(\refprop[Ex.\!]{paragraph:example-ABLPSV09})~$\Wdist$~\cite{ABLPSV09}};
		
	\node[small-NLB] at (2, 3) (P) {\normalsize$\P$};
	\node[small-NLB] at (2, 0.5) (Q) {\normalsize$\Q$};
	\node at (0., 5.3) (Alice1) {$x$};
	\node at (0.7, 4.5) (Alice2) {};
	\node at (0.7, 4.) (Alice3) {$\bullet$};
	\node at (1.205, 2.3) (Alice4) {$\bullet$};
	\node[operation] at (1.205, 1.5) (Alice5) {$\overline\oplus$};
	\node[operation] at (0.7, -0.4) (Alice6) {$\overline\oplus$};
	\node at (0.7, -0.95) (Alice7) {};
	\node at (0.7, -1.7) (Alice8) {$\overline{a_1\oplus a_2}$};
	\node at (4., 5.3) (Bob1) {$y$};
	\node at (3.3, 4.5) (Bob2) {};
	\node at (3.3, 4.) (Bob3) {$\bullet$};
	\node at (4-1.205, 2.3) (Bob4) {$\bullet$};
	\node[operation] at (4-1.205, 1.5) (Bob5) {$\wedge$};
	\node[operation] at (4-0.7, -0.4) (Bob6) {$\overline\oplus$};
	\node at (4-0.7, -0.95) (Bob7) {};
	\node at (4-0.7, -1.7) (Bob8) {$\overline{b_1\oplus b_2}$};
	
	\draw[my-arrow] (Alice1) .. controls +(right:2em) and +(up:1em).. (Alice2);
	\draw[my-line] (Alice2.center) to (Alice3.center);
	\draw[my-arrow] (Alice3.center)  .. controls  +(right:1.5em) and +(up:1em) ..  (P.160);
	\draw[my-line] (P.200) to node[left]{$a_1$} (Alice4.center);
	\draw[my-arrow] (Alice4.center) .. controls +(right:2.em) and +(right:2em) ..  (Alice5.east);
	\draw[my-arrow] (Alice5.south) to (Q.160);
	\draw[my-arrow] (Q.200) ..controls +(down:1em) and + (right:1.3em) .. node[right]{$a_2$} (Alice6.east);
	\draw[my-arrow] (Alice4.center) .. controls +(left:4em) and +(up:2em).. (Alice6.north);
	\draw[my-line-violet] (Alice3) .. controls +(left:3.em) and +(left:2em) .. (Alice5.west);
	\draw[my-arrow] (Alice3.center) .. controls +(left:3.em) and +(left:2em) .. (Alice5.west);
	\draw[my-line] (Alice6.south) to (Alice7.center);
	\draw[my-arrow] (Alice7) to (Alice8);
	
	\draw[my-arrow] (Bob1) .. controls +(left:2em) and +(up:1em).. (Bob2);
	\draw[my-line] (Bob2.center) to (Bob3.center);
	\draw[my-arrow] (Bob3.center)  .. controls  +(left:1.5em) and +(up:1em) .. (P.20);
	\draw[my-line] (P.340) to node[right]{$b_1$} (Bob4.center);
	\draw[my-arrow] (Bob4.center) .. controls +(left:2.em) and +(left:2em) ..  (Bob5.west);
	\draw[my-arrow] (Bob5.south) to (Q.20);
	\draw[my-arrow] (Q.340) ..controls +(down:1em) and + (left:1.3em) .. node[left]{$b_2$} (Bob6.west);
	\draw[my-arrow] (Bob4.center) .. controls +(right:4em) and +(up:2em)..  (Bob6.north);
	\draw[my-line-violet] (Bob3) .. controls +(right:3.em) and +(right:2em) .. (Bob5.east);
	\draw[my-arrow] (Bob3.center) .. controls +(right:3.em) and +(right:2em) .. (Bob5.east);
	\draw[my-line] (Bob6.south) to (Bob7.center);
	\draw[my-arrow] (Bob7) to (Bob8);
	
	\node[operation] at (1.205, 1.5) (Alice5) {$\overline\oplus$};
	\node[operation] at (4-1.205, 1.5) (Bob5) {$\wedge$};

\end{tikzpicture}
}
	\hspace{\hsep}
	{
\small
\begin{tikzpicture}[scale=\scalewirings, every node/.style={scale=\scalewirings},
small-NLB/.style={rectangle, draw=color-box6, fill=color-box6!35, line width=1.5pt, minimum width=60, minimum height=15, anchor=center},
big-NLB/.style={rectangle, draw=color-box6, fill=color-box6!8, line width=1.5pt, minimum width=60, minimum height=15},
operation/.style={rectangle, draw=black, line width=0.5pt},
my-arrow/.style={->, very thick, >=stealth},
my-line-violet/.style={line width=0.5em, color=color-box6!8},
my-line/.style={very thick},
]

	\draw[big-NLB] (0., -1) rectangle (4., 4.5);
	\node at (2, 5) {(\refprop[Ex.\!]{paragraph:example-ABLPSV09})~$\Wand$~\cite{ABLPSV09}};
		
	\node[small-NLB] at (2, 3) (P) {\normalsize$\P$};
	\node[small-NLB] at (2, 0.5) (Q) {\normalsize$\Q$};
	\node at (0., 5.3) (Alice1) {$x$};
	\node at (0.7, 4.5) (Alice2) {};
	\node at (0.7, 4.) (Alice3) {$\bullet$};
	\node at (1.205, 2.3) (Alice4) {};
	\node[operation] at (0.7, -0.4) (Alice6) {$\wedge$};
	\node at (0.7, -0.95) (Alice7) {};
	\node at (0.7, -1.7) (Alice8) {$a_1\wedge a_2$};
	\node at (4., 5.3) (Bob1) {$y$};
	\node at (3.3, 4.5) (Bob2) {};
	\node at (3.3, 4.) (Bob3) {$\bullet$};
	\node at (4-1.205, 2.3) (Bob4) {};
	\node[operation] at (4-0.7, -0.4) (Bob6) {$\wedge$};
	\node at (4-0.7, -0.95) (Bob7) {};
	\node at (4-0.7, -1.7) (Bob8) {$b_1\wedge b_2$};
	
	\draw[my-arrow] (Alice1) .. controls +(right:2em) and +(up:1em).. (Alice2);
	\draw[my-line] (Alice2.center) to (Alice3.center);
	\draw[my-arrow] (Alice3.center)  .. controls  +(right:1.5em) and +(up:1em) ..  (P.160);
	\draw[my-line, line cap=round] (P.200) to (Alice4.center);
	\draw[my-arrow] (Alice4.center) .. controls +(left:3.em) and +(up:2em) .. node[left]{$a_1$} (Alice6.north);
	\draw[my-arrow] (Q.200) ..controls +(down:1em) and + (right:1.3em) .. node[right]{$a_2$} (Alice6.east);
	\draw[my-line-violet] (Alice3) .. controls +(left:3.em) and +(up:2em) .. (Q.160);
	\draw[my-arrow] (Alice3.center) .. controls +(left:3.em) and +(up:2em) .. (Q.160);
	\draw[my-line] (Alice6.south) to (Alice7.center);
	\draw[my-arrow] (Alice7) to (Alice8);
	
	\draw[my-arrow] (Bob1) .. controls +(left:2em) and +(up:1em).. (Bob2);
	\draw[my-line] (Bob2.center) to (Bob3.center);
	\draw[my-arrow] (Bob3.center)  .. controls  +(left:1.5em) and +(up:1em) .. (P.20);
	\draw[my-line, line cap=round] (P.340) to (Bob4.center);
	\draw[my-arrow] (Q.340) ..controls +(down:1em) and + (left:1.3em) .. node[left]{$b_2$} (Bob6.west);
	\draw[my-arrow] (Bob4.center) .. controls +(right:3em) and +(up:2em).. node[right]{$b_1$} (Bob6.north);
	\draw[my-line-violet] (Bob3) .. controls +(right:3.em) and +(up:2em) .. (Q.20);
	\draw[my-arrow] (Bob3.center) .. controls +(right:3.em) and +(up:2em) .. (Q.20);
	\draw[my-line] (Bob6.south) to (Bob7.center);
	\draw[my-arrow] (Bob7) to (Bob8);
	
	\node[small-NLB] at (2, 3) (P) {\normalsize$\P$};
	\node[small-NLB] at (2, 0.5) (Q) {\normalsize$\Q$};

\end{tikzpicture}
}
	\hspace{\hsep}
	{
\small
\begin{tikzpicture}[scale=\scalewirings, every node/.style={scale=\scalewirings},
small-NLB/.style={rectangle, draw=color-box7, fill=color-box7!35, line width=1.5pt, minimum width=60, minimum height=15, anchor=center},
big-NLB/.style={rectangle, draw=color-box7, fill=color-box7!8, line width=1.5pt, minimum width=60, minimum height=15},
operation/.style={rectangle, draw=black, line width=0.5pt},
my-arrow/.style={->, very thick, >=stealth},
my-arrow-violet/.style={->, line width=0.5em, color=color-box7!8},
my-line/.style={very thick},
]

	\draw[big-NLB] (0., -1) rectangle (4., 4.5);
	\node at (2, 5) {\footnotesize(\refprop[Ex.\!]{paragraph:example-HR10})\,$\Wdepth$\,\cite{EWC22PRL}};
		
	\node[small-NLB] at (2, 3.5) (P) {\normalsize$\P$};
	\node[small-NLB] at (2, 1.8) (Q) {\normalsize$\Q$};
	\node[small-NLB] at (2, 0.1) (R) {\normalsize$\Rbox$};
	\node at (0., 5.3) (Alice1) {$x$};
	\node at (0.7, 4.5) (Alice2) {};
	\node at (0.7, 4.1) (Alice3) {$x$};
	\node at (0.7, 2.88) (Alice4) {$a_1$};
	\node at (0.7, 2.45) (Alice5) {$x$};
	\node at (0.7, 1.2) (Alice6) {$a_2$};
	\node at (0.6, 0.75) (Alice5bis) {\footnotesize$(\star)$};
	\node at (0.7, -0.5) (Alice6bis) {$a_3$};
	\node at (0.7, -0.95) (Alice7) {};
	\node at (0.7, -1.7) (Alice8) {$(\star\!\star\!\star)$};
	\node at (4., 5.3) (Bob1) {$y$};
	\node at (3.3, 4.5) (Bob2) {};
	\node at (3.3, 4.1) (Bob3) {$y$};
	\node at (3.3, 2.88) (Bob4) {$b_1$};
	\node at (4-0.7, 2.45) (Bob5) {$y$};
	\node at (4-0.7, 1.2) (Bob6) {$b_2$};
	\node at (4-0.6, 0.75) (Bob5bis) {$(\star\star)$};
	\node at (4-0.7, -0.5) (Bob6bis) {$b_3$};
	\node at (4-0.7, -0.95) (Bob7) {};
	\node at (4-0.7, -1.7) (Bob8) {$(\star\!\star\!\star\star)$};
	
	\draw[my-arrow] (Alice1) .. controls +(right:2em) and +(up:1em).. (Alice2);
	\draw[my-arrow] (Alice3.east)  .. controls  +(right:1em) and +(up:0.8em) ..  (P.160);
	\draw[my-arrow] (P.200)  .. controls  +(down:1em) and +(right:0.8em) ..  (Alice4.east);
	\draw[my-arrow] (Alice5.east)  .. controls  +(right:1em) and +(up:0.8em) ..  (Q.160);
	\draw[my-arrow] (Q.200)  .. controls  +(down:1em) and +(right:0.8em) ..  (Alice6.east);
	\draw[my-arrow] (Alice5bis.east)  .. controls  +(right:0.8em) and +(up:0.8em) ..  (R.160);
	\draw[my-arrow] (R.200)  .. controls  +(down:1em) and +(right:0.8em) ..  (Alice6bis.east);
	\draw[my-arrow] (Alice7) to (Alice8);
	
	\draw[my-arrow] (Bob1) .. controls +(left:2em) and +(up:1em).. (Bob2);
	\draw[my-arrow] (Bob3.west)  .. controls  +(left:1em) and +(up:0.8em) .. (P.20);
	\draw[my-arrow] (P.340)  .. controls  +(down:1em) and +(left:0.8em) ..  (Bob4.west);
	\draw[my-arrow] (Bob5.west)  .. controls  +(left:1em) and +(up:0.8em) .. (Q.20);
	\draw[my-arrow] (Q.340)  .. controls  +(down:1em) and +(left:0.8em) ..  (Bob6.west);
	\draw[my-arrow] (Bob5bis.west)  .. controls  +(left:0.8em) and +(up:0.8em) .. (R.20);
	\draw[my-arrow] (R.340)  .. controls  +(down:1em) and +(left:0.8em) ..  (Bob6bis.west);
	\draw[my-arrow] (Bob7) to (Bob8);

\end{tikzpicture}
}
	\hspace{\hsep}
	{
\small
\begin{tikzpicture}[scale=\scalewirings, every node/.style={scale=\scalewirings},
small-NLB/.style={rectangle, draw=color-box8, fill=color-box8!35, line width=1.5pt, minimum width=60, minimum height=15, anchor=center},
big-NLB/.style={rectangle, draw=color-box8, fill=color-box8!8, line width=1.5pt, minimum width=60, minimum height=15},
operation/.style={rectangle, draw=black, line width=0.5pt},
my-arrow/.style={->, very thick, >=stealth},
my-line-violet/.style={line width=0.5em, color=color-box8!8},
my-line/.style={very thick},
]

	\draw[big-NLB] (0., -1) rectangle (4., 4.5);
	\node at (2, 5) {(\refprop[Ex.\!]{paragraph:example-NSSRRB22PRL})~$\Worand$~\cite{NSSRRB22PRL}};
		
	\node[small-NLB] at (2, 3) (P) {\normalsize$\P$};
	\node[small-NLB] at (2, 0.5) (Q) {\normalsize$\Q$};
	\node at (0., 5.3) (Alice1) {$x$};
	\node at (0.7, 4.5) (Alice2) {};
	\node at (0.7, 4.) (Alice3) {$\bullet$};
	\node at (1.205, 2.3) (Alice4) {};
	\node[operation] at (0.7, -0.4) (Alice6) {$\vee$};
	\node at (0.7, -0.95) (Alice7) {};
	\node at (0.7, -1.7) (Alice8) {$\max\{a_1, a_2\}$};
	\node at (4., 5.3) (Bob1) {$y$};
	\node at (3.3, 4.5) (Bob2) {};
	\node at (3.3, 4.) (Bob3) {$\bullet$};
	\node at (4-1.205, 2.3) (Bob4) {};
	\node[operation] at (4-0.7, -0.4) (Bob6) {$\wedge$};
	\node at (4-0.7, -0.95) (Bob7) {};
	\node at (4-0.7, -1.7) (Bob8) {$\min\{b_1, b_2\}$};
	
	\draw[my-arrow] (Alice1) .. controls +(right:2em) and +(up:1em).. (Alice2);
	\draw[my-line] (Alice2.center) to (Alice3.center);
	\draw[my-arrow] (Alice3.center)  .. controls  +(right:1.5em) and +(up:1em) ..  (P.160);
	\draw[my-line, line cap=round] (P.200) to (Alice4.center);
	\draw[my-arrow] (Alice4.center) .. controls +(left:3.em) and +(up:2em) .. node[left]{$a_1$} (Alice6.north);
	\draw[my-arrow] (Q.200) ..controls +(down:1em) and + (right:1.3em) .. node[right]{$a_2$} (Alice6.east);
	\draw[my-line-violet] (Alice3) .. controls +(left:3.em) and +(up:2em) .. (Q.160);
	\draw[my-arrow] (Alice3.center) .. controls +(left:3.em) and +(up:2em) .. (Q.160);
	\draw[my-line] (Alice6.south) to (Alice7.center);
	\draw[my-arrow] (Alice7) to (Alice8);
	
	\draw[my-arrow] (Bob1) .. controls +(left:2em) and +(up:1em).. (Bob2);
	\draw[my-line] (Bob2.center) to (Bob3.center);
	\draw[my-arrow] (Bob3.center)  .. controls  +(left:1.5em) and +(up:1em) .. (P.20);
	\draw[my-line, line cap=round] (P.340) to (Bob4.center);
	\draw[my-arrow] (Q.340) ..controls +(down:1em) and + (left:1.3em) .. node[left]{$b_2$} (Bob6.west);
	\draw[my-arrow] (Bob4.center) .. controls +(right:3em) and +(up:2em).. node[right]{$b_1$} (Bob6.north);
	\draw[my-line-violet] (Bob3) .. controls +(right:3.em) and +(up:2em) .. (Q.20);
	\draw[my-arrow] (Bob3.center) .. controls +(right:3.em) and +(up:2em) .. (Q.20);
	\draw[my-line] (Bob6.south) to (Bob7.center);
	\draw[my-arrow] (Bob7) to (Bob8);
	
	\node[small-NLB] at (2, 3) (P) {\normalsize$\P$};
	\node[small-NLB] at (2, 0.5) (Q) {\normalsize$\Q$};

\end{tikzpicture}
}
	\centering
	\caption{Typical examples of wirings.
	Wirings with the same color have similar internal structures.
	The overline bar is the NOT gate: $\overline x=x\oplus 1$.
	The symbol $(\star)$ stands for $xa_2\vee x\overline{a_1}\vee \overline{x} \overline{a_2} a_1$, 
	and $(\star\star)$ for $yb_2 \vee y\overline{b_1}$,
	and $(\star\!\star\!\star)$ for $a_3a_2 \vee a_3\overline{a_1} \vee \overline{a_3}\overline{a_2}a_1$,
	and $(\star\!\star\!\star\star)$ for $b_3b_2 \vee b_3\overline{b_1} \vee \overline{b_3}\overline{b_2}b_1$.
	}
	\label{fig: Drawings of the wirings}
\end{figure}

\setcounter{example}{-1}

\begin{example} \label{paragraph:example-trivial}
	The trivial wiring $\Wtriv$ is defined as the wiring that does ``nothing", in the sense that it outputs exactly the global inputs: $(a,b) = (x,y)$.
	Similarly, the linear wiring $\Wlin$ simply connects the output of a box to the input of the box immediately below.
\end{example}

\begin{example} \label{paragraph:example-FWW09}
	In~\cite{FWW09}, Forster \etal introduce a wiring $\Woplus$ in order to distill nonlocality.
	It consists in setting boxes in parallel and in taking the sum mod $2$ of the outputs.
\end{example}
	
\begin{example} \label{paragraph:example-BS09}
	In~\cite{BS09}, Brunner and Skrzypczyk enhance the wiring from \refprop[Example]{paragraph:example-FWW09} in order to obtain a better distillation protocol of nonlocality. 
	Their wiring $\Wbs$ is \emph{adaptive}, in the sense that boxes are no longer in parallel: the second box's inputs $a_2,b_2$ are not simply equal to the previous box's outputs $a_1,b_1$, but they equal the the latter multiplied by the general inputs $x,y$.
	Their new protocol is so powerful that it allows to arbitrarily reduce the noise of any \emph{correlated box} (defined as convex combinations of $\PR$ and $\SR$) so that the $\PR$ box is almost perfectly simulated. As a consequence,  communication complexity collapses; see the next section for more details.
\end{example}	
	
\begin{example}\label{paragraph:example-ABLPSV09}
	In~\cite{ABLPSV09}, Allcock \etal study two variants of the previous wirings.
	First, their ``distillation wiring" $\Wdist$ is similar to the one in \refprop[Example]{paragraph:example-BS09}: it is also adaptive and it also distills all correlated boxes.
	Second, their ``AND wiring" $\Wand$  resembles the one in \refprop[Example]{paragraph:example-FWW09}: boxes are set in parallel, but we take the product of the outputs instead of the sum.
\end{example}	

\begin{example}\label{paragraph:example-HR10}
	In~\cite{HR10}, H{\o}yer and Rashid
	study the depth-$k$ generalizations of the wirings from \refprop[Examples]{paragraph:example-FWW09} and~\refprop{paragraph:example-BS09} and~\refprop{paragraph:example-ABLPSV09}: they wire $k$ boxes instead of only two.
	They also give an example of a depth-$3$ protocol that extends the known region of distillable boxes. In \cite{EWC22PRL} this idea is improved upon, by constructing genuine depth-$3$ protocols, such as  $\Wdepth$ drawn above. These protocols are strictly better than depth-$2$ protocols in terms of the collapse of CC, in the sense that they prove the existence of nonlocal boxes that are shown to collapse CC using this wiring but that cannot be distilled using any depth-$2$ wiring.
	To do so, they use their algorithm to single out the area of nonlocal boxes that are distillable by means of depth-$2$ wirings, and they show that their example $\Wdepth$ is out of this region.\footnote{We thank Mirjam Weilenmann for feedback and for pointing out that~\cite{BCSS11} states that the wirings $\Woplus$ and $\Wdist$ are sufficient to characterize the distillable region of the slice $\PR$--$\SR$--$\I$.}
	Note that in our work, the study is limited to depth-$2$ wirings.
\end{example}
	
\begin{example}\label{paragraph:example-NSSRRB22PRL}
	More recently, in~\cite{NSSRRB22PRL}, Naik \etal defined the ``OR-AND wiring" $\Worand$ in order to distill the nonlocality of quantum correlations.
	That wiring is a mix of the ones in \refprop[Examples]{paragraph:example-FWW09} and \refprop[\!]{paragraph:example-ABLPSV09}: it consists in setting boxes in parallel and in taking the maximum (the ``OR") of Alice's outputs and the minimum (the ``AND") of Bob's outputs.
\end{example}

		\subsection{Algebra of Boxes Induced by a Wiring}

Let $\B$ be the vector space of all the functions $\{0,1\}^4\to\R$,
and consider a mixed wiring $\W$. As defined in \eqrefprop[Equation]{eq: P times Q with mixed wiring}, the operation $\boxtimes_\W$ is bilinear, so the vector space $\B$ equipped with the product $\boxtimes_\W$ is actually an algebra, which we denote by $\B_\W$ for that specific wiring $\W\in\WW$.
Its dimension is $\dim\B_\W=2^4=16$.
Note that the (affine) dimension of the non-signalling polytope $\NS\subseteq \B_\W$ is $\dim\NS = 16-8=8$ because there are $8$ dependent variables in the affine conditions~\eqref{eq: non-signalling is a joint proba},~\eqref{eq: non-signalling 1},~\eqref{eq: non-signalling 2}; see~\cite{BLMPPR05} for a more general expression.

\paragraph{Multiplication Table.}
In order to better understand the behavior of the box product $\boxtimes$, it is interesting to compute the product of some basic boxes:
for instance the boxes $\PR$, $\P_0$, $\P_1$, $\I$ defined in \eqrefprop[Equation]{eq:definition-of-PR-P0-P1-I}.
In \refprop[Figure]{fig: multiplication table}, we present the multiplication table for the wiring $\Wbs$ from~\cite{BS09}.
By bilinearity of the box multiplication, this table shows that the convex hull $\Conv\{\PR, \P_0, \P_1\}$ is stable under $\boxtimes$.
On the contrary, observe that the convex hull $\Conv\{\PR, \P_0, \P_1,\I\}$ is not stable under $\boxtimes$: the product $\I\boxtimes\PR$ gives $\Q_1:= \frac14\,\PR-\frac{1}{8}\big(\P_0+\P_1\big) + \I$ which is out of the convex hull (nevertheless the affine hull $\Aff\{\PR, \P_0, \P_1,\I\}$ is stable under $\boxtimes$).
Notice that we show in \refprop[Proposition]{prop:PR-P0-P1-is-a-face-of-NS} that actually $\Conv\{\PR, \P_0, \P_1\}=\NS\cap\Aff\{\PR, \P_0, \P_1\}$.
From this table, one may postulate that $\P_0$ is a right identity in the sense that $\P\boxtimes\P_0=\P$ for all $\P$ in $\NS$, and it is indeed true as a simple consequence of formula \eqref{eq: definition of the box product}.
One may similarly verify that $\I$ is a right fixed point, in the sense that $\P\boxtimes\I=\I$ for all $\P$ in $\NS$, as it is possible to guess from the table.
See all the multiplication tables of the typical depth-$2$ wirings in \refprop[Appendix]{ap:the-multiplication-tables}.

\newlength{\length}
\setlength{\length}{0.3cm}
\begin{figure}[h]
	\centering
	{ \footnotesize
	\renewcommand{\arraystretch}{1.5}
	\newcommand{\mycommand}{\normalsize}
	\begin{tabular}{| c ||c|c|c|c|c|c|c|c|}
		\hline
		{\small\diagbox{$\P$}{$\Q$}} & \hspace{\length}{\mycommand$\PR$}\hspace{\length} & \hspace{\length}{\mycommand$\P_0$}\hspace{\length} & \hspace{\length}{\mycommand$\P_1$}\hspace{\length} & \hspace{\length}{\mycommand$\I$}\hspace{\length} \\
		\hline\hline
		{\mycommand$\PR$} &  $\PR$ & $\PR$ & $\PR$ & $\I$  \\ \hline
		{\mycommand$\P_0$} & $\frac{1}{2}\big(\P_0 + \P_1\big)$  & $\P_0$  & $\P_1$ & $\I$  \\
		\hline
		{\mycommand$\P_1$}  & $\PR$ &  $\P_1$ & $\P_0$ & $\I$  \\ \hline
		{\mycommand$\I$}  & $\Q_1$ &  $\I$ & $\I$ & $\I$  \\ \hline
	\end{tabular}
	\renewcommand{\arraystretch}{1}
	}
	\caption{Multiplication table of the operation $\boxtimes_{\Wbs}$ induced by the wiring from~\cite{BS09}. Each cell displays the result of $\P\boxtimes\Q$.
	The box $\Q_1$ at the bottom left is
	$\Q_1 := \frac14\,\PR-\frac{1}{8}\big(\P_0+\P_1\big) + \I$.
	Further multiplication tables are available in \refprop[Appendix]{ap:the-multiplication-tables}.
	}
	\label{fig: multiplication table}
\end{figure}

\paragraph{Non-Commutativity and Non-Associativity.}
A direct consequence of the multiplication table in \refprop[Figure]{fig: multiplication table} is that the algebra $\B_{\Wbs}$ induced by the wiring $\Wbs$ is non-commutative ($\P_0\boxtimes\PR\neq\PR\boxtimes\P_0$) and non-associative ($(\P_0\boxtimes\P_1)\boxtimes\PR \neq \P_0\boxtimes(\P_1\boxtimes\PR)$).
This non-associativity is at the root of interesting remarks, see drawings of the orbit of a box in the next section, \refprop[Figure]{figure: orbit of a box}.
Similarly, the algebra induced by the wiring $\Wdist$ is both non-commutative and non-associative, but
on the contrary, the algebras induced by $\W\in\{\Wtriv, \Woplus, \Wand, \Worand\}$ are both commutative and associative.
One may wonder if there exist induced algebras that are associative but not commutative, and the converse.
To that end, here is a characterization of commutativity and associativity in a simple case where boxes are set in parallel and with the same input functions:

\begin{proposition}[Characterization of commutativity and associativity]   \label{prop:Characterization-of-commutativity-and-associativity}
	Assume $\W$ is a wiring such that $f_1=f_2=f(x)$ and $g_1=g_2=g(y)$. Then:
	\begin{enumerate}[label=\normalfont(\roman*)]
		\item \label{prop:characterization-commutativity} $\B_\W$ is commutative \iff the functions $f_3(x,a_1,a_2)$ and $g_3(y,b_1,b_2)$ are ``symmetric" in the last two variables, in the sense that $f_3(x,a_1,a_2) = f_3(x,a_2,a_1)$ for all $x,a_1,a_2$, and similarly for $g_3$.
	\end{enumerate}
	If in addition $f(x)=x$ and $g(y)=y$:
	\begin{enumerate}[label=\normalfont(\roman*), start=2]
		\item \label{prop:characterization-associativity} $\B_\W$ is associative \iff the functions $f_3(x,a_1,a_2)$ and $g_3(y,b_1,b_2)$ are ``associative" in the last two variables, in the sense that $f_3(x,a_1,f_3(x,a_2,a_3)) = f_3(x,f_3(x,a_1,a_2),a_3)$ for all $x,a_1,a_2,a_3$, and similarly for $g_3$.
	\end{enumerate}
\end{proposition}

\begin{proof}
	\ref{prop:characterization-commutativity}~First, from the expression \eqref{eq: definition of the box product}, see that for all bits $a,b,x,y$ and any boxes $\P,\Q$ in $\B_\W$, we have:
	\begin{multline*}
		\P\boxtimesW\Q(a,b\,|\,x,y) - \Q\boxtimesW\P(a,b\,|\,x,y) \\
		\,=\,
		\sum_{a_1, a_2, b_1, b_2}
		\P\big(a_1, b_1\,|\, f(x),g(y)\big)\times \Q\big(a_2, b_2\,|\, f(x),g(y)\big)
		\times \Big[ \1_{a=f_3(x,a_1,a_2)} \1_{b=g_3(y,b_1,b_2)}
		\\
		- \1_{a=f_3(x,a_2,a_1)} \1_{b=g_3(y,b_2,b_1)} \Big]\,.
	\end{multline*}
	Hence, if $f_3$ and $g_3$ are both symmetric in the last two variables, then the difference is null and the algebra is commutative.
	Conversely, suppose that $\B_\W$ is commutative, so that the left-hand side is null. Taking probability distributions $\P$ and $\Q$ that are always positive (such as $\I$), we have that the difference in the right-hand side has to be null for all $x,y,a,b,a_1,a_2,b_1,b_2$.
	Fix $x,a_1,a_2$ and consider $a:=f_3(x,a_1,a_2)$, and similarly
	fix $y,b_1,b_2$ and consider $b:=g_3(y,b_1,b_2)$.
	We obtain $1 - \1_{a=f_3(x,a_2,a_1)} \1_{b=g_3(y,b_2,b_1)} = 0$, which means that both indicator functions are equal to $1$, and therefore both subscript equalities hold. Hence, this being true for any fixed $x,a_1,a_2$ and $y,b_1,b_2$, we obtain that $f_3$ and $g_3$ are symmetric as wanted.
	
	\ref{prop:characterization-associativity}~From \eqref{eq: definition of the box product} again, we have for all bits $a,b,x,y$ and any boxes $\P,\Q,\Rbox$ in $\B_\W$:
	\begin{multline*}
		\P\boxtimesW(\Q\boxtimesW\Rbox)(a,b\,|\,x,y) - (\P\boxtimesW\Q)\boxtimesW\Rbox(a,b\,|\,x,y) \\
		\,=\,\hspace{-0.6cm}
		\sum_{a_1, a_2, a_3, b_1, b_2, b_3} \hspace{-0.4cm}
		\P\big(a_1, b_1\,|\, x,y\big)\times \Q\big(a_2, b_2\,|\, x,y\big) \times \Rbox\big(a_3, b_3\,|\, x, y\big)
		\times \Big[ \1_{a=f_3(x,a_1,f_3(x,a_2,a_3))} \1_{b=g_3(y,b_1,g_3(y,b_2,b_3))} \\- \1_{a=f_3(x,f_3(x,a_1,a_2),a_3)} \1_{b=g_3(y,g_3(y,b_1,b_2),b_3)} \Big]\,.
	\end{multline*}
	A similar proof with double implication as in~\ref{prop:characterization-commutativity} applies, hence the associativity criterion follows.
\end{proof}

Now, it is easier to build an associative non-commutative induced algebra $\B_{\W'}$. Consider the wiring $\W'$ given by $f_1(x,a_2) = f_2(x,a_1) = x$, and $g_1(y,b_2) = g_2(y,b_1) = y$, and $f_3(x,a_1,a_2) := a_1$, and $g_3(y,b_1,b_2) := b_1$. This wiring satisfies the condition~\ref{prop:characterization-associativity} of the proposition and does not satisfy the condition~\ref{prop:characterization-commutativity}, hence it is as wanted.
Conversely, with similar arguments, a commutative non-associative algebra $\B_{\W''}$ is induced by the wiring $\W''$ defined by the same $f_1, f_2, g_1,g_2$ and $f_3(x,a_1,a_2) := a_1a_2\oplus1$ and $g_3(y,b_1,b_2) := b_1b_2\oplus1$. Therefore, we obtain the table in \refprop[Figure]{fig: commutativity and associativity of the induced algebras}.

\begin{figure}[h]
	\begin{tabular}{|c|c|c|}
		\hline
			& Associativity & Non-associativity \\
		\hline
		Commutativity & $\Wtriv, \Woplus, \Wand, \Worand$ & $\W''$ \\
		\hline
		Non-commutativity & $\W'$ & $\Wbs, \Wdist$ \\
		\hline
	\end{tabular}
	\centering
	\caption{Associativity and commutativity of the induced algebra $\B_\W$, depending on the wiring $\W$ displayed in the table cell.}
	\label{fig: commutativity and associativity of the induced algebras}
\end{figure}

\bib

	\section{Orbit of a Box}
	\label{section:Orbit-of-a-box}
	
In this section, we study the set of all boxes that can be generated given many copies of a starting box $\P$ and a wiring $\W$.
After introducing the \emph{orbit} of a box, we provide some consequences to communication complexity. Subsequently, we study a particular example, $\Wbs$, with which we find collapsing boxes in \refprop[Subsection]{subsection:analytical-results}, and then we give some general remarks about other orbits.
Finally, we conclude this section by giving the technical proof of the theorem stating that the ``best" parenthesization is the multiplication on the right.

		\subsection{Definition}
		
Given multiple copies of a non-signalling box $\P\in\NS$ and of a (mixed) wiring $\W$, Alice and Bob can produce many other boxes, \eg $(\P\boxtimes_\W\P)\boxtimes_\W\P$ or $\P\boxtimes_\W(\P\boxtimes_\W\P)$. All of these new boxes are again non-signalling because $\NS$ is closed under wirings, see \refprop[Lemma]{lemma: NS closed under mixed wirings}. We call \emph{orbit} of the box~$\P$ (induced by the wiring $\W$) the set of all of these possible new boxes:
\begin{align*}
	\Orbit_\W(\P)
	\,:=\,&
	\Big\{
	\text{\footnotesize boxes $\Q\in\NS$ that can be produced by using finitely many times the box $\P$ and the wiring $\W$}
	\Big\} \\
	\,=\,&
	\bigcup_{k\geq1} \Orbit_\W^{(k)}(\P) \subseteq\NS
	\,,
\end{align*}
where $\Orbit^{(k)}(\P)$ is called the \emph{orbit of depth $k$} of $\P$ (or simply \emph{$k$-orbit}), defined as:
\[
	\Orbit_\W^{(k)}(\P)
	\,:=\,
	\Big\{
	\text{all possible products with $k$ times the term $\P$, using the multiplication $\boxtimes_\W$}
	\Big\}
	\,.
\]
When the context is clear, we overload the notation and write $\Orbit$ and $\Orbit^{(k)}$ respectively.
In general, these $k$-orbits are not singletons for $k\geq 3$ since
the algebra $\B_\W$ induced by $\W$ is not necessarily associative and commutative (see \refprop[Figure]{fig: commutativity and associativity of the induced algebras}). Actually, up to multiplicity, the cardinal $\#\Orbit^{(k)}$ is exactly the number of parenthesizations with $k$ terms, which is the Catalan number $\frac{1}{k}\tiny\begin{pmatrix} 2k-2 \\k-1 \end{pmatrix}$, which grows exponentially fast.
Here are the $3$- and $4$- orbits:
\begin{align*}
	&
	\Orbit^{(3)}(\P)
	\,=\,
	\Big\{
	(\P\boxtimes\P)\boxtimes\P,\,
	\P\boxtimes(\P\boxtimes\P)
	\Big\}\,,
	\\
	&
	\Orbit^{(4)}(\P)
	\,=\,
	\Big\{
	\text{\footnotesize$
	\big((\P\boxtimes\P)\boxtimes\P\big)\boxtimes\P,\,
	\big(\P\boxtimes(\P\boxtimes\P)\big)\boxtimes\P,\,
	(\P\boxtimes\P)\boxtimes(\P\boxtimes\P),\,
	\P\boxtimes\big((\P\boxtimes\P)\boxtimes\P\big),\,
	\P\boxtimes\big(\P\boxtimes(\P\boxtimes\P)\big)
	$}
	\Big\}\,.
\end{align*}
Note that a $k$-orbit ($k\geq2$) can be inductively computed using orbits with lower depth:
\[
	\Orbit^{(k)}
	\,=\,
	\bigcup_{1\leq\ell\leq k-1}
	\Orbit^{(\ell)} \boxtimes \Orbit^{(k-\ell)}
	\,,
\]
which is the same recurrence relation as that of Catalan numbers.

		\subsection{Consequences to Communication Complexity}
		\label{subseq:ConsequencestoCommunicationComplexity}
		
Assume Alice and Bob are given infinitely many copies of a nonlocal box $\P$, and assume they want to distantly compute (in finite time) the value of a Boolean function $f(X,Y)$, where $X,Y\in\{0,1\}^n$ are strings that are known by Alice and Bob respectively.
Among all the possible protocols they can try to do in order to succeed, they can wire their copies of $\P$ in order to produce a ``better" box. For example, starting from a noisy box $\P$, Alice and Bob can try to produce a box that is closer to the ``perfect box" $\PR$ which satisfies $a\oplus b = xy$ without noise.
Such a protocol is called a \emph{distillation protocol}~\cite{BS09}. 
We call \emph{collapsing box} a nonlocal box that collapses CC.

\paragraph{Find Collapsing Boxes Using the Orbit.}
Imagine Alice and Bob are able to produce a collapsing box~$\Q$ after applying wirings to copies of a starting box $\P$. Then they can use that new box~$\Q$ to distantly compute the value $f(X,Y)$, which means that they have a protocol to collapse communication complexity and therefore that $\P$ is collapsing.
This point of view is particularly interesting since it implies that it is sufficient to find a single collapsing box in the union $\bigcup_\W \Orbit_\W(\P)$ to deduce that $\P$ is collapsing as well.
See an illustration in \refprop[Figure]{fig: collapsing orbits} (a).

\paragraph{Find Collapsing Boxes Using a Cone.}
Once we find a collapsing box $\P$, we can deduce many other collapsing boxes: there is a convex cone taking origin at $\P$ that is collapsing as well. More precisely, given a box $\P$, denote $\C_\P$ the convex cone of boxes $\Rbox$ for which there exists a local correlation $\L\in \LL$ such that $\P=\lambda\,\Rbox + (1-\lambda)\,\L$, with $\lambda\in[0,1]$.
We claim that if $\P$ is collapsing, then any $\Rbox\in\C_\P$ is collapsing as well.
Indeed, assume Alice and Bob are given copies of a box~$\Rbox$.
Then, they can use shared randomness to produce the wanted box $\L$ and the wanted convex coefficient $\lambda$, so that they can generate the box $\P$ with the relation $\P=\lambda\,\Rbox + (1-\lambda)\,\L$. Now, as $\P$ is collapsing, they have a protocol that collapses communication complexity, hence $\Rbox$ is collapsing as well.
See an illustration in \refprop[Figure]{fig: collapsing orbits}~(b).
In the study of collapsing boxes, notice that it is standard to assume that shared randomness is a ``free" resource; for instance Brassard \emph{et. al.} made that choice in~\cite{BBLMTU06} in their collapsing protocol.

\paragraph{}
Combining arguments from these last two paragraphs and by the fact that Alice and Bob can make a convex combination of boxes using shared randomness, we deduce a sufficient criterion for a box to collapse communication complexity:

\begin{proposition}[Collapsing orbit]   \label{prop: collapsing orbit}
	Let $\P$ be a box in $\NS$. If there exists a box $\Q\in \Conv\big(\LL\cup \bigcup_\W\Orbit_\W(\P) \big)$ that collapses communication complexity, then $\P$ is collapsing as well.  See \mbox{\href{fig: collapsing orbits}{Figure~\ref{fig: collapsing orbits}}~{\normalfont(c)}.}
\end{proposition}

\newcommand{\TEXTsize}{0.87}
\newcommand{\FIGUREsize}{1.03}
\newcommand{\myspace}{-0.7cm}
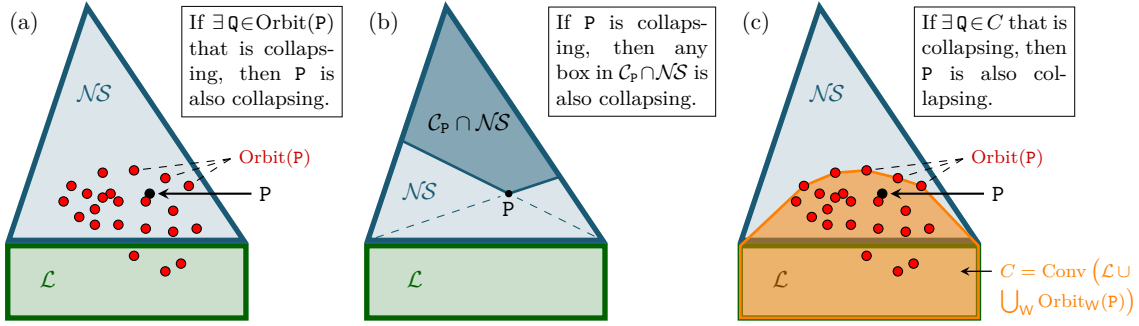
\begin{figure}[h]
	\definecolor{blueNS}{rgb}{0.1, 0.35, 0.45}
\definecolor{greenL}{rgb}{0., 0.39, 0.}
{
\begin{tikzpicture}[every node/.style={scale=\TEXTsize}, scale=\FIGUREsize,
	NS/.style={line width=2pt, color=blueNS, fill=blueNS, fill opacity=0.15},
	L/.style={line width=2pt, color=greenL, fill=greenL, fill opacity=0.20},
	anti-cone/.style={dashed, color=blueNS},
	cone/.style={line width=1pt, color=blueNS, fill=blueNS, fill opacity=0.45},
]

	\node at (-0.1,3.8) [right]{(a)};

	\node at (1, 4) (PR) {};
	\node at (0, 1) (p1) {};
	\node at (3, 1) (p2) {};
	\node at (0-0.013, 0.93) (p1bis) {};
	\node at (3+0.0295, 0.93) (p2bis) {};
	\node at (0-0.013, 0) (p3) {};
	\node at (3+0.0295, 0) (p4) {};
	
	\draw[NS] (PR.center) -- (p1.center) -- (p2.center) -- cycle;
	\draw[L] (p1bis.center) -- (p2bis.center) -- (p4.center) -- (p3.center) -- cycle;
	
	\node[color=blueNS] at (1.1, 2.9) (NS) {$\NS$};
	\node[color=greenL] at (0.5, 0.5) (L) {$\LL$};
	\node at (1.8, 1.6) (P) {};

	\draw[fill=black] (P) circle (0.06);
	\node (labelP) at (3.1, 1.6) [right]{$\P$};
	\draw[->, line width=0.8pt, >=stealth] (labelP) -- (P.east);
	
	\definecolor{my-red}{RGB}{200, 0, 0}
	\node[color=my-red] (OrbitP) at (3.4, 2.05) {\footnotesize $\Orbit(\P)$};
	\foreach \point in {
		(2.3, 1.7),
		(2, 1.8), 
		(1.6, 1.9)
		}{
		\draw[dashed] (OrbitP.west) -- \point;
		}
	\foreach \point in {
		(0.5, 0.1),
		(0.2,0.2), 
		(-0.2, 0.3),
		(0.3,-0.22),
		(-0.4, -0.4),
		(-0.6, 0.27),
		(-0.5, 0),
		(-1, 0.1),
		(-0.9, -0.3),
		(-0.8, 0),
		(-0.7, -0.2),
		(-0.6, -0.05),
		(-0.4, -0.1),
		(-1.1, -0.1),
		(-0.7, -0.4),
		(-0.2, -0.8),
		(0.2, -1),
		(0.4, -0.9),
		(-0.05, -0.45),
		(0.6, -0.45),
		(-0.05, -0.1),
		(0.3, -0.49)
		}{
		\draw[fill=red] (P)+\point circle (0.06);
		}

	\node [right] at (2.1, 3.3) {\fbox{\parbox{2.2cm}{\small If $\exists\,\Q\!\!\in\!\!\Orbit(\P)$ that is collapsing, then $\P$ is also collapsing.}}};

\end{tikzpicture}
}
	\hspace{\myspace}
	\definecolor{blueNS}{rgb}{0.1, 0.35, 0.45}
\definecolor{greenL}{rgb}{0., 0.39, 0.}
{
\begin{tikzpicture}[every node/.style={scale=\TEXTsize}, scale=\FIGUREsize,
	NS/.style={line width=2pt, color=blueNS, fill=blueNS, fill opacity=0.15},
	L/.style={line width=2pt, color=greenL, fill=greenL, fill opacity=0.20},
	anti-cone/.style={dashed, color=blueNS},
	cone/.style={line width=1pt, color=blueNS, fill=blueNS, fill opacity=0.45},
]

	\node at (-0.1,3.8) [right]{(b)};

	\node at (1, 4) (PR) {};
	\node at (0, 1) (p1) {};
	\node at (3, 1) (p2) {};
	\node at (0-0.013, 0.93) (p1bis) {};
	\node at (3+0.0295, 0.93) (p2bis) {};
	\node at (0-0.013, 0) (p3) {};
	\node at (3+0.0295, 0) (p4) {};
	
	\draw[NS] (PR.center) -- (p1.center) -- (p2.center) -- cycle;
	\draw[L] (p1bis.center) -- (p2bis.center) -- (p4.center) -- (p3.center) -- cycle;
	
	\node[color=blueNS] at (0.67, 1.63) (NS) {$\NS$};
	\node[color=greenL] at (0.6, 0.5) (L) {$\LL$};
	\node at (1.8, 1.6) (P) {};
	\node at (2.46, 1.82) (Aux1) {};
	\node at (0.42, 2.29) (Aux2) {};
	
	\draw[cone] (P.center) -- (Aux1.center) -- (PR.center) -- (Aux2.center) -- cycle;
	\draw[fill=black] (P) circle (0.04) node[below] {$\P$};
	\draw[anti-cone] (p1) -- (P);
	\draw[anti-cone] (p2) -- (P);
	
	\node at (1.3, 2.5) {$\C_\P\cap\NS$};
	
	\node [right] at (2.2, 3.3) {\fbox{\parbox{2.3cm}{\small If $\P$ is collapsing, then any box in $\C_\P\cap\NS$ is also collapsing.}}};

\end{tikzpicture}
}
	\hspace{\myspace}
	\definecolor{blueNS}{rgb}{0.1, 0.35, 0.45}
\definecolor{greenL}{rgb}{0., 0.39, 0.}
{
\begin{tikzpicture}[every node/.style={scale=\TEXTsize}, scale=\FIGUREsize,
	NS/.style={line width=2pt, color=blueNS, fill=blueNS, fill opacity=0.15},
	L/.style={line width=2pt, color=greenL, fill=greenL, fill opacity=0.15},
	anti-cone/.style={dashed, color=blueNS},
	cone/.style={line width=1pt, color=blueNS, fill=blueNS, fill opacity=0.45},
	collapsing/.style={line width=1pt, color=orange, fill=orange, fill opacity=0.5},
	>=stealth
]

	\node at (-0.1,3.8) [right]{(c)};

	\node at (1, 4) (PR) {};
	\node at (0, 1) (p1) {};
	\node at (3, 1) (p2) {};
	\node at (0-0.013, 0.93) (p1bis) {};
	\node at (3+0.0295, 0.93) (p2bis) {};
	\node at (0-0.013, 0) (p3) {};
	\node at (3+0.0295, 0) (p4) {};
	
	\draw[NS] (PR.center) -- (p1.center) -- (p2.center) -- cycle;
	\draw[L] (p1bis.center) -- (p2bis.center) -- (p4.center) -- (p3.center) -- cycle;
	
	\node[color=blueNS] at (1.1, 2.9) (NS) {$\NS$};
	\node[color=black] at (0.5, 0.5) (L) {$\LL$};
	\node at (1.8, 1.6) (P) {};
	
	\draw[collapsing] (p2bis.center) -- (2.3, 1.7) -- (2, 1.8) -- (1.6, 1.9) -- (1.2, 1.87) -- (0.8, 1.7) -- (p1bis.center) -- (p3.center) -- (p4.center) -- cycle;

	\draw[fill=black] (P) circle (0.06);
	\node (labelP) at (3.1, 1.6) [right]{$\P$};
	\draw[->, line width=0.8pt] (labelP) -- (P.east);
	
	\definecolor{my-red}{RGB}{200, 0, 0}
	\node[color=my-red] (OrbitP) at (3.4, 2.05) {\footnotesize $\Orbit(\P)$};
	\foreach \point in {
		(2.3, 1.7),
		(2, 1.8), 
		(1.6, 1.9)
		}{
		\draw[dashed] (OrbitP.west) -- \point;
		}
	\foreach \point in {
		(0.5, 0.1),
		(0.2,0.2), 
		(-0.2, 0.3),
		(0.3,-0.22),
		(-0.4, -0.4),
		(-0.6, 0.27),
		(-0.5, 0),
		(-1, 0.1),
		(-0.9, -0.3),
		(-0.8, 0),
		(-0.7, -0.2),
		(-0.6, -0.05),
		(-0.4, -0.1),
		(-1.1, -0.1),
		(-0.7, -0.4),
		(-0.2, -0.8),
		(0.2, -1),
		(0.4, -0.9),
		(-0.05, -0.45),
		(0.6, -0.45),
		(-0.05, -0.1),
		(0.3, -0.49)
		}{
		\draw[fill=red] (P)+\point circle (0.06);
		}
	
	\node[color=orange] (conv) at (3.2, 0.4) [right]{\parbox{2cm}{\footnotesize$C=\Conv\big(\LL\,\cup$\\$\bigcup_\W \Orbit_\W(\P)\big)$}};
	\draw[line width=0.5pt, ->] (conv.168) -- (2.8, 0.6);

	\node [right] at (2.1, 3.3) {\fbox{\parbox{2.2cm}{\small If $\exists\,\Q\!\in\!C$ that is collapsing, then $\P$ is also collapsing.}}};

\end{tikzpicture}
}
	\centering
	\caption{Orbits that collapse communication complexity.}
	\label{fig: collapsing orbits}
\end{figure}

		\subsection{Case Study: Orbit of $\Wbs$}
		\label{subsec:case-study}

In this subsection, we focus our attention on the wiring $\Wbs$ inspired by Brunner and Skrzypczyk~\cite{BS09}. Denote $\boxtimes$ the corresponding box multiplication. Define the \emph{shared randomness} box as $\SR:=(\P_0+\P_1)/2$; it is designed to output a couple $(a,b)$ such that $a=b$ uniformly and independently of the inputs. From the multiplication table in \refprop[Figure]{fig: multiplication table}, one can see that the $2$-dimensional affine space $\A:=\Aff\{\PR, \SR, \I\}$ is stable under $\boxtimes$.
As a consequence, the orbit $\Orbit(\P)$ of any box $\P$ in $\A$ is itself included in $\A$, and as $\A$ is two-dimensional, it is particularly easy to draw the orbit of a box in that case.
We represent an orbit in \refprop[Figure]{figure: orbit of a box}.

\begin{figure}[h]
	\hspace{-1cm}
	\includegraphics[width=9.5cm]{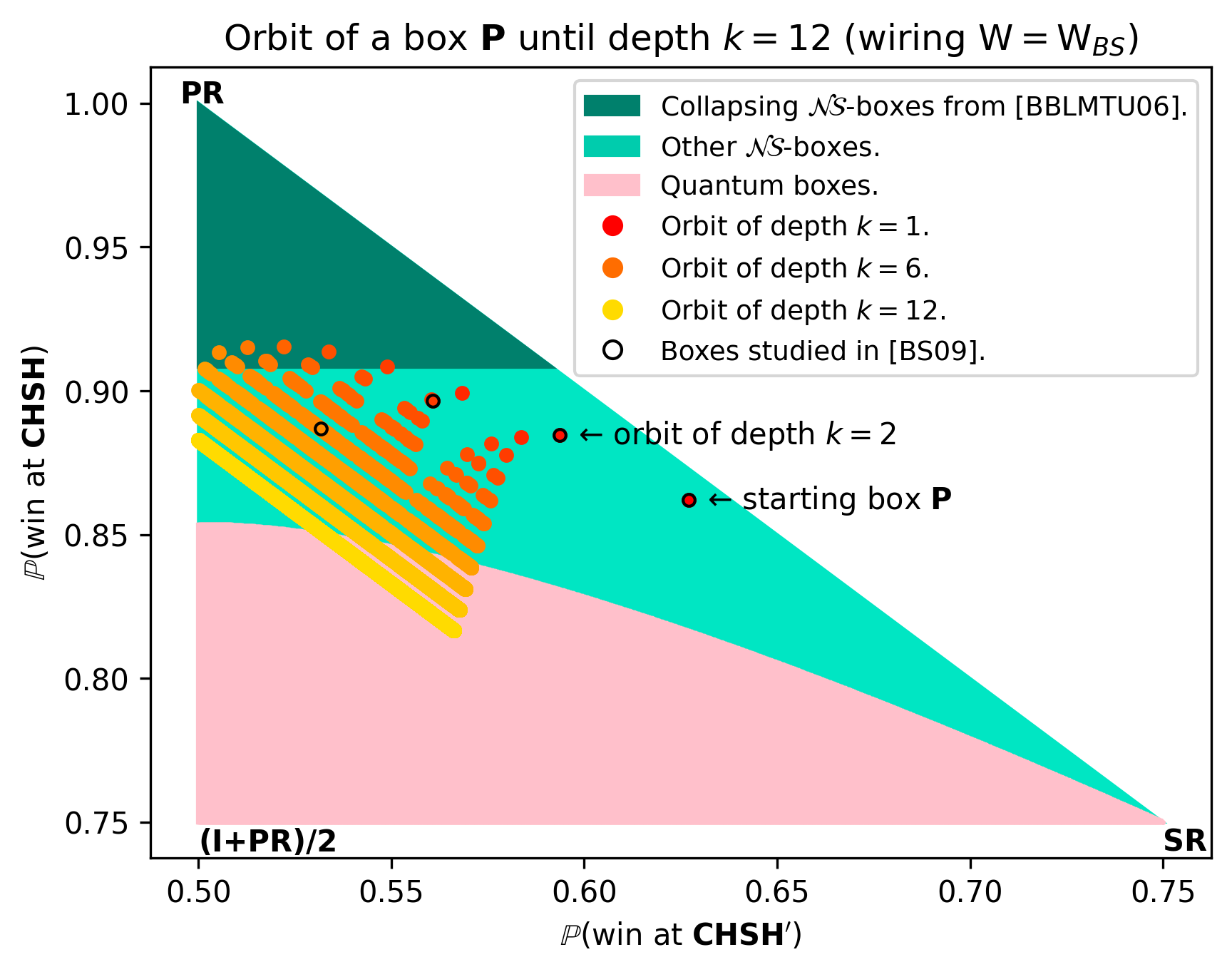}
	\centering
	\caption{Example of a box orbit, drawn for depth up to $k=12$.
	The quantum area $\QQ$ (in pink) is drawn using formulas from~\cite{Masanes03}.
	Dark green represents the collapsing area that was found by Brassard \emph{et. al.} in~\cite{BBLMTU06}, which consists of all the boxes with $\CHSH$-value higher than $\frac{3+\sqrt{6}}{6}\approx0.91$.
	The orbit is drawn in yellow and orange dots --- observe that it intersects the collapsing area in dark green, so \refprop[Proposition]{prop: collapsing orbit} tells us that the starting box $\P$ is collapsing.
The black circles represent the boxes that were studied in~\cite{BS09}, doing ``pairwise" multiplications: $\P$, $\P\boxtimes\P$, $(\P\boxtimes\P)\boxtimes(\P\boxtimes\P)$, \emph{etc}... Each iteration is the wiring of two copies of the previous iteration, it gives a subset of our orbit. As displayed in the drawing and detailed in the proof of \refprop[Proposition]{theo: new known collapsing boxes}, our method allows us to find a larger set of boxes $\P$ that are collapsing.
	}
	\label{figure: orbit of a box}
\end{figure}

By definition of the affine space $\A$, any box $\Abox\in\A$ can be uniquely written as $\Abox=c_1(\Abox)\,\PR + c_2(\Abox)\,\SR + c_3(\Abox)\,\I$ for some real coefficients $c_i(\Abox)$ that sum to $1$, called \emph{convex coordinates} of $\Abox$ in the affine basis $\{\PR, \SR, \I\}$. An interesting aspect of considering convex coordinates is that it gives a simple characterization of the parallelism property of lines:
\begin{equation}   \label{eq:equivalent-condition-of-being-parallel}
	\forall \Abox,\Bbox\in\A,\quad\quad
	\Aff\{\Abox, \Bbox\}\, ||\, \Aff\{\PR, \SR\}
	\quad\Longleftrightarrow\quad
	c_3(\Abox) = c_3(\Bbox)\,.~~\footnote{Indeed, for $\Abox\neq\Bbox\in\A$ whose convex coefficients are respectively $a_1, a_2, a_3$ and $b_1, b_2, b_3$, saying that the line $\Aff\{\Abox, \Bbox\}$ is parallel to the line $\Aff\{\PR, \SR\}$ is equivalent to knowing that there exists a scalar $\lambda\in\R^*$ such that $\Abox-\Bbox = \lambda\,(\PR-\SR)$, \ie there exists $\lambda\in\R^*$ such that $a_1-b_1=\lambda$ and $a_2-b_2 = -\lambda$ and $a_3-b_3=0$, \ie we have two equations: $a_1+a_2=b_1+b_2$ and $a_3=b_3$. Finally, using the normalization condition $\sum_i a_i = \sum_j b_j = 1$, we see that these two equations are equivalent to simply imposing $a_3 = b_3$, as claimed.}
\end{equation}
Moreover, in our case, we have an additional interesting property of the third convex coordinate:

\begin{lemma}    \label{lem:multiplicativity-of-the-third-coef}
	The function $1-c_3(\cdot)$ is multiplicative:
	\[
		\forall \Abox,\Bbox\in\A,\quad\quad
		1-c_3(\Abox\boxtimes\Bbox) = \big(1-c_3(\Abox)\big)\,\big(1-c_3(\Bbox)\big)\,.
	\]
\end{lemma}

\begin{proof}
The multiplication table induced by the wiring $\Wbs$~\cite{BS09} is:
\begin{equation}   \label{eq:multiplication-table-PR-SR-I}
	\text{
	\setlength{\length}{0.6cm}
	\small
	\renewcommand{\arraystretch}{1.5}
	\newcommand{\mycommand}{\normalsize}
	\begin{tabular}{| c ||c|c|c|c|c|c|c|c|}
		\hline
		{\small\diagbox{$\P$}{$\Q$}} & \hspace{\length}{\mycommand$\PR$}\hspace{\length} & \hspace{\length}{\mycommand$\SR$}\hspace{\length} & \hspace{\length}{\mycommand$\I$}\hspace{\length} \\
		\hline\hline
		{\mycommand$\PR$} &  $\PR$ & $\PR$& $\I$  \\ \hline
		{\mycommand$\SR$} & $\textstyle\frac{1}{2}\PR + \frac12\SR$  & $\SR$ & $\I$  \\ \hline
		{\mycommand$\I$}  & $\textstyle\frac14\PR - \frac14\SR + \I$ &  $\I$ & $\I$  \\ \hline
	\end{tabular}
	\renewcommand{\arraystretch}{1}
	}\,,
\end{equation}
where each cell displays the result of $\P\boxtimes\Q$.
For $\Abox, \Bbox\in\A$ whose coefficients $c_i$ are denoted $a_1, a_2, a_3$ and $b_1, b_2, b_3$ for the sake of readability,
we use the bilinearity of the product~$\boxtimes$ and we get:
\begin{align*}
	\Abox\boxtimes\Bbox
	\,=\,
	&
	\Big[ \textstyle a_1b_1 + a_1b_2 + \frac12 a_2 b_1 + \frac14 a_3 b_1 \Big]\,\PR
	\,+
	\Big[ \textstyle\frac12 a_2 b_1 + a_2 b_2 - \frac14 a_3 b_1 \Big]\,\SR
	\\
	& \quad +
	\Big[ a_1b_3 + a_2 b_3 + a_3 b_1 + a_3 b_2 + a_3 b_3 \Big]\,\I\,.
\end{align*}
Hence, using the normalization property of coefficients $\sum_i a_i = \sum_j b_j = 1$, the third coefficient simplifies as $b_3 + a_3(1-b_3)$, which is equal to $1-(1-a_3)(1-b_3)$ as wanted.
\end{proof}

Now, interestingly, we observe that the points of a given $k$-orbit are all aligned, and we even know the equation of the line:

\begin{theorem}[Alignment]
	\label{theo: alignment and parallelism}
	For any $k\geq1$ and $\P\in\mathcal \mathcal \A$, the points of $\Orbit^{(k)}(\P)$ are all aligned on a line $\line_k$ whose expression in convex coordinates is given by:
	\[
		\line_k
		\,:=\,
		\Big\{ \Abox\in\A\,:\, c_3(\Abox) = 1 - \big(1-c_3(\P)\big)^k \Big\}\,.
	\]
\end{theorem}

\begin{proof}
	We prove by induction on $k\geq1$ that $\Orbit^{(k)}\subseteq\line_k$.
	For $k=1$, the $1$-orbit contains only one element, namely $\P$, which obviously satisfies $c_3(\P) = 1 - \big(1-c_3(\P)\big)$, so $\P$ indeed belongs to $\line_1$.
	Now, assume the result holds \emph{until} some integer $k\geq1$, and let $\Q\in\Orbit^{(k)}$. By definition, the box~$\Q$ decomposes as $\Q=\Q_1\boxtimes\Q_2$, for some $\Q_1\in\Orbit^{(\ell)}$ and $\Q_2\in\Orbit^{(k-\ell)}$ for some $1\leq \ell \leq k-1$. By the induction hypothesis, we know that
	$
		c_3(\Q_1) = 1 - \big( 1-c_3(\P) \big)^\ell$ and
		$c_3(\Q_2) = 1 - \big( 1-c_3(\P) \big)^{k-\ell}\,.
	$
	Then using \refprop[Lemma]{lem:multiplicativity-of-the-third-coef}, we obtain:
	\begin{align*}
		c_3(\Q)
		\,=\, &
		1 - \big( 1-c_3(\Q_1) \big)\,\big( 1-c_3(\Q_2) \big)
		\\
		\,=\, &
		1 - \big( 1-c_3(\P) \big)^\ell\,\big( 1-c_3(\P) \big)^{k-\ell}
		\\
		\,=\, &
		1 - \big( 1-c_3(\P) \big)^k\,,
	\end{align*}
	which means that $\Q$ belongs to the line $\line_k$.
\end{proof}

As a consequence, we see that all the points of the $k$-orbit have the same third convex coefficient, so using the equivalence given in \eqrefprop[Equation]{eq:equivalent-condition-of-being-parallel}, we obtain:

\begin{corollary}[Parallelism]   \label{coro:parallelism}
	The supporting line $\line_k$ of all the orbits $\Orbit^{(k)}$ are parallel to the diagonal line $\line_D:=\Aff\{\PR, \SR\}$:
	\[
		\forall k\geq1, \quad\quad \Orbit^{(k)}||\,\line_D\,.
	\]
	In particular, all the orbits are parallel to each other:
	\begin{equation*}
		\forall k,\ell\geq1,\quad \Orbit^{(k)}||\Orbit^{(\ell)}\,.
		\tag*{\qed}
	\end{equation*}
\end{corollary}

Moreover, looking closely at the sequence of coefficients $1 - \big(1-c_3(\P)\big)^k$ and noticing that the diagonal line $\line_D$ is defined by the equation $c_3(\Abox) = 0$, we see that:

\begin{corollary}[Orbits move to the left]  \label{coro:orbits-move-to-the-left}
	Assume $\P\notin\line_D$.
	Then the orbits are more and more distant from the diagonal line as $k$ grows.
	Moreover, the sequence of lines $(\line_k)_k$ tends to the line $\line_\infty$ defined by the equation $c_3(\Abox)=1$, which is exactly the line passing through $\I$ and parallel to the diagonal $\line_D$.
	\qed
\end{corollary}

It takes a lot of computational time to draw $k$-orbits of a box $\P$ as $k$ grows, since it requires to compute $\frac{1}{k}\tiny\begin{pmatrix} 2k-2 \\k-1 \end{pmatrix}$ elements (Catalan number), which grows exponentially.
However, our goal is not to compute the whole orbit, but simply to determine whether or not the orbit intersects the known collapsing area (dark green). To that end, one may notice that it is enough to compute the ``highest" box of each $k$-orbit in the $y$-coordinate (see \refprop[Figure]{figure: orbit of a box}) and to check whether those ``highest" boxes intersect the collapsing area (dark green area).
This is the purpose of the following proposition, which displays a simple expression of the ``highest" box of each $k$-orbit, and which allows much faster tests of a box $\P$ being collapsing or not without computing all the points of the orbit.
We prove this result only in a subset of the orbit, that we call \emph{tilted orbit}, which is easier to manipulate in inductions, and which is defined by $\Orbittilde{}^{(1)}(\P)
	\,:=\,
	\{\P\}$ and for $k\geq2$:
\[
	\Orbittilde{}^{(k)}
	\,:=\,
	\Big(\P \boxtimes \Orbittilde{}^{(k-1)}  \Big) \cup \Big( \Orbittilde{}^{(k-1)} \boxtimes \P  \Big)
	\,=\,
	\bigcup_{\ell\in\{1, k-1\}}
	\Orbittilde{}^{(\ell)} \boxtimes \Orbittilde{}^{(k-\ell)}
	\subseteq
	\Orbit^{(k)}
	\,.
\]
Note that the cardinality of that set is $\#\Orbittilde{}^{(k+1)}=2^k$, up to multiplicity.
We call $\CHSH$-value the $y$-coordinate, indicating how ``high" is a box:
$$
	\CHSH(\P)
	\,:=\,
	\mathbb P\big( \text{win at $\CHSH$} \big)
	\,=\,
	\frac14 \sum_{a\oplus b=xy} \P(a,b\,|\,x,y)\,.
$$
We say that a tilted orbit \emph{distills the $\CHSH$-value} if it contains a box $\Q$ such that $\CHSH(\Q)\geq\CHSH(\P)$. {In the following theorem, we present the expression of the best parenthesization in terms of $\CHSH$-value, which explains the numerical observation reported in~\cite[Supplementary Material, II]{EWC22PRL}}:

\begin{theorem}[Highest box]  \label{theo: highest box}
Let $\P\in\Atilde$ be a box, and let $k\geq2$ an integer such that the tilted $(k-1)$-orbit distills the $\CHSH$-value.
	Then the highest $\CHSH$-value of $\Orbittilde{}^{(k)}(\P)$ is achieved at a box whose expression is the product of $k$ times $\P$ \emph{on the right}:
\[
	\P^{\boxtimes k}
	\,:=\,
	\Big(\big(\text{\small$(\P\boxtimes\P)$}\boxtimes \P\big)\cdots\Big)\boxtimes\P
	\,\,\in
	\underset{\Q\in\Orbittilde{}^{(k)}(\P)}{\argmax} \,\,\,\CHSH(\Q)
	\,.
\]
\end{theorem}

\begin{proof}
	See \refprop[Subsection]{appendix: proof of Highest Box proposition}.
\end{proof}

\begin{remark}
	In the Ph.D. thesis of Giorgos Eftaxias~\cite{Eftaxias22}, the author presents three types of architectures of wirings in Subsection 5.5.1: the exponential architecture, that we call here pair-wise multiplication, used in~\cite{BS09}; the linear architecture, that is the same as the one used in the theorem above; and the Fibonacci architecture. They present some subsets of $\NS$ for which the linear architecture seems to be the best one among the three (Remark 1), and some other subsets of $\NS$ for which it is the Fibonacci architecture (Subsection 5.F.2).
\end{remark}

\begin{conjecture}[Dyck paths]
	We conjecture that the same result actually holds without the tilde, \ie the right multiplication $\P^{\boxtimes k}$ gives the highest $\CHSH$-value of $\Orbit^{(k)}(\P)$, as observed numerically.
	An idea of the proof could be to use Dyck paths. Each time we open/close a parenthesis, the path goes up/down respectively, which produces a certain Dyck path.
	The statement to be proved is that each time we convert a $\vee$ into a $\wedge$, the $\CHSH$-value is non-decreasing. Then, we would have that the best Dyck path is necessarily the one that always goes up first and then always goes down, which corresponds to the multiplication of boxes on the right.
\end{conjecture}

{As previously mentioned, the next theorem is concurrent and independent of the work of~\cite{EWC22PRL}:}

\begin{theorem}[New collapsing boxes]
	\label{theo: new known collapsing boxes}
	The techniques described in \refprop[Subsection]{subseq:ConsequencestoCommunicationComplexity} allow the discovery of new collapsing boxes.
	See new collapsing areas in \refprop[Figure]{fig:numerical-new-collapsing-regions}.
\end{theorem}

\begin{proof}
	See \refprop[Figure]{figure: orbit of a box} for an intuition of the proof.
	Take the starting box $\P$ with coordinates $(0.627, 0.862)$ in the affine plane $\A=\Aff\{\PR, \SR, \I\}$, where the coordinate system is given by the $\CHSH'$- and $\CHSH$- values of $\P$.
	On the one hand, the tilted orbit of $\P$ intersects the collapsing area that was found in~\cite{BBLMTU06} (in dark green), since for instance $\CHSH(\P^{\boxtimes5}) \approx 0.913 > 0.908 \approx \frac{3+\sqrt{6}}{6}$, so $\P$ is collapsing by \refprop[Proposition]{prop: collapsing orbit}.
	On the other hand, this box $\P$ does not lie in any of the previously-known collapsing areas from~\cite{vD99, BBLMTU06, BS09, BMRC19, BBP23} {(to the best of our knowledge, these five references are the only previous results showing a collapse of communication complexity, in addition to~\cite{EWC22PRL} which concurrently and independently found a similar result to ours as mentioned before)}.
	Indeed, it is not in the collapsing areas from~\cite{vD99, BBLMTU06} since $\CHSH(\P)=0.862<0.908$, nor is it in the collapsing area from~\cite{BBP23} since $A+B \approx 14.13 < 16$ (using the authors' notation).
	{The box $\P$ neither is in any of the collapsing regions found in~\cite{BMRC19} since it does not belong to the boundary $\partial\NS$ of the non-signalling set.}
	The last area to check is the one from~\cite{BS09}, which was numerically found. From a box $\P$, they define a sequence of boxes using ``pairwise" multiplications: $\Q_1 := \P$ and $\Q_{n+1}:=\Q_n\boxtimes\Q_n$ for $n\geq1$, and they check whether or not there exists an integer $n$ such that $\CHSH(\Q_n)>\frac{3+\sqrt{6}}{6}$. But, for our starting box $\P$, none of the $\Q_n$ satisfy this inequality: indeed, for $1\leq n\leq 5$, it possible to check it by hand, for $n=6$ we have $\Q_6\in \LL$, and for $n\geq7$ we also have $\Q_n\in\LL$ since $\LL$ is closed under wirings~\cite{ABLPSV09}.
	Hence our example $\P$ is a new collapsing box.
\end{proof}

		\subsection{Some Other Orbits}
		\label{subsec:some-other-orbits}

In the previous subsection, we studied specifically the orbit of the wiring $\Wbs$ in the slice of $\NS$ passing through the boxes $\PR$, $\SR$, $\I$.
Here we comment on some examples of other orbits in three different ways:
(i)~it is possible to study the same wiring $\Wbs$ but in different slices of $\NS$;
(ii)~it is possible to study another wiring than $\Wbs$ but to keep the same slice as in the previous subsection;
(iii)~it is possible to change both the wiring and the slice.
See \refprop[Appendix]{sec:drawing-of-some-orbits} for many drawings.

(i) We keep the wiring $\Wbs$ and we consider the slice of $\NS$ passing through $\PR$, $\P_0$, $\P_1$.
Notice that we prove in \refprop[Proposition]{prop:PR-P0-P1-is-a-face-of-NS} that this slice is actually precisely the convex combination of the three points.
We draw two examples of such an orbit in \refprop[Figure]{figure:orbit-of-Wbs-in-a-different-slice}, with two different starting boxes. We observe that both of them seem to recover the alignment and parallelism properties that we showed in \refprop[Theorem]{theo: alignment and parallelism} and its corollary. Again, these lines seem all parallel to what we called previously the diagonal line $\line_D$, which is defined as the line passing through $\PR$ and $\SR=\frac12(\P_0+\P_1)$.
Notice that we show in \refprop[Theorem]{thm: the triangle PR-P0-P1 is collapsing} that all the boxes of this triangle are actually collapsing, except the ones in the segment $\Conv\{\P_0, \P_1\}$, drawn in pink.

\begin{figure}[h]
	\includegraphics[height=5.5cm]{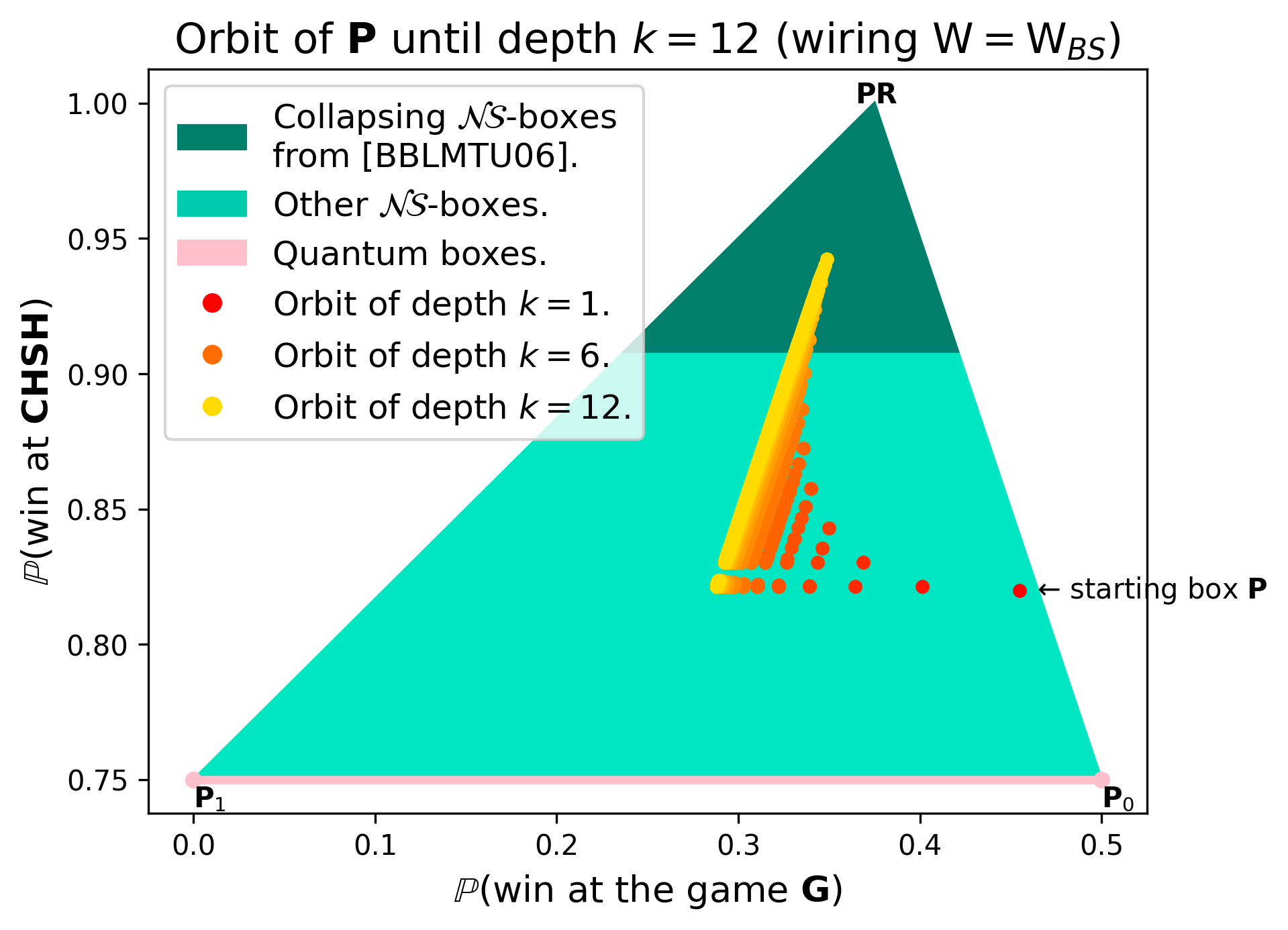}
	\includegraphics[height=5.5cm]{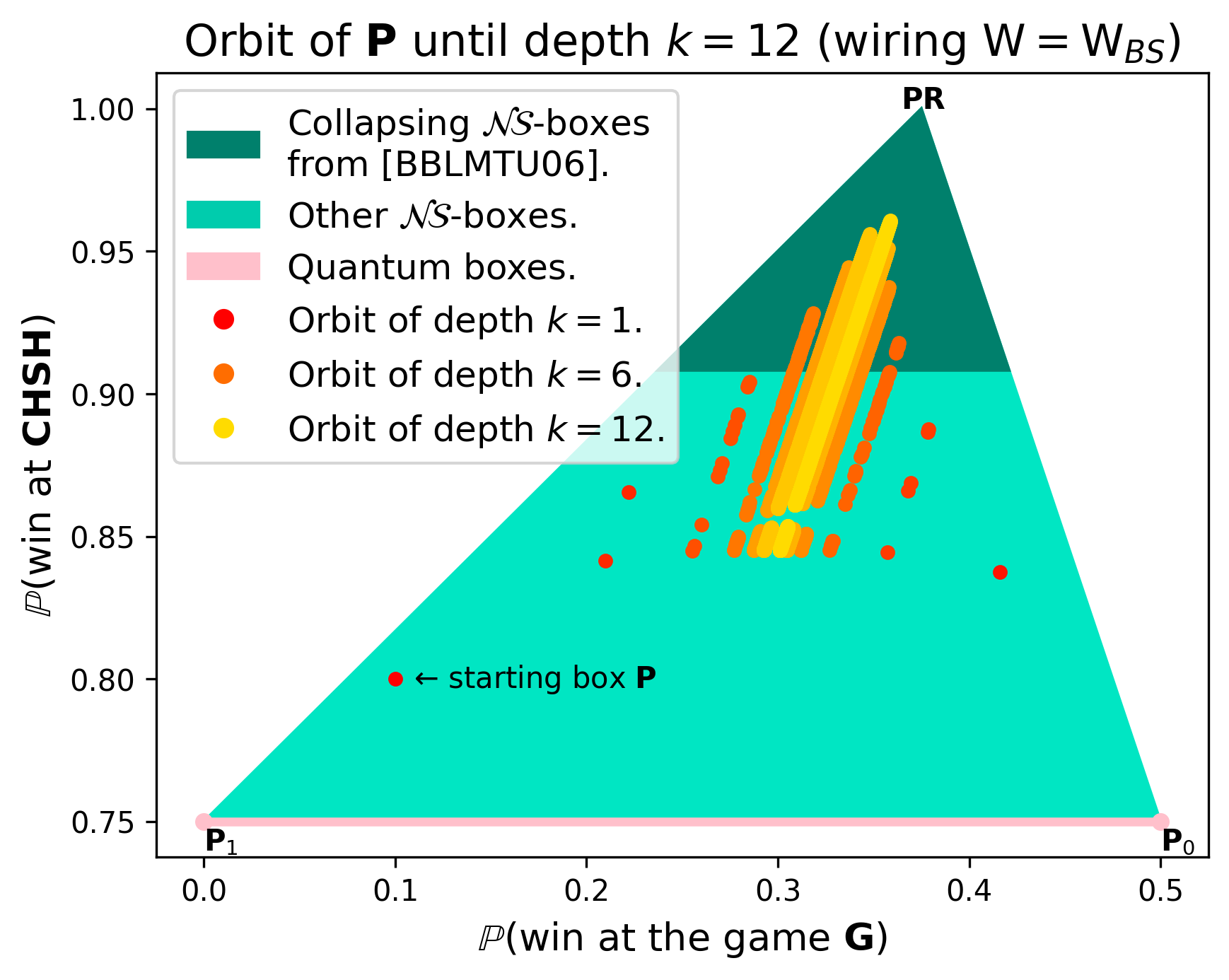}
	\centering
	\caption{Orbit of $\Wbs$ in a different slice than in \refprop[Subsection]{subsec:case-study}: here we consider the slice of $\NS$ passing through $\PR$, $\P_0$, and $\P_1$. We represent the orbit with two different starting boxes.
	Each orbit is drawn with depth going until $k=12$.
	The game $\G$ is defined by the winning rule $a=0$ and $b=y$.
	Notice that we give a proof based on CC that this triangle $\Conv\{\PR,\P_0,\P_1\}$ is a quantum void in \refprop[Corollary]{coro:quantum-boxes-in-the-boundary-of-NS}, which is why the only quantum boxes in this triangle are actually local boxes.
	}
	\label{figure:orbit-of-Wbs-in-a-different-slice}
\end{figure}

(ii) Among the ``typical" wirings defined in \refprop[Subsection]{subsec:typical-examples-of-wirings}, the only ones that stabilize the plane $\Aff\{\PR, \SR, \I\}$ are $\Woplus$ and $\Wbs$, see \refprop[Appendix]{ap:the-multiplication-tables}.
This is why, for these two wirings, we can conveniently draw the orbits in a plane.
The orbit of $\Woplus$ is drawn in \refprop[Appendix]{sec:drawing-of-some-orbits}~(a). We observe that each $k$-orbit contains only one element, which is not surprising since we know from \refprop[Figure]{fig: commutativity and associativity of the induced algebras} that its induced algebra is associative, meaning that the choice of parenthesization does not lead to a different result.
In the same appendix, we also draw the orbit for three other wirings. Surprisingly, we observe that the three new orbits look the same as the orbit of $\Wbs$.

(iii) We add the slice $\PR$, $\P_0$, $\P_1$ for each wiring of \refprop[Appendix]{sec:drawing-of-some-orbits}. We observe that the alignment and parallelism properties seem to still hold in those cases, as before. Moreover, we see that the example~(d) distills the $\CHSH$-value better than the other examples in that slice.

		\subsection{Proof of~\refprop[Theorem]{theo: highest box}}
		\label{appendix: proof of Highest Box proposition}
	
Recall that $\SR := (\P_0+\P_1)/2$ is the \emph{shared randomness} box.
Given a non-signalling box $\P\in\NS$, its $\CHSH$- and $\CHSH'$- values are defined as follows:
\[
	\CHSH(\P)
	\,:=\,
	\frac14
	\sum_{a\oplus b=xy} \P(a,b\,|\,x,y)\,,
	\quad\quad \text{and} \quad\quad
	\CHSH'(\P)
	\,:=\,
	\frac14
	\hspace{-0.4cm}
	\sum_{\tiny\begin{array}{c}a\oplus b\\=(x\oplus1)(y\oplus1)\end{array}}\hspace{-0.4cm} \P(a,b\,|\,x,y)\,.
\]
For example, we have $\CHSH(\PR) = 1$ and $\CHSH(\SR)=\frac34$ and $\CHSH(\I) = \frac12$.
Denote $\A$ the affine space $\A:=\Aff\{\PR, \SR, \I\}$, and denote $\Atilde\subseteq \A$ the set of boxes $\P$ in the convex hull $\Conv\{\PR, \SR, \I\}$ whose $\CHSH$-value is $\geq3/4$.
We will prove our results in $\Atilde$; by the symmetry of the problem, similar results also hold in other areas, such as $2\I-\Atilde$ the symmetric of $\Atilde$ by $\I$.
\vspace{-0.1cm}

\begin{lemma}[Multiplying by $\P$ preserves the $\CHSH$-value order]
\label{lem: Multiplication by P preserves the CHSH order}
Let $\P\in\Atilde$, and let $\Q\neq\Rbox\in\A$ such that the line $\Aff\{ \Q, \Rbox\}$ is parallel to the diagonal line $\line_D:=\Aff\{\PR, \SR\}$.
We have:
\[
	\CHSH(\Q)\geq\CHSH(\Rbox)
	\quad\quad \Longrightarrow \quad\quad
	\left\{
	\begin{array}{c}
		\CHSH(\Q\boxtimes \P)\geq\CHSH(\Rbox\boxtimes \P)\,,\\
		\CHSH(\P\boxtimes \Q)\geq\CHSH(\P\boxtimes \Rbox)\,.
	\end{array}
	\right.
\]
\end{lemma}
\vspace{-0.3cm}

\begin{proof}
	As the box $\P$ lies in $\Atilde$, it is of the form $\P=p_1\PR + p_2\SR + (1-p_1-p_2)\I$ for some coefficients $p_1,p_2\geq0$ such that $p_1+p_2\leq1$.
	Rewrite it as $\P = \tiny \begin{pmatrix} p_1 \\ p_2 \end{pmatrix}$, and similarly denote $\Q = \tiny \begin{pmatrix} q_1 \\ q_2 \end{pmatrix}$ and $\Rbox = \tiny \begin{pmatrix} r_1 \\ r_2 \end{pmatrix}$ for some coefficients $q_i, r_j\in\R$.
	By the parallelism assumption, vectors $\Q-\Rbox$ and $\PR-\SR$ have to be colinear, \ie~there must exist some $\lambda\in\R^*$ such that $\Q-\Rbox = \lambda(\PR-\SR) = \lambda \tiny \begin{pmatrix} 1 \\ -1 \end{pmatrix}$, so we may rewrite the second coefficient of $\Rbox$ as $r_2 = q_1 + q_2 - r_1$.
	With this notation, we can use the linearity of the function $\CHSH(\cdot)$ to see that condition $\CHSH(\Q)\geq\CHSH(\Rbox)$ simplifies to $(q_1-r_1)\geq0$:
	\begin{align*}
		\small
		\CHSH(\Q)\!-\!\CHSH(\Rbox)
		&=\,
		\small
		(q_1\!-\!r_1) \!\times\! \CHSH(\PR)
		+ (q_2\!-\!r_2)\!\times\! \CHSH(\SR)
		+ \Big(\!(1\!-\!q_1\!-\!q_2) \!-\! (1\!-\!r_1\!-\!r_2)\!\Big)\!\times\! \CHSH(\I)
		\\
		&=\,
		(q_1-r_1) \times 1
		- \textstyle (q_1-r_1)\times\frac34
		+ 0 \times \frac12
		\\
		&=\,
		\textstyle
		\frac14(q_1-r_1)\,.
	\end{align*}
	Now, using the multiplication table from \refprop[Figure]{fig: multiplication table} and bilinearity of $\boxtimes$, we may compute the following expressions:
	\begin{align*}
		\CHSH(\Q\boxtimes \P)-\CHSH(\Rbox\boxtimes \P)
		&\,=\,
		\textstyle\frac{1}{8} (p_1+2 p_2) (q_1-r_1)\geq0\,,
		\\
		\CHSH(\P\boxtimes \Q)-\CHSH(\P\boxtimes \Rbox)
		&\,=\,
		\textstyle\frac{1}{16} (1-p_1+p_2) (q_1-r_1)\geq0\,,
	\end{align*}
	which gives the desired result.
\end{proof}
\vspace{-0.50cm}

\begin{lemma}[Right multiplication gives better $\CHSH$-value]
\label{lem: Multiplication to the right is higher}
For any $\P\in\Atilde$ and $\Q\in\Orbittilde(\P)$, we have:
\[
	\CHSH(\Q)\geq\CHSH(\P)
	\quad \Longrightarrow \quad
	\CHSH\Big( \Q\boxtimes\P \Big)
	\,\geq\,
	\CHSH\Big( \P\boxtimes\Q \Big).
\]
\end{lemma}

\begin{proof}
	Use the coordinate system $(x,y)$ given by the $\CHSH'$- and $\CHSH$- values respectively in order to write $\P$ and $\Q$ as taking coordinates $(x_\P, y_\P)$ and $(x_\Q, y_\Q)$.
For instance we have $\PR:(\frac12,1)$ and $\SR:(\frac34,\frac34)$ and $\I:(\frac12, \frac12)$.
	Use the multiplication table from \eqrefprop[Equation]{eq:multiplication-table-PR-SR-I} and apply the bilinearity of $\boxtimes$ in order to obtain the following expression:
	\[
		\CHSH\big( \Q\boxtimes\P \big)
		-
		\CHSH\big( \P\boxtimes\Q \big)
		\,=\,
		\textstyle \frac{1}{8} (12 x_\P y_\Q-12 y_\P x_\Q-7 x_\P+7 y_\P+7 x_\Q-7 y_\Q)
		\,=:\,
		f_\P(x_\Q, y_\Q)\,.
	\]
	For any fixed $\P\in\Atilde$, we want to show that $f_\P(x_\Q, y_\Q)\geq0$.
	By construction, we know that $\P\in\line_1$ and $\Q\in\line_k$ for some $k\geq1$, so by \refprop[Corollary]{coro:orbits-move-to-the-left} we have $x_\Q + y_\Q \leq x_\P + y_\P$, which we may rewrite as $x_\Q \leq x_\P + y_\P - y_\Q$.
	As $\P$ lies in $\Atilde$, we have $y_\P\geq \frac34$, so
	the first partial derivative is non-positive:
	$\frac{\partial}{\partial x_\Q} f_\P(x_\Q, y_\Q) = \frac18(7 - 12 y_\P) \leq -1/4\leq0$, which means that the function $f_\P(\cdot, y_\Q)$ is decreasing over $\R$ for any fixed $y_\Q$.
	It yields the following inequalities:
	\[
		f_\P(x_\Q, y_\Q)
		\,\geq\,
		f_\P(x_\P + y_\P - y_\Q, y_\Q)
		\,=\,
		\textstyle\frac{3}{2}\big(y_\Q - y_\P\big) \big(x_\P + y_\P - \frac76\big)
		\,\geq\,
		0\,,
	\]
	since both factors are non-negative:
	the first one is non-negative using the hypothesis $\CHSH(\Q)\geq\CHSH(\P)$,
	and the second one is non-negative using $x_\P\geq1/2$ and $y_\P\geq3/4$ since $\P\in\Atilde$. Hence $f_\P$ is non-negative and we obtain the wanted result.
\end{proof}

Recall that the set $\Orbittilde{}^{(k)}(\P)$ is called the tilted $k$-orbit of the box $\P$ and contains some boxes~$\Q$ that are generated by applying a wiring to copies of~$\P$. We say that this tilted $k$-orbit \emph{distills the $\CHSH$-value} if it contains a box $\Q$ such that $\CHSH(\Q)\geq\CHSH(\P)$. In that distilling scenario, we can compute the expression of a box achieving the best $\CHSH$-value:
\vspace{-0.20cm}

\begin{statement}[Theorem~\ref{theo: highest box}]
	Let $\P\in\Atilde$ be a box, and let $k\geq2$ be an integer such that the tilted $(k-1)$-orbit distills the $\CHSH$-value.
	Then the highest $\CHSH$-value of $\Orbittilde{}^{(k)}(\P)$ is achieved at a box whose expression is the product of $k$ times $\P$ \emph{on the right}:
\[
	\P^{\boxtimes k}
	\,:=\,
	\Big(\big(\text{\small$(\P\boxtimes\P)$}\boxtimes \P\big)\cdots\Big)\boxtimes\P
	\,\,\in
	\underset{\Q\in\Orbittilde{}^{(k)}(\P)}{\argmax} \,\,\,\CHSH(\Q)
	\,.
\]
\end{statement}
\vspace{-0.50cm}

\begin{proof}
	We prove the result by induction on $k\geq2$.
	It is obviously true for $k=2$ since $\Orbittilde{}^{(2)}(\P)$ only contains $\P\boxtimes\P$.
	Now, fix $k\geq2$ and assume $\CHSH(\P^{\boxtimes k})\geq\CHSH(\Q)$ for any $\Q$ in the tilted $k$-orbit (induction hypothesis). Assume as well that $\CHSH(\P^{\boxtimes k})\geq\CHSH(\P)$ (distillation hypothesis).
	We want to show that:
	\[
		\CHSH(\P^{\boxtimes k+1}) \geq \CHSH(\Q\boxtimes\P)
		\quad\quad\text{and}\quad\quad
		\CHSH(\P^{\boxtimes k+1}) \geq \CHSH(\P\boxtimes\Q)\,,
	\]
	for all $\Q$ in the tilted $k$-orbit.
	The first inequality follows from \refprop[Lemma]{lem: Multiplication by P preserves the CHSH order} using the relation $\P^{\boxtimes k+1} = \P^{\boxtimes k} \boxtimes \P$ and the induction assumption.
	For the other inequality, start from $\CHSH(\P^{\boxtimes k+1}) = \CHSH(\P^{\boxtimes k}\boxtimes\P)$ and apply \refprop[Lemma]{lem: Multiplication to the right is higher}
	in order to get $\geq \CHSH(\P\boxtimes \P^{\boxtimes k})$.
	Then conclude using \refprop[Lemma]{lem: Multiplication by P preserves the CHSH order} and the induction hypothesis in order to obtain $\geq \CHSH(\P\boxtimes \Q)$
	for any $\Q$ in the tilted $k$-orbit.
\end{proof}

\bib

\section{Numerical Optimization on the Set of Wirings}
	\label{sec:numerical-approach}

We saw in the previous section that, given a non-signalling box $\P$, there may exist a wiring $\W$ that sufficiently distills the box $\P$ in order to collapse communication complexity.
The question we address in this section is the following: if the box $\P$ is fixed, how to find a wiring $\W$ good enough to collapse communication complexity (when it is possible)?
The difficulty is that, for each input $x,y\in\{0,1\}$, there are $82$ possible deterministic wirings~\cite{SPG06}, leading to a total number of $82^4\approx10^8$ possible deterministic wirings. So a naive discrete optimization over deterministic wirings seems inefficient. 
To that end, we present two optimization algorithms:
(i) an algorithm that tests many different combinations of wirings and that is suitable for numerical simulations, and
(ii) another one that finds a ``uniform" collapsing wiring $\W$ in a whole region of boxes, which is appropriate for deriving an analytical proof (see the next section).
This section might be skipped at first reading as it is more technical.
See our GitHub page for the details of the algorithms~\cite{GitHub-algebra-of-boxes}.

\begin{remark}[Comparison with~\cite{EWC22PRL, BMRC19}]  \label{rem:two-other-existing-methods}
	We now compare and contrast our methods with two recent works that also study optimization over wirings:
\begin{enumerate}[label=(\roman*)]
	\item In~\cite{BMRC19}, the authors suggest reducing the $82^4$ possible deterministic wirings for Alice and Bob to only $3152$ by simply considering the ones that preserve the $\PR$ box, \ie wirings $\W$ such that $\PR\boxtimes_\W\PR = \PR$, and then doing a discrete optimization over that smaller set. This smaller set encompasses for instance the wirings $\Wbs$, $\Wdist$
	but discards $\Woplus$, $\Wand$, $\Worand$
	(see definitions in \refprop[Subsection]{subsec:typical-examples-of-wirings}); see also \cite[Supplementary Material, I]{EWC22PRL} which mentions that even some optimal wirings are discarded.
	This technique allows them to analytically prove that many new areas of boxes are collapsing.
	
	\item In~\cite{EWC22PRL}, the authors use a mix of exhaustive search and linear programming. For each of Bob's $82^2$ extremal half-wirings, they apply linear programming to optimize Alice's half-wiring, and then they select the best pair of half-wirings. This allows them to numerically find optimal wirings for any pair of boxes, which leads them to discover new collapsing boxes.
	
	\item In our work, we use an efficient variant of the Gradient Descent algorithm, based on Line Search methods, frequent resets, and parallel descents.
	A limitation in the method from~\cite{BMRC19} could come from the fact that many wirings are discarded; this is why we choose to take our feasible set to be the entire set of mixed wirings $\WW\subseteq[0,1]^{32}$.
	In~\cite[Supplementary Material, II]{EWC22PRL}, the authors implement a sequence of different optimal wirings, which we do similarly in what we call later Task A, but we also implement a uniform version of it in Task B, which allows us to find a single optimal wiring for a whole region of boxes (instead of a sequence of wirings) and then to prove by hand the collapse of communication complexity for those boxes.
	In this manner, we recover both the numerical results of~\cite{EWC22PRL} (see details in \refprop[Subsection]{subsec:numerical-results}) and the analytical results of~\cite{BMRC19} (see details in \refprop[Subsection]{subsection:analytical-results}).
\end{enumerate}
\end{remark}

		\subsection{Goals of the Algorithms}

\paragraph{Task~A.}
To prove that a box $\P$ is collapsing, a particular case of \refprop[Proposition]{prop: collapsing orbit} says that it is enough to find a finite sequence of wirings $(\W_1, \dots, \W_N)$ such that the following box is collapsing:
\[
	\P_{N+1}:=\Big(\big((\P\underset{\,\W_1}{\boxtimes} \P) \underset{\,\W_2}{\boxtimes} \P\big) \underset{\,\W_3}{\boxtimes} \dots \Big) \underset{\,\W_N}{\boxtimes} \P\,.
\]
Note that we need to specify the parenthetization because the different products $\boxtimes_{\W_i}$ are potentially non-associative. Among the numerous possibilities, we choose the parenthesization on the left because it is easy to implement and because it is the best one when the wiring is $\Wbs$, see \refprop[Theorem]{theo: highest box}.
This algorithm will consist in an iterative construction of the sequence $(\W_i)_i$: first, find a wiring $\W_1$ such that the $\CHSH$-value of the box $\P_2:= \P\boxtimes_{\W_1}\P$ is the highest possible, then find $\W_2$ such that the $\CHSH$-value of the box $\P_3:=\P_2\boxtimes_{\W_2}\P$ is the highest possible, so on and so forth until the $N$-th iteration.
If the $\CHSH$-value of the box $\P_{N+1}$ is above the threshold $\frac{3+\sqrt{6}}6\approx 0.91\%$, we know that communication complexity collapses~\cite{BBLMTU06}, so the starting box $\P$ is collapsing as well. Otherwise, we cannot conclude whether $\P$ is collapsing or not.

\paragraph{Task~B.}
The goal of this algorithm is essentially the same as the first one, but we add a strong constraint: we want all the $\W_i$ to be the same wiring $\W$:
\[
	\Big(\big((\P\underset{\,\W}{\boxtimes} \P) \underset{\,\W}{\boxtimes} \P\big) \underset{\,\W}{\boxtimes} \dots \Big) \underset{\,\W}{\boxtimes} \P
	\,\, =: \,\,
	\P^{\boxtimes_\W N+1}\,.
\]
In that sense, this is a ``uniform" version of the first algorithm.
The interest of this algorithm is that it helps to give analytical proofs (see \refprop[Section]{section: Analytical Proofs of the New Collapsing Regions}): if the value $\CHSH(\P^{\boxtimes_\W N})$ is above the threshold $\frac{3+\sqrt{6}}6\approx 0.91\%$ for some $N$, then by continuity of $\boxtimes_\W$, there is an open neighborhood around $\P$ such that for any $\Q$ close enough to $\P$ we also have that $\CHSH(\Q^{\boxtimes_\W N})$ is above the threshold, and therefore the whole neighborhood of $\P$ is collapsing.
This technique will help to discover wide collapsing areas and to provide analytical proofs by hand.

		\subsection{Toy Example ($N=1$)}
		\label{subsec:toy-example-case}

In this subsection, we treat the case when there is only one product $\boxtimes_\W$ between two boxes $\Q,\P\in\NS$.
We detail the maximization algorithm we use: a Projected Gradient Descent.
The optimization problem consists in finding $\W^*$ as follows:
\begin{equation}    \label{eq:minimization-problem}
	\W^*
	\,=\,
	\underset{\W\in\WW}\argmax\hspace{0.3cm} \Phi(\W)\,.
\end{equation}
where the objective function is $\Phi(\W) := \CHSH(\Q\boxtimes_\W\P)$ for some fixed non-signalling boxes $\Q,\P$,
and where $\WW$ is the set of mixed wirings introduced in \refprop[Definition]{def: mixed wiring}, which we recall below.

\paragraph{The Constraint $\W\in\WW$.}
Recall that a \emph{mixed wiring} $\W$ between two boxes $\Q,\P\in\NS$ is the data of six functions $f_1, f_2, g_1, g_2:\{0,1\}^2\to[0,1]$ and $f_3, g_3:\{0,1\}^3\to[0,1]$ satisfying the following \emph{non-cyclicity conditions}:
	\begin{align}
		\forall x, \quad\quad     \label{eq: condition 1 for valid wiring 2}
		&\big( f_1(x,0) - f_1(x,1) \big) \big( f_2(x,0) - f_2(x,1) \big) \,=\,0\,,\\
		\forall y, \quad\quad	\label{eq: condition 2 for valid wiring 2}
		&\big( g_1(y,0) - g_1(y,1) \big) \big( g_2(y,0) - g_2(y,1) \big) \,=\,0\,.
	\end{align}
	Recall that the corresponding diagram can be found in \refprop[Figure]{figure: general wiring} (b), and that
	mixed wirings form a set that we denote $\WW$.
	In our algorithms, we view $\W$ a real vector with $4\times 2^2 + 2 \times 2^3 = 32$ variables.
	This vector stores each value of each function:
	\begin{equation}   \label{eq:wiring-as-a-vector}
		\W = [f_1(0,0)\quad f_1(0,1)\quad f_1(1,0)\quad f_1(1,1)\quad g_1(0,0)\quad\dots]^\top\in\R^{32}\,.
	\end{equation}
	To satisfy the normalization constraint that the $f_i,g_j$ take value in $[0,1]$, and the non-cyclicity conditions \eqref{eq: condition 1 for valid wiring 2} and \eqref{eq: condition 2 for valid wiring 2} (which are non-linear conditions), we implement a projection function $\proj:\R^{32} \to\R^{32}$ in \refprop[Algorithm]{alg:proj-function-on-feasible-set}.
	Notice that our real code is written in a vectorized fashion and is difficult to read as such, so we only present the idea here.
	Moreover, we use the package PyTorch~\cite{PGCCYDLDAL17} for automatic differentiation.

\begin{algorithm}
\caption{Projection function $\proj$ on the feasible set $\WW$. Vectorized version in our GitHub page~\cite{GitHub-algebra-of-boxes}.}
\label{alg:proj-function-on-feasible-set}
\KwData{$\W = \begin{bmatrix} w_1 & \dots & w_{32} \end{bmatrix} = \small \begin{bmatrix} f_1(0,0) & f_1(0,1) & f_1(1,0) & f_1(1,1) & g_1(0,0) & \dots & g_3(1,1,1) \end{bmatrix}\in\R^{32}.$}
\KwResult{$\proj(\W)\in\WW.$}
\vspace{0.2cm}
$\W \gets \begin{bmatrix} \max\{w_1, 0\} & \dots & \max\{w_{32}, 0\} \end{bmatrix}$ \;
$\W \gets \begin{bmatrix} \min\{w_1, 1\} & \dots & \min\{w_{32}, 1\} \end{bmatrix}$ \;
\For{$x\in\{0,1\}$}{
	\leIf{\small$|f_1(x,0) - f_1(x,1)| \leq |f_2(x,0) - f_2(x,1)|$}{
		\small$f_1(x,0), f_1(x,1) \gets \big(f_1(x,0) + f_1(x,1)\big)/2$ \;
	}{
		\small$f_2(x,0), f_2(x,1) \gets \big(f_2(x,0) + f_2(x,1)\big)/2$
	}
}
\For{$y\in\{0,1\}$}{
	\leIf{\small$|g_1(y,0) - g_1(y,1)| \leq |g_2(y,0) - g_2(y,1)|$}{\small$g_1(y,0), g_1(y,1) \gets \big(g_1(y,0) + g_1(y,1)\big)/2$ \;
	}{
		\small $g_2(y,0), g_2(y,1) \gets \big(g_2(y,0) + g_2(y,1)\big)/2$
	}
}
\Return{$\W$}
\end{algorithm}

\paragraph{The Objective Function $\Phi(\W):= \CHSH(\Q\boxtimes_\W\P)$.}
In our algorithms, we view a box $\P$ as a $2\times2\times2\times2$-tensor: $\P[a,b,x,y]:=\P(a,b\,|\,x,y)$ with $a,b,x,y\in\{0,1\}$, whose entries are float numbers between $0$ and $1$.
Two things need to be computed separately: $\Q\boxtimes_\W\P$ and then $\CHSH(\cdot)$.
On the one hand, the product $\P\boxtimes_\W\P$ is computed using \eqrefprop[Equation]{eq: P times Q with mixed wiring}, which we vectorized in our algorithm using five types of operations: tensor transposition $\top$, tensor sum $+$, tensor product $\otimes$, contraction of tensors $\cdot$, and entry-wise multiplication $*$; see details in the pdf document of our GitHub page~\cite{GitHub-algebra-of-boxes}.
On the other hand, the function $\CHSH(\cdot)$ is a linear function that computes the $\CHSH$-value of a box, implemented with a dot product as follows:
\[
	\CHSH(\Rbox)
	:=
	\frac14
	\sum_{a\oplus b=xy} \Rbox(a,b\,|\,x,y)
	\,=\,
	\langle \Rbox, \T \rangle\,,
\]
where $\T$ is the $2\times2\times2\times2$-tensor defined as $\T[a,b,x,y]=\nicefrac14$ if $a\oplus b=xy$, and $=0$ otherwise.

	\subsubsection{Naive Gradient Descent}
	
To gain insight into the complexity of the optimization problem, we begin by studying a basic algorithm, the Projected Gradient Descent, with a small learning rate ($\alpha\ll1$) and a lot of iterations ($K\gg1$).
We will obtain a histogram of the frequency of the different results we obtain, see \refprop[Figure]{fig:histogram}~(a).

\paragraph{Projected Gradient Descent.}
We implement a ``projected" version of the Gradient Descent algorithm in order to satisfy the constraint $\W\in\WW$ at each step.
It simply means that each iteration is projected on the feasible set:
\[
	\W^{k+1} = \proj\big(\W^k + \alpha\, \nabla \Phi(\W^k)\big)\,,
\]
where $\alpha\in\R$ is the learning rate.
Our implementation can be found in \refprop[Algorithm]{alg:projected-gradient-descent}.
We compute the gradient of the objective function using the automatic differentiation Python package \texttt{torch.autograd} that provides us with the commands \texttt{backward} and \texttt{grad}.
As we do not have a good intuition of what could be a good wiring $\W$ in $\WW$ to start with given a fixed box $\P$, we take a random initialization: $\W^0$ is uniformly generated in the hypercube $[0,1]^{32}$.
As such, the vector $\W^0$ is not necessarily a well-defined mixed wiring since it does not necessarily satisfy the non-cyclicity conditions~\eqref{eq: condition 1 for valid wiring 2} and~\eqref{eq: condition 2 for valid wiring 2}, but this problem is fixed after one iteration in the Projected Gradient Descent algorithm since the wiring is then projected.
Otherwise, one can also directly apply $\proj$ to $\W^0$.
The notation $\W\sim\mathcal U(X)$ means that we uniformly generate $\W$ in the set $X$.

\begin{algorithm}
\caption{Projected Gradient Descent. More details on our GitHub page~\cite{GitHub-algebra-of-boxes}.}
\label{alg:projected-gradient-descent}
\KwData{
	$\Phi:\R^{32}\to\R$ objective function,
	$\alpha\in\R$ learning rate,
	$K\in\N$ number of iterations,
	$\varepsilon>0$ tolerance.
}
\KwResult{$\W_{\text{out}}\approx\argmax_\W\Phi(\W)\in\WW\subseteq \R^{32}$.}
\vspace{0.2cm}
$\W \sim \mathcal U ([0,1]^{32})$ \;
\For{$k \in \{0, \dots, K-1\}$}{
	$\Wold \gets \W$ \;
	$\W \gets \proj\big( \W + \alpha\, \nabla \Phi(\W) \big)$ using \refprop[Algorithm]{alg:proj-function-on-feasible-set} \;
	\lIf{$|| \W-\Wold||_\infty <\varepsilon$}{
		break
	}
}
\Return{$\Wout:=\W\in\R^{32}$.}
\end{algorithm}

\begin{figure}[h]
	(a)
	\includegraphics[trim={0 7 0 6}, clip, height=4.6cm]{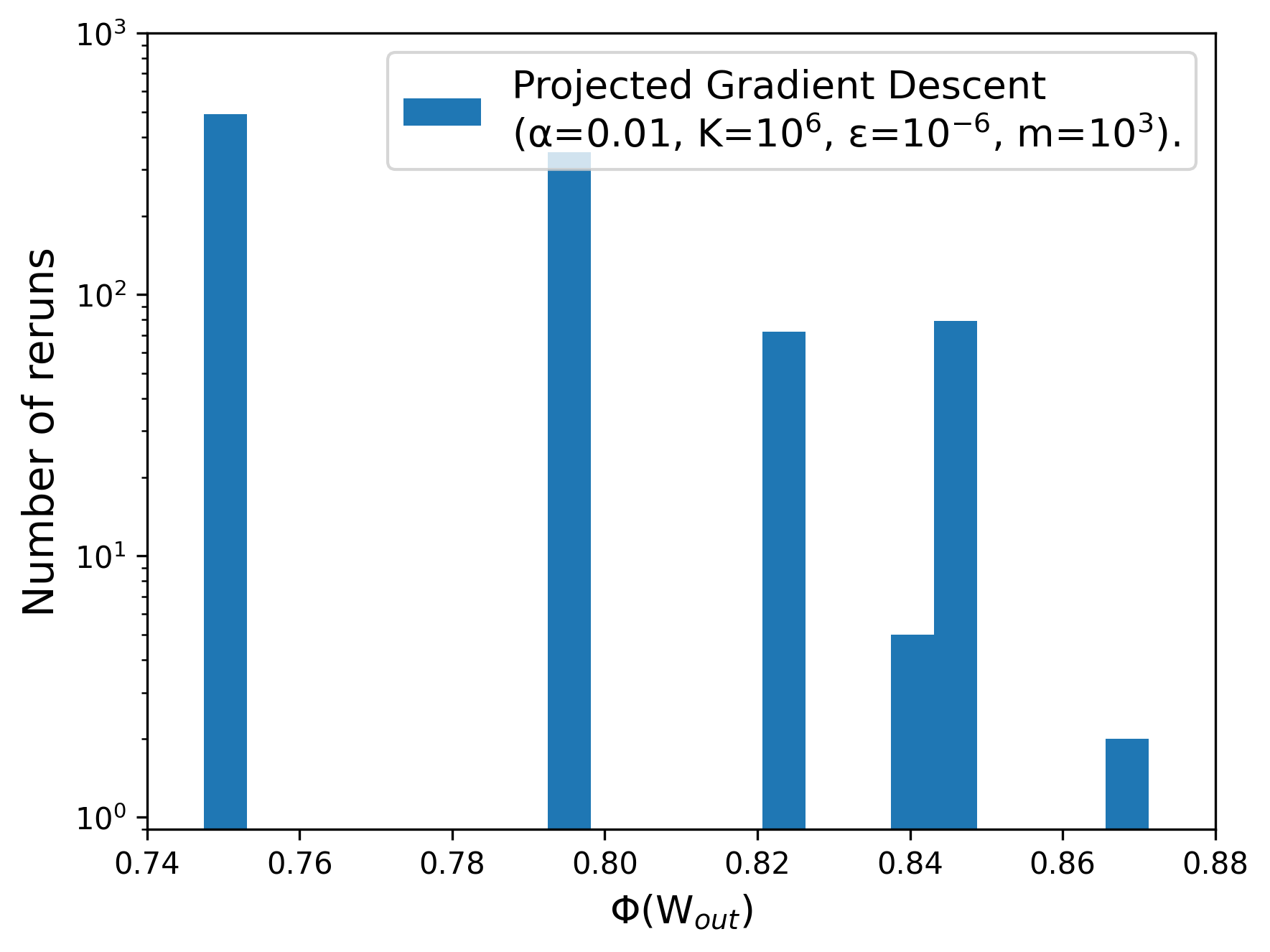}
	~~~(b)
	\includegraphics[trim={0 7 0 6}, clip, height=4.6cm]{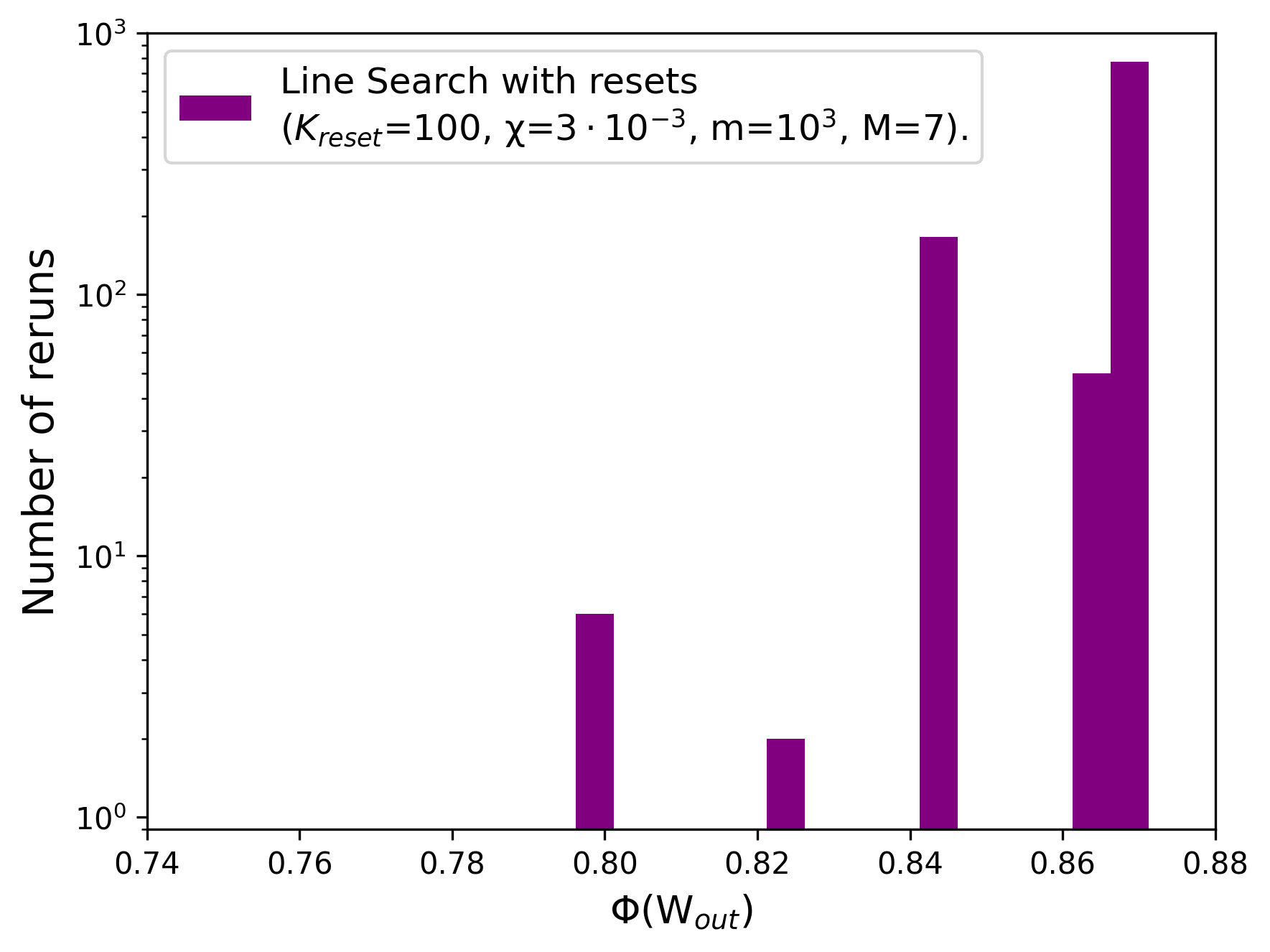}
	\centering
	\caption{
	Histogram of the evaluations of the objective function $\Phi$ applied at the output $\Wout$ of (a)~\href{{alg:projected-gradient-descent}}{Algorithm~\ref{alg:projected-gradient-descent}}
	and (b)~\refprop[Algorithm]{alg:line-search}.
	As expected, we observe that the latter is more efficient than the former in maximizing $\Phi$, for equivalent computation duration.
	}
	\label{fig:histogram}
\end{figure}

\paragraph{Estimating the Proportion of ``Good" Outputs.}
We use \refprop[Algorithm]{alg:projected-gradient-descent} with a learning rate $\alpha=0.01$, a number of iterations of $K=10^6$, a tolerance of $\varepsilon=10^{-6}$, and we obtain the histogram presented in \refprop[Figure]{fig:histogram}~(a).
Recall that the objective function is $\Phi(\W) := \CHSH(\Q\boxtimes_\W\P)$; this histogram is drawn with $\Q=\P=p\PR+q\SR + (1-p-q)\I$, where $p=0.39$ and $q=0.6$.
The number of reruns is $m=10^3$, done simultaneously in parallel, which is faster than doing $m$ descents one after another\footnote{In order to do $m$ gradient descents in parallel efficiently, we ``parallelize" \refprop[Algorithm]{alg:projected-gradient-descent}: instead of viewing $\W$ as a $32$-vector, we view it as a $(32\times m)$-matrix, where each column represents a different wiring.
Comparing this method with a naive FOR loop, we observe a factor of $100$ in the speed gain.
Notice that when the descent is done, one can post-select the best wiring among the $m$ columns of $\W_{\text{out}}$. See our GitHub page~\cite{GitHub-algebra-of-boxes}.}.
We observe that the results concentrate on certain discrete values.
These values correspond to different attractive points in different basins of attraction (recall that the initial $\W$ is taken uniformly at random in $[0,1]^{32}$).
As we want to maximize $\Phi$, we are interested in the highest concentrated value $\approx0.87$. In that example, we observe that the proportion of starting wirings such that $\W_{\text{out}}$ is only $\chi \approx \nicefrac{2}{10^3} = 0.2\%$ using this basic Gradient Descent algorithm. This information tells us that the function $\Phi$ is difficult to maximize, which is why we present a more efficient algorithm in the following subsection.

	\subsubsection{More Efficient Algorithm: Line Search with Resets}
	
In this subsubsection, we present a variant of the Gradient Descent algorithm called Line Search, which we enhance with frequent resets of bad outcomes. See~\cite{NW99} for a standard reference book in numerical optimization. The idea of this algorithm is, instead of always keeping the same $\alpha$, to estimate the best coefficient $\alpha_k$ at each step of the descent:
\[
\left\{
\begin{array}{l}
	\alpha_{k} = \argmax_{\alpha\in\R} \Phi\big(\W^k + \alpha \nabla \Phi(\W^k)\big)\,,\\
	\W^{k+1} = \proj\big(\W^k + \alpha_k\, \nabla \Phi(\W^k)\big)\,.
\end{array}
\right.
\]
As we observed in the previous subsection, the proportion $\chi$ of ``good" starting wirings is very weak, which is why we apply frequent resets: we do $m=10^3$ descents in parallel but only $K_{\text{reset}}=100$ steps, then we keep only the best $m\cdot\chi$ wirings and we reset all the others to a new random initialization.
Then we repeat that procedure but we reset fewer wirings (say, at the $j$-th repetition, keep for instance the best $j\cdot m\cdot \chi$ wirings),
and we repeat this procedure $\nicefrac1\chi$ times.
In the end, most of the wirings should be in the good basin of attraction, so we can apply one final run of line search, with many more steps so that it converges to the attractor.
See \refprop[Algorithm]{alg:line-search}, and
we obtain results of \refprop[Figure]{fig:histogram}~(b).

\begin{algorithm}[h]
\caption{Line Search with Resets. More details on our GitHub page~\cite{GitHub-algebra-of-boxes}.}
\label{alg:line-search}
\KwData{
	$\Phi:\R^{32}\to\R$ objective function,
	$K_{\text{reset}}\in\N$ number of iterations before reset,
	$\chi\in[0,1]$ proportion of ``good" random wirings,
	$m\in\N$ number of descents in parallel,
	$M\in\N$ number of iterations to compute the best learning rate $\alpha^*$.
}
\KwResult{$\W_{\text{out}}\in\R^{32\times m}$, where each column is $\approx\argmax_\W\Phi(\W)$.}
\vspace{0.2cm}
$\W=$ ($32\times m$ null matrix), whose columns are denoted $\W_i$ \;
\For{$j\in\{0,\dots,\lfloor\nicefrac{1}{\chi}\rfloor-1\}$}{
$\W\gets$ among the $m$ columns of $\W$, keep the $j\cdot m\cdot \chi$ ones giving the highest values for the objective function $\Phi$, and reset all the other columns randomly with $\mathcal U ([0,1]^{32})$ \;
\lIf{$j=\lfloor\nicefrac{1}{\chi}\rfloor-1$}{$K_{\normalfont\text{reset}}\gets10\cdot K_{\normalfont\text{reset}}$}
\For{$k \in \{0, \dots, K_{\normalfont\text{reset}}-1\}$}{
	\lFor{$i\in\{0,\dots, m-1\}$}{$\alpha_i^* \gets \argmax_{\alpha>0} \Phi\big(\W_i + \alpha \nabla \Phi(\W_i)\big)$ using $M$ iterations }
	$\W \gets \Big[ \proj\big( \W_i + \alpha_i^*\, \nabla \Phi(\W_i) \big) \Big]_i$ using \refprop[Algorithm]{alg:proj-function-on-feasible-set} \;
}
}
\Return{\normalfont$\Wout=\W\in\R^{32\times m}$.}
\end{algorithm}

		\subsection{Task~A}
		
Algorithm A is presented in \refprop[Algorithm]{alg:algo-A}; it simply consists in applying the toy case $\Q\boxtimes\P$ from the previous subsection recursively $N$ times.
We want to find a sequence of wirings $\W_1, ..., \W_N$ such that the $\CHSH$-value is above the following threshold:
\[
	\CHSH \bigg(
	\underbrace{\Big(\big((\P\underset{\,\W_1}{\boxtimes} \P) \underset{\,\W_2}{\boxtimes} \P\big) \underset{\,\W_3}{\boxtimes} \dots \Big) \underset{\,\W_N}{\boxtimes} \P\,}_{=:\,\P_{N+1}}
	\bigg)
	\,>\,
	\frac{3+\sqrt{6}}6\,.
\]
A consequence is that the box $\P$ collapses communication complexity~\cite{BBLMTU06}. Notice that for some boxes $\P\in\NS$, it might not be possible to find such a sequence of wirings because it is impossible to distill them by any means.
This algorithm is used in \refprop[Subsection]{subsec:numerical-results} in order to plot the new regions of collapsing nonlocal boxes.

\begin{algorithm}
\caption{Task~A.
More details on our GitHub page~\cite{GitHub-algebra-of-boxes}.}
\label{alg:algo-A}
\KwData{
	$\P\in [0,1]^{2+2+2+2}$ box,
	$N\in\N$ number of box products, \\
	\hspace{0.9cm}$(K_{\normalfont\text{reset}}, \chi, m, M)$ parameters for \refprop[Algorithm]{alg:line-search}.
}
\KwResult{$[\W_1^*, \dots, \W_n^*]\in\WW^n$ for some $n\leq N$.}
\vspace{0.2cm}
$\P_1\gets \P$ \;
\For{$n \in \{1, \dots, N\}$}{
	$\W_n^*\in[0,1]^{32} \gets \argmax_\W \CHSH(\P_n\boxtimes_\W\P)$ by picking the best column among the $m$ columns of the output $\Wout\in\R^{32\times m}$ of \refprop[Algorithm]{alg:line-search} \;

	$\P_{n+1} \gets \P_n\boxtimes_{\W_n^*} \P$ \;
	\lIf{$\CHSH(\P_{n+1})>\frac{3+\sqrt{6}}6$}{
		\Return{$[\W_1^*, \dots, \W_n^*]$}
	}
}
\Return{``Nothing found."}
\end{algorithm}

\begin{remark}
	Going further, once we find $(\W_1^*, ..., \W_N^*)$, it is possible to do a ``backward" process: for all $i\in\{1, ..., N\}$, fix $\W_j^*$ for $j\neq i$, optimize the function $\W_i\mapsto \CHSH(\P_{N+1})$ and update $\W_i^*$.
	It is also possible to use neural network methods to optimize all the $\W_i$ ``at the same time".
\end{remark}

		\subsection{Task~B}

Task~B is a ``uniform" version of task~A, in the sense that we want all the $\W_i$'s to be the same.
In order words, we want to find a wiring $\W$ and an integer $N$ such that:
\[
	\CHSH \bigg(
	\underbrace{\Big(\big((\P\underset{\,\W}{\boxtimes} \P) \underset{\,\W}{\boxtimes} \P\big) \underset{\,\W}{\boxtimes} \dots \Big) \underset{\,\W}{\boxtimes} \P\,}_{=:\,\P^{\boxtimes_\W (N+1)}}
	\bigg)
	\,>\,
	\frac{3+\sqrt{6}}6\,.
\]

This algorithm is used in the proof of \refprop[Theorem]{thm:new-collapsing-triangles} in order to find appropriate collapsing wirings for the analytical proof.

\paragraph{Idea of the Algorithm.}
First, find a wiring $\W_1^* = \argmax_\W \CHSH(\P\boxtimes_\W\P)$ with a Gradient Descent algorithm, and then evaluate the powers of $\P$ with that wiring $\W_1^*$ until $N+1$, \ie compute $\CHSH\big(\P^{\boxtimes_{\W_{\text{\fontsize{3}{60}$1$}}^*} n}\big)$ for $n=1, \dots, N+1$. If one of those evaluations is greater than the threshold $(3+\sqrt{6})/6$ from~\cite{BBLMTU06}, then we can stop the algorithm, it means that the wiring $\W_1^*$ achieves the goal.
Otherwise, compute $\W_2^* = \argmax_\W \CHSH\big(\P^{\boxtimes_\W3}\big)$ and repeat the same evaluation process of the powers of $\P$ as in the previous step.
Proceed like this until computing $\W_M^*$, where $M\in\N$ is some hyper-parameter.
Typically, we take $M\leq N$ because it is a lot faster to evaluate the $N$-th power of $\P$ than to optimize the $N$-th power of $\P$.
See the details in \refprop[Algorithm]{alg:algo-B}.

\begin{algorithm}
\caption{Task~B.
More details on our GitHub page~\cite{GitHub-algebra-of-boxes}.}
\label{alg:algo-B}
\KwData{
	$\P\in [0,1]^{2+2+2+2}$ box,
	$N\in\N$ maximal tested box power,
	$L\in\N$ maximal optimized box power,
	$(K_{\normalfont\text{reset}}, \chi, m, M)$ parameters for \refprop[Algorithm]{alg:line-search}.
}
\KwResult{$\W^*\in\WW$.}
\vspace{0.2cm}
\For{$\ell\in\{1, \dots, L\}$}{
	$\W_\ell^* \gets \argmax_\W \CHSH\big(\P^{\boxtimes_\W \ell+1}\big)$ using \refprop[Algorithm]{alg:line-search} \;
	\For{$n\in\{1, \dots, N+1\}$}{
		\lIf{$\CHSH\big(\P^{\boxtimes_{\W_\ell^*} n}\big)>\frac{3+\sqrt{6}}6$}{\Return{$\W_\ell^*$} }
	}
}
\Return{``Nothing found."}
\end{algorithm}

\bib

	\section{New Collapsing Boxes}
	\label{section: Analytical Proofs of the New Collapsing Regions}
	
In this section, we present collapsing boxes found in two different ways. (i)~First with a numerical approach, using the algorithms (\refprop[Section]{sec:numerical-approach}). (ii)~Then with an analytical approach, using the algebra of boxes (\refprop[Section]{section:Algebra-of-boxes}) and the orbit of a box (\refprop[Section]{section:Orbit-of-a-box}).

		\subsection{Numerical Results}
		\label{subsec:numerical-results}

Using \refprop[Algorithm]{alg:algo-A} that addresses Task~A, we obtain many collapsing boxes. Some samples are drawn in \refprop[Figure]{fig:numerical-new-collapsing-regions} on some slices of the non-signalling set $\NS$, but note that this algorithm also applies more generally to any desired slice. As previously mentioned, this work is concurrent and independent of the work of~\cite{EWC22PRL}.
In the drawings, some boxes are denoted $\P_{\text{L}}$ and $\P_{\text{NL}}$, let us recall their definition here.
The local set $\LL$ and the non-signalling set $\NS$ are polytopes, \ie the convex hull of a finite number of extremal points. The first set $\LL$ admits exactly $16$ extremal points, called \emph{local extreme points} and denoted $\P_{\text{L}}^{\mu,\nu, \sigma, \tau}$, where $\mu,\nu, \sigma, \tau\in\{0,1\}$. These $16$ points are as well extremal points of $\NS$, together with $8$ additional extremal points, called \emph{non-local extreme points} and denoted $\P_{\text{NL}}^{\mu, \nu, \sigma}$.
They are defined as follows~\cite{BLMPPR05,ABPS09}:
\definecolor{blue2}{rgb}{0.14, 0.44, 0.64}
\definecolor{red2}{rgb}{0.66, 0.20, 0.15}
\definecolor{quantumviolet}{HTML}{53257F}
\begin{align} \label{eq: extremal points of NS}
	\begin{array}{lrl}
	\text{$\bullet$ Local extremal points:} &
	\hspace{-0.5cm}
	\P_{\text{L}}^{{\color{quantumviolet!80}\mu},{\color{quantumviolet!80}\nu}, {\color{quantumviolet!80}\sigma}, {\color{quantumviolet!80}\tau}}(a, b\,|\,x,y)
	\,&:=\,
	\left\{
	\begin{array}{cl}
		1 & \text{\small if $a={\color{quantumviolet!80}\mu}\,x\oplus {\color{quantumviolet!80}\nu}$ \ and \ $b={\color{quantumviolet!80}\sigma}\,y \oplus {\color{quantumviolet!80}\tau}$},\\
		0 & \text{\small otherwise},
	\end{array}
	\right.\\
	\text{$\bullet$ Nonlocal extremal points:} &
	\P_{\text{NL}}^{{\color{quantumviolet!80}\mu},{\color{quantumviolet!80}\nu}, {\color{quantumviolet!80}\sigma}}(a, b\,|\,x,y)
	\,&:=\,
	\left\{
	\begin{array}{cl}
		1/2 & \text{\small if $a\oplus b=x y\oplus{\color{quantumviolet!80}\mu}\,x \oplus {\color{quantumviolet!80}\nu}\,y\oplus{\color{quantumviolet!80}\sigma}$},\\
		0 & \text{\small otherwise}.
	\end{array}
	\right.
	\end{array}
\end{align}
Note that $\PR = \P_{\text{NL}}^{000}$ and $\P_0 = \P_{\text{L}}^{0000}$ and $\P_1 = \P_{\text{L}}^{0101}$ (we remove the commas in the superscripts for the sake of simplicity of notations).

\newlength{\heyhey}
\setlength{\heyhey}{4.7cm} 
\begin{figure}[h]
	\centering
	\includegraphics[width=9.5cm]{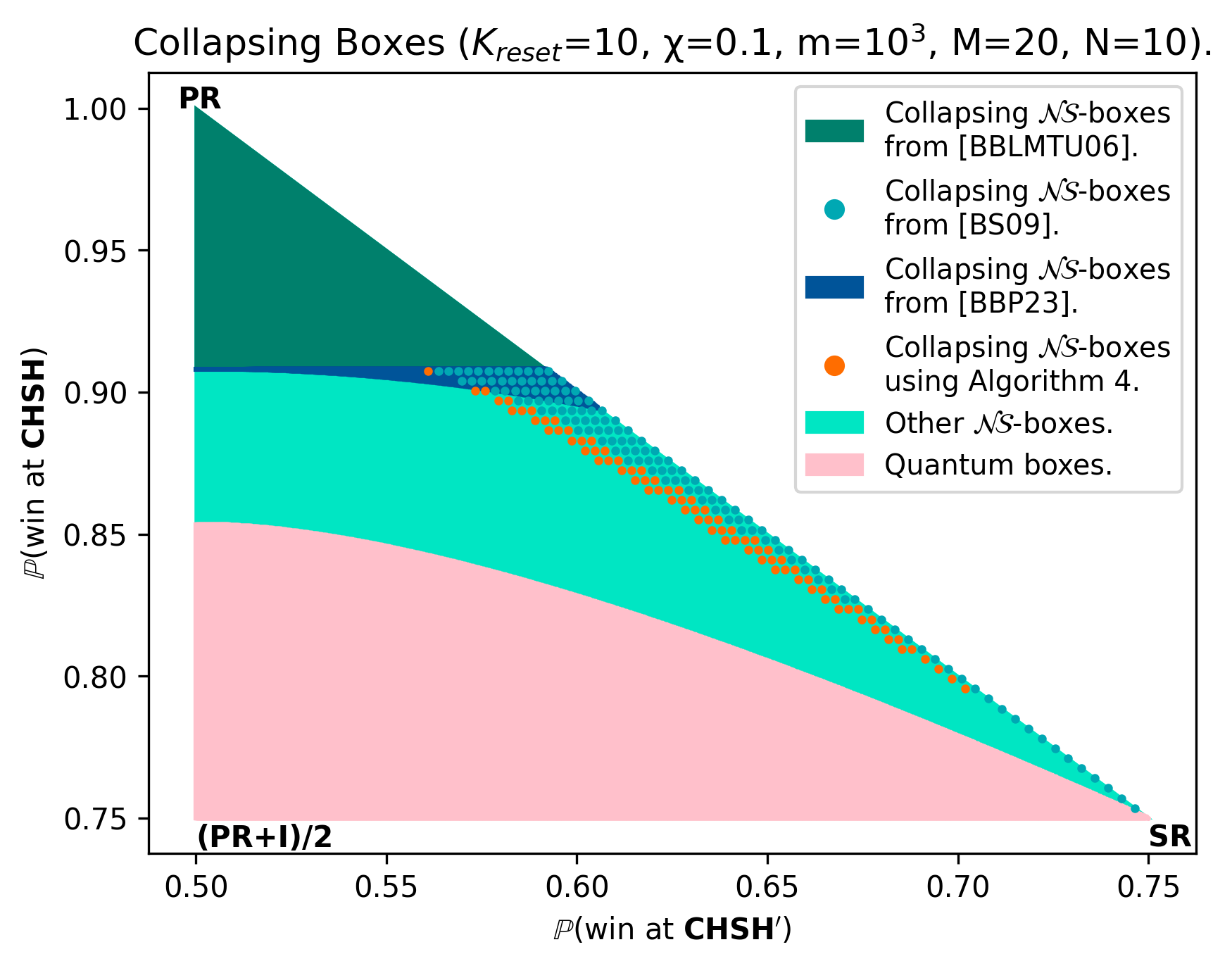} 
	\begin{tabular}{c}
		\vspace{-7.55cm}\\
		\includegraphics[width=\heyhey]{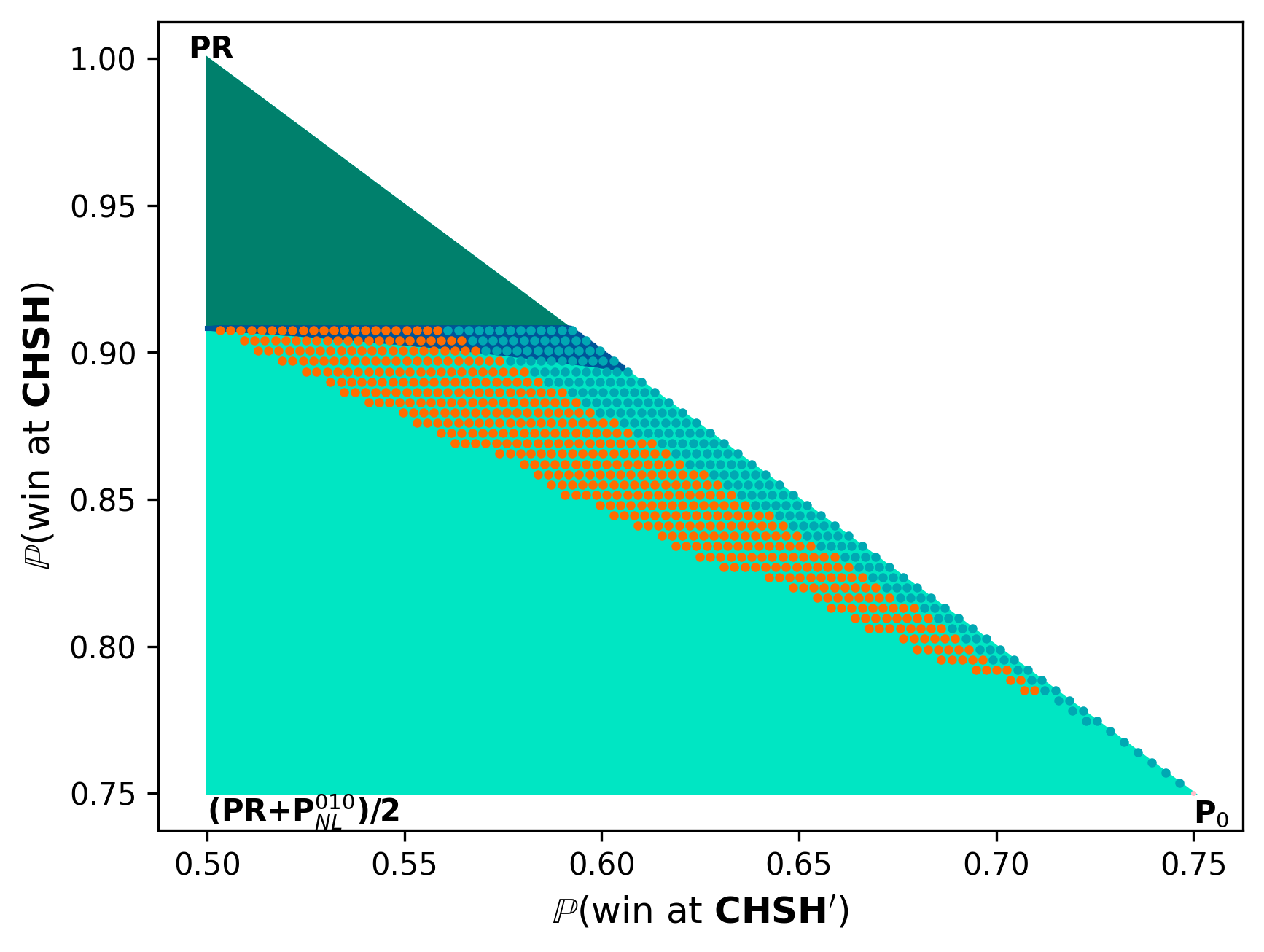} \vspace{-0.5cm}\\
		\hspace{-0.1cm}\includegraphics[width=\heyhey]{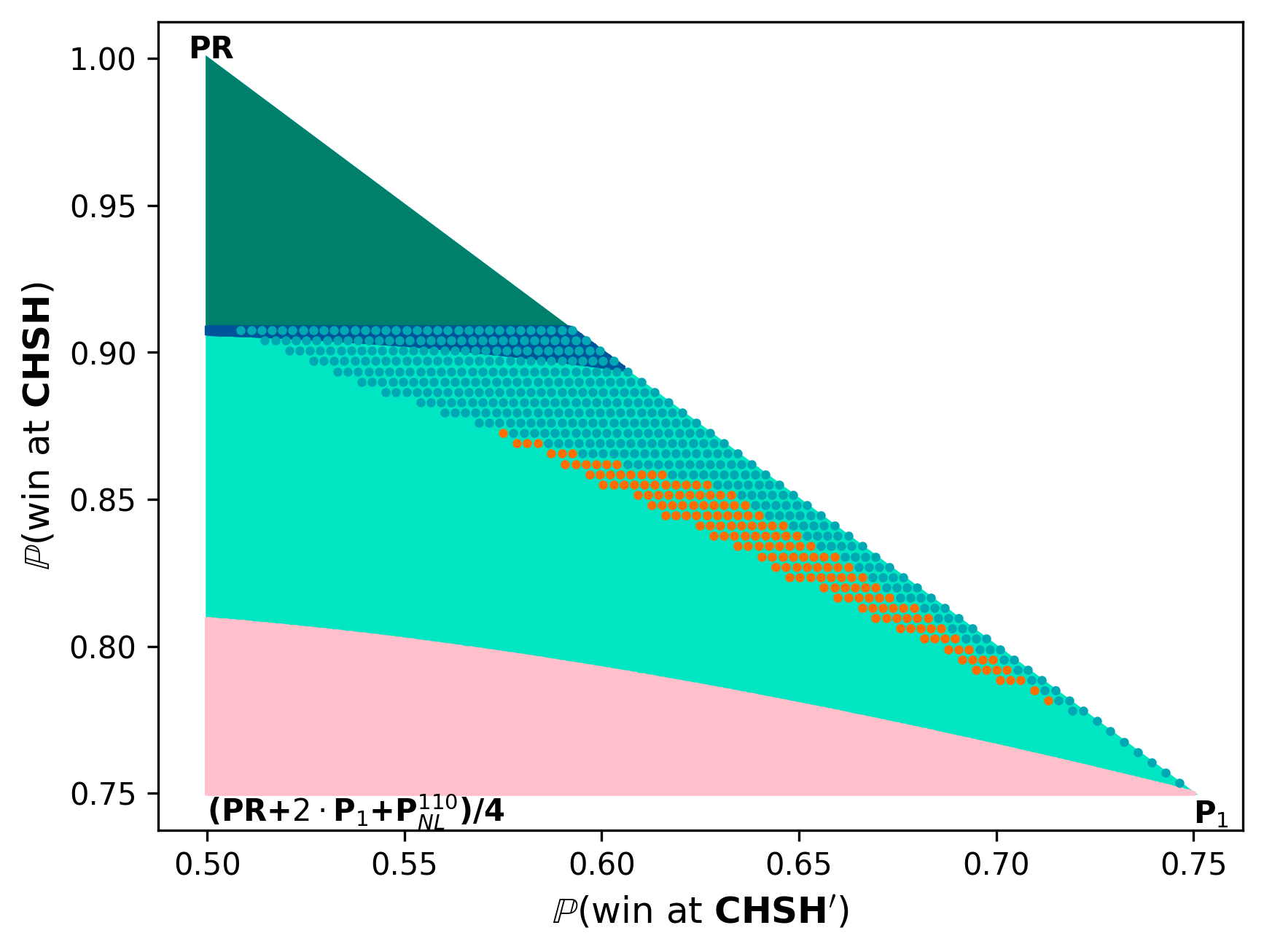}
	\end{tabular}
	\caption{In orange are drawn the collapsing boxes outputted by \refprop[Algorithm]{alg:algo-A}.
	Each drawing represents a different slice of $\NS$; the extreme points of the triangles indicate which slice is drawn and the definition of the boxes $\PNL$ can be found in \eqrefprop[Equation]{eq: extremal points of NS}.  The three graphs have the same color legend, displayed at the center, and they are all configured with the same algorithm parameters $(K_{\normalfont\text{reset}}, \chi, m, M, N)$, detailed at the top.
	We adopt the following convention:
	(i)~boxes that are numerically determined are drawn with dots,
	(ii)~boxes that are analytically determined are drawn in plain regions (there exist explicit equations describing those regions).
	Notice that the left drawing is similar to~\cite[Figure 3]{EWC22PRL}, which was found using a different algorithm as detailed in \refprop[Remark]{rem:two-other-existing-methods}.
	The quantum set $\QQ$ (in pink) is drawn using formulas from~\cite{Masanes03}.
	References: [BBLMTU06]=\cite{BBLMTU06}, [BS09]=\cite{BS09}, [BBP23]=\cite{BBP23}.
	}
	\label{fig:numerical-new-collapsing-regions}
\end{figure}

Observe in \refprop[Figure]{fig:numerical-new-collapsing-regions} that, depending on the chosen slice, the collapsing area does not always have the same ``shape" nor the same ``area".
Moreover, notice in the graphs that there seems to exist a collapsing area in the neighborhood below the diagonal segments joining $\PR$ and respectively $\SR$, $\P_0$, $\P_1$. This is actually true. Indeed, we analytically show below in \refprop[Theorem]{thm: the triangle PR-P0-P1 is collapsing} that those three segments are collapsing, and we also know that the box product $\P\boxtimes_\W\Q$ is continuous in $\P$ and $\Q$ for any $\W$ (it is even bilinear, recall the expression in~\eqref{eq: P times Q with mixed wiring}), so distillation protocols are continuous and in some sense the orbits are also ``continuous", hence there exists an open neighborhood below these diagonal segments that collapses communication complexity.

\begin{remark}[Continuous extension of a finite collapsing set]
The algorithm only provides us with finitely many collapsing boxes, but we can still deduce a continuous ``extension" of that collapsing set.
Indeed, as explained in \refprop[Subsection]{subseq:ConsequencestoCommunicationComplexity}, if we know that a box $\P\in\NS$ collapses communication complexity, then we also know that the cone $\C_\P$ is collapsing, where $\C_\P$ denotes a certain cone taking origin at $\P$ represented in \refprop[Figure]{fig: collapsing orbits}~(b). As a consequence, if we list the collapsing boxes $\{\P_k\}_{1\leq k\leq K}$ obtained by numerical means, we can deduce what follows:
\[
	\text{The union of cones }
	\bigcup_{k=1}^K \C_{\P_k}
	\text{ is a collapsing set.}
\]

\end{remark}

		\subsection{Analytical Results}
		\label{subsection:analytical-results}
			
			\subsubsection{The Triangle $\PR$, $\P_0$, $\P_1$ is Collapsing}
	
In this subsection, we extend the result~\cite{BS09} from Brunner and Skrzypczyk, who showed that any box in the segment joining the boxes $\PR$ and $\SR:=(\P_0+\P_1)/2$ is collapsing (except the box $\SR$, which is classical).
In the following theorem, {we recover a result from~\cite{BMRC19} stating that any box in the triangle joining the boxes $\PR, \P_0, \P_1$ is collapsing (except the boxes in the segment joining $\P_0$ and $\P_1$, which are classical), with a new proof, based on the algebra of boxes}.
Recall that $\PR$ is the non-signalling box that outputs $(a,b)$ such that $a\oplus b=xy$ when $(x,y)$ is inputted, and $\P_0$ and $\P_1$ are respectively the deterministic boxes that output $(0,0)$ and $(1,1)$ independently of the inputs.
Recall also that the \emph{convex hull} of a set $\{\Q_1, \dots, \Q_N\}$ is the set of all possible convex combinations of these $\Q_i$:
\[
	\Conv\{\Q_1, \dots, \Q_N\}
	\,:=\,
	\left\{ \sum_{i=1}^N q_i\,\Q_i \text{ such that } q_i\geq0\text{ and } \textstyle\sum_{i} q_i=1 \right\}\,,
\]
and the \emph{affine hull} of $\{\Q_1, \dots, \Q_N\}$ has the same definition but without the non-negativity constraint:
\[
	\Aff\{\Q_1, \dots, \Q_N\}
	\,:=\,
	\left\{ \sum_{i=1}^N q_i\,\Q_i \text{ such that } q_i\in\R\text{ and }\textstyle\sum_i q_i=1 \right\}
	\supseteq \Conv\{\Q_1, \dots, \Q_N\}\,.
\]
\vspace{-0.2cm}

\begin{proposition}  \label{prop:PR-P0-P1-is-a-face-of-NS}
	The triangle $C:=\Conv\{\PR, \P_0, \P_1\}$ is a face on the boundary of $\NS$.
\end{proposition}
\vspace{-0.2cm}

\begin{proof}
	First, we prove the equality between the sets $C$ and $A:=\NS \cap \Aff\{\PR, \P_0, \P_1\}$, meaning that the convex hull $C$ is actually a slice of $\NS$.
	The first inclusion $C\subseteq A$ is trivial because the three points $\PR, \P_0, \P_1$ are in $\NS$ and because $\NS$ is a polytope so it is stable under taking convex combination.
	Conversely, recall that by definition $\CHSH(\P):=\frac14 \sum_{a\oplus b=xy} \P(a,b\,|\,x,y)$, and $\CHSH'(\cdot)$ and $\CHSH''(\cdot)$ are defined similarly but with respective summand conditions $a\oplus b=(x\oplus 1)\cdot(y\oplus1)$ and $a\oplus b=x\cdot(y\oplus1)$.
	As these three functions are linear, they preserve alignment and convexity.
	It means that if a box $\P$ is of the form $\P=\sum_iq_i\,\Q_i$ for some reals $q_i$ and boxes $\Q_i$, then $\CHSH(\P)=\sum_iq_i\,\CHSH(\Q_i)$, and similarly with $\CHSH'$ and $\CHSH''$.
	We apply the preservation of alignment in \refprop[Figure]{fig:CHSH-and-CHSHprime-values-of-extremal-points-of-NS} representing the $24$ extremal points of $\NS$~\cite{BLMPPR05},
	 and we obtain the following inclusions:
	 \begin{align*}
	 	&\text{(a)}\quad\quad A\subseteq \Conv\{\PR, \P_0, \P_1, \P_L^{1011}, \P_L^{1110}, \P_{NL}^{111}\}\,;\\
		&\text{(b)}\quad\quad A\subseteq \Conv\{\PR, \P_0, \P_1, \P_L^{1000}, \P_L^{1101}, \P_{NL}^{100}\}\,.
	\end{align*}
	Now, taking the intersection, we obtain $A\subseteq C$, which yields the wanted equality $A=C$, and $C$ is indeed a slice of $\NS$.
	
\newcommand{\scaleGRAPH}{5} 
\newcommand{\scaleTEXT}{0.7} 
	
\begin{figure}
	\centering
	\definecolor{my-gray}{rgb}{0.7529411764705882,0.7529411764705882,0.7529411764705882}
{
\begin{tikzpicture}[scale=\scaleGRAPH, every node/.style={scale=\scaleTEXT},
my-axis/.style={->,color=black, >=triangle 45, line width=0.8pt},
]
	
	\draw [color=my-gray, xstep=0.25, ystep=0.25, dashed] (-0., -0.) grid (1.1, 1.1);
	\draw[my-axis] (-0.1, 0.) -- (1.1,0.);
	\draw[my-axis] (0, -0.1) -- (0, 1.1);
	
	\foreach \xy in {0.25, 0.5, 0.75, 1}{
		\draw[shift={(\xy,0)},color=black] (0pt,1pt) -- (0pt,-1pt) node[below]{$\xy$};
		\draw[shift={(0, \xy)},color=black] (1pt, 0) -- (-1pt, 0) node[left]{$\xy$};
	}
	\node at (-0.7pt, -0.7pt) [below left] {$0$};
	\node at (0.01, 1) [above right] {\large$\CHSH$-value};
	\node at (1.1, 0.03) [above left] {\large$\CHSH'$-value};
	
	\foreach \x in {0, 0.25, 0.5}{
		\foreach \dx in {0, 0.25, 0.5}{
			\draw[line width=0.7pt, fill=black!30] (\x+\dx, 0.5+\x-\dx) circle (0.4pt);
		}
	}
	
	\node at (0.5, 1) [above right] {\fbox{$\PR$}};
	
	\node at (0.75, 0.75) [above left] {\fbox{$\P_0$}};
	\node at (0.75, 0.75) [above right] {\fbox{$\P_1$}};
	\node at (0.75, 0.75) [below left] {$\P_{\text{L}}^{1011}$};
	\node at (0.75, 0.75) [below right] {$\P_{\text{L}}^{1110}$};
	
	\node at (1, 0.5) [above right] {$\P_{\text{NL}}^{111}$};
	
	\node at (0.25, 0.75) [above left] {$\P_{\text{L}}^{0010}$};
	\node at (0.25, 0.75) [above right] {$\P_{\text{L}}^{0111}$};
	\node at (0.25, 0.75) [below left] {$\P_{\text{L}}^{1000}$};
	\node at (0.25, 0.75) [below right] {$\P_{\text{L}}^{1101}$};
	
	\node at (0.5, 0.5) [above left] {$\P_{\text{NL}}^{010}$};
	\node at (0.5, 0.5) [above right] {$\P_{\text{NL}}^{011}$};
	\node at (0.5, 0.5) [below left] {$\P_{\text{NL}}^{100}$};
	\node at (0.5, 0.5) [below right] {$\P_{\text{NL}}^{101}$};
	
	\node at (0.75, 0.25) [above left] {$\P_{\text{L}}^{0011}$};
	\node at (0.75, 0.25) [above right] {$\P_{\text{L}}^{0110}$};
	\node at (0.75, 0.25) [below left] {$\P_{\text{L}}^{1001}$};
	\node at (0.75, 0.25) [below right] {$\P_{\text{L}}^{1100}$};
	
	\node at (0, 0.5) [above right] {$\P_{\text{NL}}^{110}$};
	
	\node at (0.25, 0.25) [above left] {$\P_{\text{L}}^{0001}$};
	\node at (0.25, 0.25) [above right] {$\P_{\text{L}}^{0100}$};
	\node at (0.25, 0.25) [below left] {$\P_{\text{L}}^{1010}$};
	\node at (0.25, 0.25) [below right] {$\P_{\text{L}}^{1111}$};
	
	\node at (0.5, 0) [above right] {$\P_{\text{NL}}^{001}$};

\end{tikzpicture}
}
	 \definecolor{my-gray}{rgb}{0.7529411764705882,0.7529411764705882,0.7529411764705882}
{
\begin{tikzpicture}[scale=\scaleGRAPH, every node/.style={scale=\scaleTEXT},
my-axis/.style={->,color=black, >=triangle 45, line width=0.8pt},
]
	
	\draw [color=my-gray, xstep=0.25, ystep=0.25, dashed] (-0., -0.) grid (1.1, 1.1);
	\draw[my-axis] (-0.1, 0.) -- (1.1,0.);
	\draw[my-axis] (0, -0.1) -- (0, 1.1);
	
	\foreach \xy in {0.25, 0.5, 0.75, 1}{
		\draw[shift={(\xy,0)},color=black] (0pt,1pt) -- (0pt,-1pt) node[below]{$\xy$};
		\draw[shift={(0, \xy)},color=black] (1pt, 0) -- (-1pt, 0) node[left]{$\xy$};
	}
	\node at (-0.7pt, -0.7pt) [below left] {$0$};
	\node at (0.01, 1) [above right] {\large$\CHSH$-value};
	\node at (1.1, 0.03) [above left] {\large$\CHSH''$-value};
	
	\foreach \x in {0, 0.25, 0.5}{
		\foreach \dx in {0, 0.25, 0.5}{
			\draw[line width=0.7pt, fill=black!30] (\x+\dx, 0.5+\x-\dx) circle (0.4pt);
		}
	}
	
	\node at (0.5, 1) [above right] {\fbox{$\PR$}};
	
	\node at (0.75, 0.75) [above left] {\fbox{$\P_0$}};
	\node at (0.75, 0.75) [above right] {\fbox{$\P_1$}};
	\node at (0.75, 0.75) [below left] {$\P_{\text{L}}^{1000}$};
	\node at (0.75, 0.75) [below right] {$\P_{\text{L}}^{1101}$};
	
	\node at (1, 0.5) [above right] {$\P_{\text{NL}}^{100}$};
	
	\node at (0.25, 0.75) [above left] {$\P_{\text{L}}^{0010}$};
	\node at (0.25, 0.75) [above right] {$\P_{\text{L}}^{0111}$};
	\node at (0.25, 0.75) [below left] {$\P_{\text{L}}^{1011}$};
	\node at (0.25, 0.75) [below right] {$\P_{\text{L}}^{1110}$};
	
	\node at (0.5, 0.5) [above left] {$\P_{\text{NL}}^{010}$};
	\node at (0.5, 0.5) [above right] {$\P_{\text{NL}}^{011}$};
	\node at (0.5, 0.5) [below left] {$\P_{\text{NL}}^{110}$};
	\node at (0.5, 0.5) [below right] {$\P_{\text{NL}}^{111}$};
	
	\node at (0.75, 0.25) [above left] {$\P_{\text{L}}^{0011}$};
	\node at (0.75, 0.25) [above right] {$\P_{\text{L}}^{0110}$};
	\node at (0.75, 0.25) [below left] {$\P_{\text{L}}^{1010}$};
	\node at (0.75, 0.25) [below right] {$\P_{\text{L}}^{1111}$};
	
	\node at (0, 0.5) [above right] {$\P_{\text{NL}}^{101}$};
	
	\node at (0.25, 0.25) [above left] {$\P_{\text{L}}^{0001}$};
	\node at (0.25, 0.25) [above right] {$\P_{\text{L}}^{0100}$};
	\node at (0.25, 0.25) [below left] {$\P_{\text{L}}^{1001}$};
	\node at (0.25, 0.25) [below right] {$\P_{\text{L}}^{1100}$};
	
	\node at (0.5, 0) [above right] {$\P_{\text{NL}}^{001}$};

\end{tikzpicture}
}
	\caption{
	Computation of the $\CHSH$-, $\CHSH'$- and $\CHSH''$- values of the $24$ extremal points of $\NS$~\cite{BLMPPR05},
	where by definition $\CHSH(\P):=\frac14 \sum_{a\oplus b=xy} \P(a,b\,|\,x,y)$, and $\CHSH'$ and $\CHSH''$ are defined similarly but with respective summand conditions $a\oplus b=(x\oplus 1)\cdot(y\oplus1)$ and $a\oplus b=x\cdot(y\oplus1)$.
	}
	\label{fig:CHSH-and-CHSHprime-values-of-extremal-points-of-NS}
\end{figure}
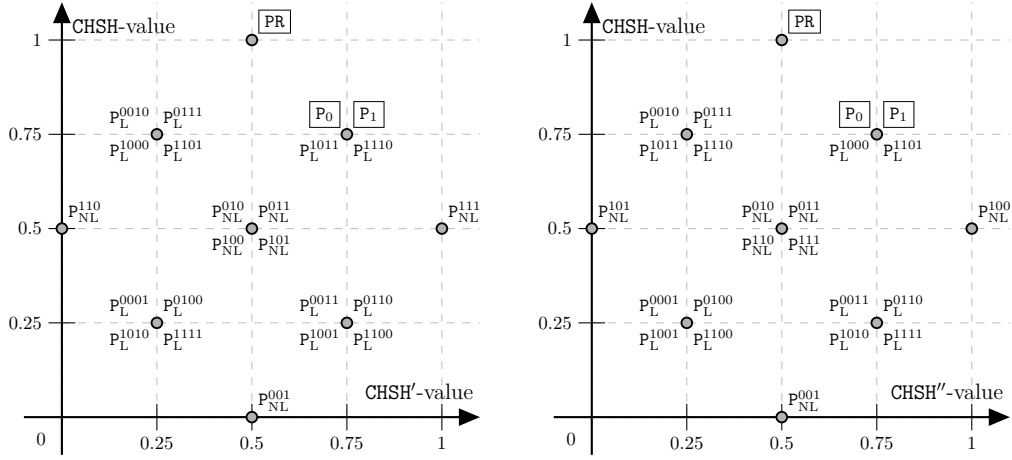
	
	 It remains to show that the slice $C$ is included in the boundary $\partial\NS$, so that it is indeed a face.
	 Assume that there is a point $\P$ in $C$ of the form $\P=q\Q_1+(1-q)\Q_2$ with $q>0$ and $\Q_1,\Q_2\in\NS$.
	 Applying the convexity preservation property of $\CHSH(\cdot)$, $\CHSH'(\cdot)$, $\CHSH''(\cdot)$ in \refprop[Figure]{fig:CHSH-and-CHSHprime-values-of-extremal-points-of-NS}, we obtain the following two necessary conditions:
	  \begin{align*}
	 	&\text{(a)}\quad\quad \Q_1,\Q_2\in \Conv\{\PR, \P_0, \P_1, \P_L^{1011}, \P_L^{1110}, \P_{NL}^{111}\}\,;\\
		&\text{(b)}\quad\quad \Q_1,\Q_2\in \Conv\{\PR, \P_0, \P_1, \P_L^{1000}, \P_L^{1101}, \P_{NL}^{100}\}\,.
	\end{align*}
	Then, taking the intersection, we get $\Q_1,\Q_2\in C$ and therefore $C\subseteq \partial\NS$ as wanted.
\end{proof}
\vspace{-0.2cm}

\begin{theorem}   \label{thm: the triangle PR-P0-P1 is collapsing}
	The face $C$ of $\NS$ is collapsing, except in the segment $\Conv\{\P_0, \P_1\}$.
\end{theorem}
\vspace{-0.2cm}

\begin{figure}[h]
	\centering
	\begin{tikzpicture}[line cap=round,line join=round,>=triangle 45,x=1.0cm,y=1.0cm, 
my-point/.style={fill=black},
every node/.style={scale=0.8}, 
scale=1]

	\node at (0,3) (PR) {};
	\node at (4,2) (P0) {};
	\node at (4,4) (P1) {};
	\node at (2.55, 3) (CC) {Collapsing triangle};
	\node[anchor = west] at (4.1, 3) {$\leftarrow$\small (Non-collapsing segment)};
	
	\fill[line width=2.2pt, color=quantumviolet, fill=quantumviolet, fill opacity=0.15] (P0.center) -- (PR.center) -- (P1.center) -- cycle;
	\draw [line width=2.2pt,color=quantumviolet] (P0.center) -- (PR.center) -- (P1.center);
	\draw [line width=2.pt,color=black, dashed] (P0) -- (P1);
	
	\draw[fill=quantumviolet] (PR) circle (0.07) node[anchor = east] {$\PR$\,};
	\draw[my-point] (P0) circle (0.07) node[anchor = west] {\,$\P_0$};
	\draw[my-point] (P1) circle (0.07) node[anchor = west] {\,$\P_1$};

\end{tikzpicture}
	\caption{Illustration of \refprop[Theorem]{thm: the triangle PR-P0-P1 is collapsing}.}
\end{figure}
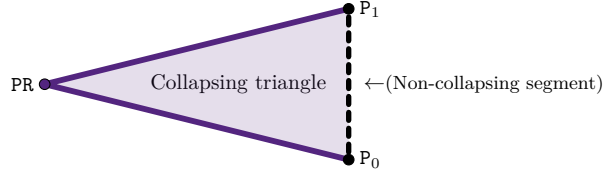

\begin{proof}
	Denote $\T:=\Conv\{\PR, \P_0, \P_1\}\backslash\Conv\{\P_0, \P_1\}$,
	and consider a convex combination of the form $\P_{\alpha, \beta} := \alpha\PR + \beta\P_0 + (1-\alpha-\beta)\P_1\in \T$ with $\alpha, \beta\geq0$ and $\alpha\neq0$ and $\alpha+\beta\leq1$. Similarly fix other convex coefficients $(\alpha_0, \beta_0)=:u_0$ with the same conditions.
	We want to build a sequence $(u_k)_k=\big((\alpha_k, \beta_k)\big)_k$ such that $(\P_{u_k})_k$ tends to the $\PR$ box.
	Denote $\boxtimes$ the box product induced by the wiring $\Wbs$ (see definition in \refprop[Subsection]{subsec:typical-examples-of-wirings}).
	By bilinearity of $\boxtimes$ and using the multiplication table in \refprop[Figure]{fig: multiplication table}, computations lead to $\P_{\alpha, \beta} \boxtimes \P_{\alpha_0, \beta_0} = \P_{\tilde\alpha, \tilde\beta}$ where:
	\begin{equation*}
		\begin{bmatrix}
			\tilde\alpha\\
			\tilde\beta
		\end{bmatrix}
		\,=\,
		A
		 \begin{bmatrix}
			\alpha\\
			\beta
		\end{bmatrix}
		+
		b\,,
		\quad\quad
		A:=\begin{bmatrix}
			1 - \alpha_0 & - \alpha_0\\
			-1+ \alpha_0+ \beta_0  & -1 +  \frac32 \alpha_0  + 2\beta_0
		\end{bmatrix}\,,
		\quad\quad
		b:=\begin{bmatrix}
			\alpha_0\\
			1 - \alpha_0 -\beta_0
		\end{bmatrix}\,.
	\end{equation*}
	From this remark, we define the following sequence:
	\[
		u_0:=(\alpha_0, \beta_0)\,,
		\quad\quad\quad\quad
		u_{k+1} = A\, u_k + b\,.
	\]
	We easily identify that $\ell:=(1, 0)$ is a fixed point of $x\mapsto A\,x+b$, so it yields:
	\[
		u_{k+1} - \ell
		\,=\,
		\big(A\,u_k+b\big) - \big(A\ell + b\big)
		\,=\,
		A\,(u_k-\ell)
		\,=\,
		A^{k+1}\,(u_0-\ell)\,,
	\]
	where the last equality follows from an induction on $k$. But the matrix $A$ admits exactly two distinct\footnote{The eigenvalues $\lambda_1$ and $\lambda_2$ are distinct because otherwise we would have $2=\frac32(a+b)+\frac b2$, which is achieved only if both $a+b=1$ and $b=1$, which is equivalent to $a=0$ and $b=1$ and which contradicts the assumption $a\neq0$.} eigenvalues $\lambda_1=1-a/2$ and $\lambda_2=-1+a+2b$. So $A$ is diagonalizable, and its power $A^{k}=P{\tiny \begin{bmatrix} \lambda_1^{k} & 0 \\ 0 & \lambda_2^k \end{bmatrix}} P^{-1}$ tends to the null matrix because $|\lambda_1|,|\lambda_2|<1$, where $P$ is an invertible matrix.
	Hence, from the above equation, the sequence $(u_k)_k$ tends to $\ell$, and by continuity
	we have that the sequence of boxes $(\P_{u_k})_k\subseteq \R^{16}$ converges to $\P_{\ell}=\PR$.
	But Brassard \emph{et al.} showed that there is an open neighbor around $\PR$ that collapses communication complexity~\cite{BBLMTU06}.
	Therefore, we know that the sequence $(\P_{u_k})_k$ reaches this collapsing neighbor for some $k$ large enough, and using \refprop[Proposition]{prop: collapsing orbit} we conclude that any starting box $\P_{u_0}\in \T$ is indeed collapsing.
\end{proof}
\vspace{-0.2cm}

\begin{remark}[Why is $\Conv\{\P_0, \P_1\}$ non-collapsing?]
	It is not surprising that the boundary segment $\Conv\{\P_0, \P_1\}$ of $T$ is not in the collapsing area because this segment is included in the local set $\LL$, which is itself included in the quantum set $\QQ$, for which it is known that communication complexity does not collapse~\cite{CvDNT99}.
\end{remark}
\vspace{-0.2cm}

\begin{remark}[Left multiplication does not give the same result]
	In the proof, we defined our sequence of boxes $(\P_{u_k})_k$ based on right multiplication.
	One could instead try to do the left multiplication: $\P_{u_{k+1}} = \P_{u_0}\boxtimes \P_{u_k}$. In that case, similar computations lead to:
	\[
		u_{k+1}
		\,=\,
		A'
		u_k
		+
		b'\,,
		\quad\quad
		A':=\begin{bmatrix}
			1 -\alpha_0 - \beta_0 & 0\\
			-1 + \alpha_0 +\frac32 \beta_0 & -1+ \alpha_0 + 2\beta_0
		\end{bmatrix}\,,
		\quad\quad
		b':=\begin{bmatrix}
			\alpha_0\\
			1 - \alpha_0 - \beta_0
		\end{bmatrix}\,.
	\]
	The map $x\mapsto A'\,x+b'$ admits a unique fixed point $\ell':=\big(\frac{\alpha_0}{\alpha_0+\beta_0}, \frac{\beta_0}{2(\alpha_0+\beta_0)}\big)$ (no division by $0$ since $\alpha_0>0$).
	The matrix $A'$ is already in the triangular form, its eigenvalues are $\lambda_1':=1-\alpha_0-\beta_0$ and $\lambda_2':=-1+\alpha_0+2\beta_0$, and again $|\lambda_1'|, |\lambda_2'|<1$ so $\P_{\alpha_k, \beta_k}\to \P_{\ell'}$. But in that case $\P_{\ell'}$ is not the $\PR$ box, so we cannot apply Ref.~\cite{BBLMTU06} to build a collapsing protocol from any starting box.
\end{remark}
\vspace{-0.2cm}

\begin{remark}[Pairwise multiplication gives the same result]   \label{rem: Pairwise multiplication gives the same result}
	It is also possible to try the pairwise multiplication: $\P_{u_{k+1}} = \P_{u_k}\boxtimes \P_{u_k}$, which is the way the authors of~\cite{BS09} originally proved that the segment $\Conv\{\PR, \SR\}\backslash\{\SR\}$ is collapsing.
	But this pairwise multiplication does not behave as well as with the right multiplication, iterations $u_k=(\alpha_k, \beta_k)$ are non-affine here:
	\[
		u_{k+1}
		\,=\,
		\begin{bmatrix}
			-1 & -1 & 0 \\
			1 & 5/2 & 2
		\end{bmatrix}
		\begin{bmatrix}
			\alpha_k^2\\
			\alpha_k \beta_k\\
			\beta_k^2
		\end{bmatrix}
		+
		\begin{bmatrix}
			2 & 0\\
			-2 & -2
		\end{bmatrix}
		u_k
		+
		\begin{bmatrix}
			0 \\
			1
		\end{bmatrix}
		\,.
	\]
	Nevertheless, the result still holds using this pairwise multiplication, but the proof we found is much more technical, see \refprop[Appendix]{appendix: proof of the theorem using Pairwise Multiplication}.
\end{remark}
\vspace{-0.2cm}

The intersection of the quantum set $\QQ$ with the boundary $\partial\NS$ of the non-signalling set was recently studied in~\cite{CTJMWL23}. {Moreover, the notion of the \emph{quantum void} was introduced and studied in~\cite{RDBC19, BMRC19}, which consist in a subset of $\partial\NS$ for which all quantum correlations are actually local.} A direct corollary of the previous theorem allows us to single out the quantum correlations of the face $C=\Conv\{\PR, \P_0, \P_1\}$: they are exactly the ones in the segment $\Conv\{\P_0, \P_1\}$. Indeed, on the one hand, it is known that quantum correlations do \emph{not} collapse communication complexity~\cite{CvDNT99}, and on the other hand, local correlations are particular cases of quantum correlations, {so we recover the following statement from~\cite{RDBC19} with a new proof, based on communication complexity}:
\vspace{-0.2cm}

\begin{corollary}   \label{coro:quantum-boxes-in-the-boundary-of-NS}
	{The face $C=\Conv\{\PR, \P_0, \P_1\}\subseteq\partial\NS$ is a quantum void:}
\begin{equation*}
	\QQ\cap C = \Conv\{\P_0, \P_1\}\,.
	\tag*{\qed}
\end{equation*}
\end{corollary}

	\subsubsection{Other Collapsing Triangles}
	
The fact that the triangle $\T:=\Conv\{\PR, \P_0, \P_1\}\backslash\Conv\{\P_0, \P_1\}$ is collapsing induces many further collapsing triangles. {In this subsection, we give some examples of such collapsing triangles, thus recovering some results of~\cite{BMRC19} with a new proof, based on the algebra of boxes}.
\vspace{-0.2cm}

\begin{proposition}   \label{prop:collapsing-if-good-multiplication-table}
Let $\P, \Q, \Rbox$ be three boxes. If there exists a wiring $\W\in\WW$ that induces the multiplication table below, then the triangle $\Conv\{\P, \Q, \Rbox\}\backslash\Conv\{\Q, \Rbox\}$ is collapsing.
\begin{center}
	{\footnotesize
	\renewcommand{\arraystretch}{1.5}
	\newcommand{\mycommand}{\normalsize}
	\begin{tabular}{| c ||c|c|c|c|c|c|c|c|}
		\hline
		& \hspace{\length}{\mycommand\hspace{0.2cm}$\P$\hspace{0.2cm}}\hspace{\length} & \hspace{\length}{\mycommand\hspace{0.2cm}$\Q$\hspace{0.2cm}}\hspace{\length} & \hspace{\length}{\mycommand\hspace{0.2cm}$\Rbox$\hspace{0.2cm}}\hspace{\length}  \\
		\hline\hline
		{\mycommand\hspace{0.3cm}$\P$\hspace{0.3cm}} &  $\P$ & $\P$ & $\P$  \\ \hline
		{\mycommand$\Q$} & $\frac{1}{2}\big(\Q + \Rbox\big)$  & $\Q$  & $\Rbox$  \\
		\hline
		{\mycommand$\Rbox$}  & $\P$ &  $\Rbox$ & $\Q$ \\ \hline
	\end{tabular}
	\renewcommand{\arraystretch}{1}}
\end{center}
\end{proposition}

\begin{proof}
	For any $\alpha, \beta\in\R$, consider the convex combination $\P_{\alpha, \beta} := \alpha\P + \beta\Q + (1-\alpha-\beta)\Rbox$.
	We have the equality $\P_{\alpha, \beta} \boxtimes \P_{\alpha_0, \beta_0} = \P_{\tilde\alpha, \tilde\beta}$, with the same coefficients $(\tilde\alpha, \tilde\beta)$ as in the proof of \refprop[Theorem]{thm: the triangle PR-P0-P1 is collapsing}.
	Then, applying the same proof gives the desired result.
\end{proof}
\vspace{-0.2cm}

\begin{theorem}   \label{thm:new-collapsing-triangles}
	All the triangles drawn in \refprop[Figure]{fig:new-collapsing-triangles} are collapsing.
\end{theorem}
\vspace{-0.2cm}

\definecolor{color-Wbs}{HTML}{53257F}
\definecolor{color-W1}{RGB}{191, 95, 0}
\definecolor{color-W2}{RGB}{0, 96, 81}
\definecolor{color-W3}{RGB}{96, 0, 81}
\definecolor{color-W4}{RGB}{150, 30, 30}

\newcommand{\scaleCOLLAPSINGtriangle}{0.79}
\newcommand{\scaleCOLLAPSINGwirings}{0.63}
\newcommand{\headphone}{-0.6cm}
\newlength{\camera}
\setlength{\camera}{-0.6cm}
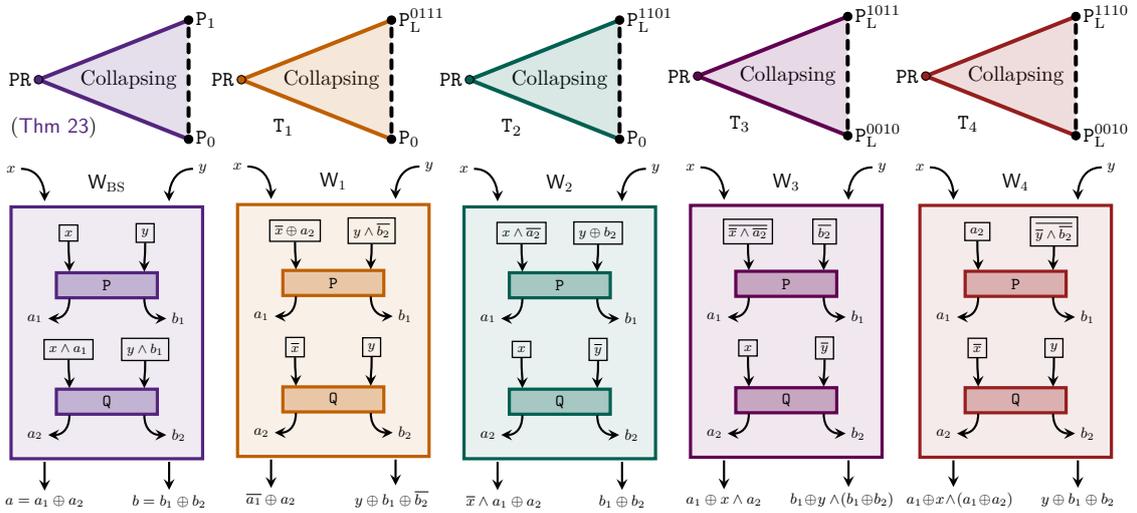
\begin{figure}[h]
	\begin{tikzpicture}[line cap=round,line join=round,>=triangle 45,x=1.0cm,y=1.0cm, 
my-point/.style={fill=black},
every node/.style={scale=\scaleCOLLAPSINGtriangle}, 
scale=\scaleCOLLAPSINGtriangle]

	\node at (0,3) (PR) {};
	\node at (2.5,2) (P0) {};
	\node at (2.5,4) (P1) {};
	\node at (1.5, 3) {Collapsing};
	\node[anchor=east] at (1.1, 2.2) {(\refprop[Thm]{thm: the triangle PR-P0-P1 is collapsing})};
	
	\fill[line width=1.5pt, color=color-Wbs, fill=color-Wbs, fill opacity=0.15] (P0.center) -- (PR.center) -- (P1.center) -- cycle;
	\draw [line width=1.5pt,color=color-Wbs] (P0.center) -- (PR.center) -- (P1.center);
	\draw [line width=1.5pt,color=black, dashed] (P0) -- (P1);
	
	\draw[fill=color-Wbs] (PR) circle (0.07) node[anchor = east] {$\PR$};
	\draw[my-point] (P0) circle (0.07) node[anchor = west] {$\P_0$};
	\draw[my-point] (P1) circle (0.07) node[anchor = west] {$\P_1$};

\end{tikzpicture}
	\hspace{\headphone}
	\begin{tikzpicture}[line cap=round,line join=round,>=triangle 45,x=1.0cm,y=1.0cm, 
my-point/.style={fill=black},
every node/.style={scale=\scaleCOLLAPSINGtriangle}, 
scale=\scaleCOLLAPSINGtriangle]

	\node at (0,3) (PR) {};
	\node at (2.5,2) (P0) {};
	\node at (2.5,4) (P1) {};
	\node at (1.5, 3) (CC) {Collapsing};
	\node[anchor=east] at (1, 2.2) {$\T_{1}$};
	
	\fill[line width=1.5pt, color=color-W1, fill=color-W1, fill opacity=0.15] (P0.center) -- (PR.center) -- (P1.center) -- cycle;
	\draw [line width=1.5pt,color=color-W1] (P0.center) -- (PR.center) -- (P1.center);
	\draw [line width=1.5pt,color=black, dashed] (P0) -- (P1);
	
	\draw[fill=color-W1] (PR) circle (0.07) node[anchor = east] {$\PR$};
	\draw[my-point] (P0) circle (0.07) node[anchor = west] {$\P_0$};
	\draw[my-point] (P1) circle (0.07) node[anchor = west] {$\P_{\text{L}}^{0111}$};

\end{tikzpicture}
	\hspace{\headphone}
	\begin{tikzpicture}[line cap=round,line join=round,>=triangle 45,x=1.0cm,y=1.0cm, 
my-point/.style={fill=black},
every node/.style={scale=\scaleCOLLAPSINGtriangle}, 
scale=\scaleCOLLAPSINGtriangle]

	\node at (0,3) (PR) {};
	\node at (2.5,2) (P0) {};
	\node at (2.5,4) (P1) {};
	\node at (1.5, 3) (CC) {Collapsing};
	\node[anchor=east] at (1, 2.2) {$\T_{2}$};
	
	\fill[line width=1.5pt, color=color-W2, fill=color-W2, fill opacity=0.15] (P0.center) -- (PR.center) -- (P1.center) -- cycle;
	\draw [line width=1.5pt,color=color-W2] (P0.center) -- (PR.center) -- (P1.center);
	\draw [line width=1.5pt,color=black, dashed] (P0) -- (P1);
	
	\draw[fill=color-W2] (PR) circle (0.07) node[anchor = east] {$\PR$};
	\draw[my-point] (P0) circle (0.07) node[anchor = west] {$\P_0$};
	\draw[my-point] (P1) circle (0.07) node[anchor = west] {$\P_{\text{L}}^{1101}$};

\end{tikzpicture}
	\hspace{\headphone}
	\begin{tikzpicture}[line cap=round,line join=round,>=triangle 45,x=1.0cm,y=1.0cm, 
my-point/.style={fill=black},
every node/.style={scale=\scaleCOLLAPSINGtriangle}, 
scale=\scaleCOLLAPSINGtriangle]

	\node at (0,3) (PR) {};
	\node at (2.5,2) (P0) {};
	\node at (2.5,4) (P1) {};
	\node at (1.5, 3) (CC) {Collapsing};
	\node[anchor=east] at (1, 2.2) {$\T_{3}$};
	
	\fill[line width=1.5pt, color=color-W3, fill=color-W3, fill opacity=0.15] (P0.center) -- (PR.center) -- (P1.center) -- cycle;
	\draw [line width=1.5pt,color=color-W3] (P0.center) -- (PR.center) -- (P1.center);
	\draw [line width=1.5pt,color=black, dashed] (P0) -- (P1);
	
	\draw[fill=color-W3] (PR) circle (0.07) node[anchor = east] {$\PR$};
	\draw[my-point] (P0) circle (0.07) node[anchor = west] {$\P_{\text{L}}^{0010}$};
	\draw[my-point] (P1) circle (0.07) node[anchor = west] {$\P_{\text{L}}^{1011}$};

\end{tikzpicture}
	\hspace{\headphone}
	\begin{tikzpicture}[line cap=round,line join=round,>=triangle 45,x=1.0cm,y=1.0cm, 
my-point/.style={fill=black},
every node/.style={scale=\scaleCOLLAPSINGtriangle}, 
scale=\scaleCOLLAPSINGtriangle]

	\node at (0,3) (PR) {};
	\node at (2.5,2) (P0) {};
	\node at (2.5,4) (P1) {};
	\node at (1.5, 3) (CC) {Collapsing};
	\node[anchor=east] at (1, 2.2) {$\T_{4}$};
	
	\fill[line width=1.5pt, color=color-W4, fill=color-W4, fill opacity=0.15] (P0.center) -- (PR.center) -- (P1.center) -- cycle;
	\draw [line width=1.5pt,color=color-W4] (P0.center) -- (PR.center) -- (P1.center);
	\draw [line width=1.5pt,color=black, dashed] (P0) -- (P1);
	
	\draw[fill=color-W4] (PR) circle (0.07) node[anchor = east] {$\PR$};
	\draw[my-point] (P0) circle (0.07) node[anchor = west] {$\P_{\text{L}}^{0010}$};
	\draw[my-point] (P1) circle (0.07) node[anchor = west] {$\P_{\text{L}}^{1110}$};

\end{tikzpicture}
	{
\newcommand{\shiftP}{-0.65}
\newcommand{\shiftQ}{-0.8}
\small
\begin{tikzpicture}[scale=\scaleCOLLAPSINGwirings, every node/.style={scale=\scaleCOLLAPSINGwirings},
small-NLB/.style={rectangle, draw=color-Wbs, fill=color-Wbs!35, line width=1.2pt, minimum width=60, minimum height=15, anchor=center},
big-NLB/.style={rectangle, draw=color-Wbs, fill=color-Wbs!9, line width=1.2pt, minimum width=60, minimum height=15},
operation/.style={rectangle, draw=black, line width=0.5pt},
my-arrow/.style={->, very thick, >=stealth, line width=0.8pt},
my-arrow-violet/.style={->, line width=0.5em, color=color-Wbs!8},
my-line/.style={very thick},
]

	\draw[big-NLB] (0., -1+0.2) rectangle (4., 4.5);
	\node at (2, 5) {\large$\Wbs$};
		
	\node[small-NLB] at (2, 3.5+\shiftP) (P) {\normalsize$\P$};
	\node[small-NLB] at (2, 1.2+\shiftQ) (Q) {\normalsize$\Q$};
	\node at (0., 5.3) (Alice1) {$x$};
	\node at (0.7, 4.5) (Alice2) {};
	\node[operation] at (1.2, 4.6+\shiftP) (Alice3) {\footnotesize$x$};
	\node at (0.5, 2.8+\shiftP) (Alice4) {$a_1$};
	\node[operation] at (1.2, 2.3+\shiftQ) (Alice5) {\footnotesize$x\wedge a_1$};
	\node at (0.5, 0.5+\shiftQ) (Alice6) {$a_2$};
	\node at (0.7, -0.95+0.2) (Alice7) {};
	\node at (0.7, -1.7) (Alice8) {$a=a_1\oplus a_2$};
	\node at (4., 5.3) (Bob1) {$y$};
	\node at (3.3, 4.5) (Bob2) {};
	\node[operation] at (4-1.2, 4.6+\shiftP) (Bob3) {\footnotesize$y$};
	\node at (4-0.5, 2.8+\shiftP) (Bob4) {$b_1$};
	\node[operation] at (4-1.2, 2.3+\shiftQ) (Bob5) {\footnotesize$y\wedge b_1$};
	\node at (4-0.5, 0.5+\shiftQ) (Bob6) {$b_2$};
	\node at (4-0.7, -0.95+0.2) (Bob7) {};
	\node at (4-0.7, -1.7) (Bob8) {$b=b_1\oplus b_2$};
	
	\draw[my-arrow] (Alice1) .. controls +(right:2em) and +(up:1em).. (Alice2);
	\draw[my-arrow] (Alice3.south)  .. controls  +(down:1em) and +(up:0.8em) ..  (P.160);
	\draw[my-arrow] (P.200)  .. controls  +(down:1em) and +(right:0.8em) ..  (Alice4.east);
	\draw[my-arrow] (Alice5.south)  .. controls  +(south:1em) and +(up:0.8em) ..  (Q.160);
	\draw[my-arrow] (Q.200)  .. controls  +(down:1em) and +(right:0.8em) ..  (Alice6.east);
	\draw[my-arrow] (Alice7) to (Alice8);
	
	\draw[my-arrow] (Bob1) .. controls +(left:2em) and +(up:1em).. (Bob2);
	\draw[my-arrow] (Bob3.south)  .. controls  +(down:1em) and +(up:0.8em) .. (P.20);
	\draw[my-arrow] (P.340)  .. controls  +(down:1em) and +(left:0.8em) ..  (Bob4.west);
	\draw[my-arrow] (Bob5.south)  .. controls  +(down:1em) and +(up:0.8em).. (Q.20);
	\draw[my-arrow] (Q.340)  .. controls  +(down:1em) and +(left:0.8em) ..  (Bob6.west);
	\draw[my-arrow] (Bob7) to (Bob8);

\end{tikzpicture}
}
	\hspace{\camera}
	{
\newcommand{\shiftP}{-0.65}
\newcommand{\shiftQ}{-0.8}
\small
\begin{tikzpicture}[scale=\scaleCOLLAPSINGwirings, every node/.style={scale=\scaleCOLLAPSINGwirings},
small-NLB/.style={rectangle, draw=color-W1, fill=color-W1!35, line width=1.2pt, minimum width=60, minimum height=15, anchor=center},
big-NLB/.style={rectangle, draw=color-W1, fill=color-W1!9, line width=1.2pt, minimum width=60, minimum height=15},
operation/.style={rectangle, draw=black, line width=0.5pt},
my-arrow/.style={->, very thick, >=stealth, line width=0.8pt},
my-arrow-violet/.style={->, line width=0.5em, color=color-W1!8},
my-line/.style={very thick},
]

	\draw[big-NLB] (0., -1+0.2) rectangle (4., 4.5);
	\node at (2, 5) {\large$\W_{1}$};
		
	\node[small-NLB] at (2, 3.5+\shiftP) (P) {\normalsize$\P$};
	\node[small-NLB] at (2, 1.2+\shiftQ) (Q) {\normalsize$\Q$};
	\node at (0., 5.3) (Alice1) {$x$};
	\node at (0.7, 4.5) (Alice2) {};
	\node[operation] at (1.2, 4.6+\shiftP) (Alice3) {\footnotesize$\overline x \oplus a_2$};
	\node at (0.5, 2.8+\shiftP) (Alice4) {$a_1$};
	\node[operation] at (1.2, 2.3+\shiftQ) (Alice5) {\footnotesize$\overline x$};
	\node at (0.5, 0.5+\shiftQ) (Alice6) {$a_2$};
	\node at (0.7, -0.95+0.2) (Alice7) {};
	\node at (0.7, -1.7) (Alice8) {$\overline{a_1}\oplus a_2$};
	\node at (4., 5.3) (Bob1) {$y$};
	\node at (3.3, 4.5) (Bob2) {};
	\node[operation] at (4-1.2, 4.6+\shiftP) (Bob3) {\footnotesize$y\wedge \overline{b_2}$};
	\node at (4-0.5, 2.8+\shiftP) (Bob4) {$b_1$};
	\node[operation] at (4-1.2, 2.3+\shiftQ) (Bob5) {\footnotesize$y$};
	\node at (4-0.5, 0.5+\shiftQ) (Bob6) {$b_2$};
	\node at (4-0.7, -0.95+0.2) (Bob7) {};
	\node at (4-0.7, -1.7) (Bob8) {\!\!\!$y\oplus b_1\oplus \overline{b_2}$};
	
	\draw[my-arrow] (Alice1) .. controls +(right:2em) and +(up:1em).. (Alice2);
	\draw[my-arrow] (Alice3.south)  .. controls  +(down:1em) and +(up:0.8em) ..  (P.160);
	\draw[my-arrow] (P.200)  .. controls  +(down:1em) and +(right:0.8em) ..  (Alice4.east);
	\draw[my-arrow] (Alice5.south)  .. controls  +(south:1em) and +(up:0.8em) ..  (Q.160);
	\draw[my-arrow] (Q.200)  .. controls  +(down:1em) and +(right:0.8em) ..  (Alice6.east);
	\draw[my-arrow] (Alice7) to (Alice8);
	
	\draw[my-arrow] (Bob1) .. controls +(left:2em) and +(up:1em).. (Bob2);
	\draw[my-arrow] (Bob3.south)  .. controls  +(down:1em) and +(up:0.8em) .. (P.20);
	\draw[my-arrow] (P.340)  .. controls  +(down:1em) and +(left:0.8em) ..  (Bob4.west);
	\draw[my-arrow] (Bob5.south)  .. controls  +(down:1em) and +(up:0.8em).. (Q.20);
	\draw[my-arrow] (Q.340)  .. controls  +(down:1em) and +(left:0.8em) ..  (Bob6.west);
	\draw[my-arrow] (Bob7) to (Bob8);

\end{tikzpicture}
}
	\hspace{\camera}
	{
\newcommand{\shiftP}{-0.65}
\newcommand{\shiftQ}{-0.8}
\small
\begin{tikzpicture}[scale=\scaleCOLLAPSINGwirings, every node/.style={scale=\scaleCOLLAPSINGwirings},
small-NLB/.style={rectangle, draw=color-W2, fill=color-W2!35, line width=1.2pt, minimum width=60, minimum height=15, anchor=center},
big-NLB/.style={rectangle, draw=color-W2, fill=color-W2!9, line width=1.2pt, minimum width=60, minimum height=15},
operation/.style={rectangle, draw=black, line width=0.5pt},
my-arrow/.style={->, very thick, >=stealth, line width=0.8pt},
my-arrow-violet/.style={->, line width=0.5em, color=color-W2!8},
my-line/.style={very thick},
]

	\draw[big-NLB] (0., -1+0.2) rectangle (4., 4.5);
	\node at (2, 5) {\large$\W_{2}$};
		
	\node[small-NLB] at (2, 3.5+\shiftP) (P) {\normalsize$\P$};
	\node[small-NLB] at (2, 1.2+\shiftQ) (Q) {\normalsize$\Q$};
	\node at (0., 5.3) (Alice1) {$x$};
	\node at (0.7, 4.5) (Alice2) {};
	\node[operation] at (1.2, 4.6+\shiftP) (Alice3) {\footnotesize$x \wedge \overline{a_2}$};
	\node at (0.5, 2.8+\shiftP) (Alice4) {$a_1$};
	\node[operation] at (1.2, 2.3+\shiftQ) (Alice5) {\footnotesize$x$};
	\node at (0.5, 0.5+\shiftQ) (Alice6) {$a_2$};
	\node at (0.7, -0.95+0.2) (Alice7) {};
	\node at (0.7, -1.7) (Alice8) {$\overline{x}\wedge {a_1}\oplus a_2$\!\!\!\!\!};
	\node at (4., 5.3) (Bob1) {$y$};
	\node at (3.3, 4.5) (Bob2) {};
	\node[operation] at (4-1.2, 4.6+\shiftP) (Bob3) {\footnotesize$y\oplus {b_2}$};
	\node at (4-0.5, 2.8+\shiftP) (Bob4) {$b_1$};
	\node[operation] at (4-1.2, 2.3+\shiftQ) (Bob5) {\footnotesize$\overline y$};
	\node at (4-0.5, 0.5+\shiftQ) (Bob6) {$b_2$};
	\node at (4-0.7, -0.95+0.2) (Bob7) {};
	\node at (4-0.7, -1.7) (Bob8) {$b_1\oplus {b_2}$};
	
	\draw[my-arrow] (Alice1) .. controls +(right:2em) and +(up:1em).. (Alice2);
	\draw[my-arrow] (Alice3.south)  .. controls  +(down:1em) and +(up:0.8em) ..  (P.160);
	\draw[my-arrow] (P.200)  .. controls  +(down:1em) and +(right:0.8em) ..  (Alice4.east);
	\draw[my-arrow] (Alice5.south)  .. controls  +(south:1em) and +(up:0.8em) ..  (Q.160);
	\draw[my-arrow] (Q.200)  .. controls  +(down:1em) and +(right:0.8em) ..  (Alice6.east);
	\draw[my-arrow] (Alice7) to (Alice8);
	
	\draw[my-arrow] (Bob1) .. controls +(left:2em) and +(up:1em).. (Bob2);
	\draw[my-arrow] (Bob3.south)  .. controls  +(down:1em) and +(up:0.8em) .. (P.20);
	\draw[my-arrow] (P.340)  .. controls  +(down:1em) and +(left:0.8em) ..  (Bob4.west);
	\draw[my-arrow] (Bob5.south)  .. controls  +(down:1em) and +(up:0.8em).. (Q.20);
	\draw[my-arrow] (Q.340)  .. controls  +(down:1em) and +(left:0.8em) ..  (Bob6.west);
	\draw[my-arrow] (Bob7) to (Bob8);

\end{tikzpicture}
}
	\hspace{\camera}
	{
\newcommand{\shiftP}{-0.65}
\newcommand{\shiftQ}{-0.8}
\small
\begin{tikzpicture}[scale=\scaleCOLLAPSINGwirings, every node/.style={scale=\scaleCOLLAPSINGwirings},
small-NLB/.style={rectangle, draw=color-W3, fill=color-W3!35, line width=1.2pt, minimum width=60, minimum height=15, anchor=center},
big-NLB/.style={rectangle, draw=color-W3, fill=color-W3!9, line width=1.2pt, minimum width=60, minimum height=15},
operation/.style={rectangle, draw=black, line width=0.5pt},
my-arrow/.style={->, very thick, >=stealth, line width=0.8pt},
my-arrow-violet/.style={->, line width=0.5em, color=color-W3!8},
my-line/.style={very thick},
]

	\draw[big-NLB] (0., -1+0.2) rectangle (4., 4.5);
	\node at (2, 5) {\large$\W_{3}$};
		
	\node[small-NLB] at (2, 3.5+\shiftP) (P) {\normalsize$\P$};
	\node[small-NLB] at (2, 1.2+\shiftQ) (Q) {\normalsize$\Q$};
	\node at (0., 5.3) (Alice1) {$x$};
	\node at (0.7, 4.5) (Alice2) {};
	\node[operation] at (1.2, 4.6+\shiftP) (Alice3) {\footnotesize$\overline{\overline{x}\wedge\overline{a_2}}$};
	\node at (0.5, 2.8+\shiftP) (Alice4) {$a_1$};
	\node[operation] at (1.2, 2.3+\shiftQ) (Alice5) {\footnotesize$x$};
	\node at (0.5, 0.5+\shiftQ) (Alice6) {$a_2$};
	\node at (0.7, -0.95+0.2) (Alice7) {};
	\node at (0.7, -1.7) (Alice8) {$a_1 \oplus x\wedge a_2$};
	\node at (4., 5.3) (Bob1) {$y$};
	\node at (3.3, 4.5) (Bob2) {};
	\node[operation] at (4-1.2, 4.6+\shiftP) (Bob3) {\footnotesize$\overline{b_2}$};
	\node at (4-0.5, 2.8+\shiftP) (Bob4) {$b_1$};
	\node[operation] at (4-1.2, 2.3+\shiftQ) (Bob5) {\footnotesize$\overline{y}$};
	\node at (4-0.5, 0.5+\shiftQ) (Bob6) {$b_2$};
	\node at (4-0.7, -0.95+0.2) (Bob7) {};
	\node at (4-0.7, -1.7) (Bob8) {\!\!\!\!\!$b_1\!\oplus\!y\wedge\!(b_1\!\oplus\!{b_2})$};
	
	\draw[my-arrow] (Alice1) .. controls +(right:2em) and +(up:1em).. (Alice2);
	\draw[my-arrow] (Alice3.south)  .. controls  +(down:1em) and +(up:0.8em) ..  (P.160);
	\draw[my-arrow] (P.200)  .. controls  +(down:1em) and +(right:0.8em) ..  (Alice4.east);
	\draw[my-arrow] (Alice5.south)  .. controls  +(south:1em) and +(up:0.8em) ..  (Q.160);
	\draw[my-arrow] (Q.200)  .. controls  +(down:1em) and +(right:0.8em) ..  (Alice6.east);
	\draw[my-arrow] (Alice7) to (Alice8);
	
	\draw[my-arrow] (Bob1) .. controls +(left:2em) and +(up:1em).. (Bob2);
	\draw[my-arrow] (Bob3.south)  .. controls  +(down:1em) and +(up:0.8em) .. (P.20);
	\draw[my-arrow] (P.340)  .. controls  +(down:1em) and +(left:0.8em) ..  (Bob4.west);
	\draw[my-arrow] (Bob5.south)  .. controls  +(down:1em) and +(up:0.8em).. (Q.20);
	\draw[my-arrow] (Q.340)  .. controls  +(down:1em) and +(left:0.8em) ..  (Bob6.west);
	\draw[my-arrow] (Bob7) to (Bob8);

\end{tikzpicture}
}
	\hspace{1.3\camera}
	{
\newcommand{\shiftP}{-0.65}
\newcommand{\shiftQ}{-0.8}
\small
\begin{tikzpicture}[scale=\scaleCOLLAPSINGwirings, every node/.style={scale=\scaleCOLLAPSINGwirings},
small-NLB/.style={rectangle, draw=color-W4, fill=color-W4!35, line width=1.2pt, minimum width=60, minimum height=15, anchor=center},
big-NLB/.style={rectangle, draw=color-W4, fill=color-W4!9, line width=1.2pt, minimum width=60, minimum height=15},
operation/.style={rectangle, draw=black, line width=0.5pt},
my-arrow/.style={->, very thick, >=stealth, line width=0.8pt},
my-arrow-violet/.style={->, line width=0.5em, color=color-W4!8},
my-line/.style={very thick},
]

	\draw[big-NLB] (0., -1+0.2) rectangle (4., 4.5);
	\node at (2, 5) {\large$\W_{4}$};
		
	\node[small-NLB] at (2, 3.5+\shiftP) (P) {\normalsize$\P$};
	\node[small-NLB] at (2, 1.2+\shiftQ) (Q) {\normalsize$\Q$};
	\node at (0., 5.3) (Alice1) {$x$};
	\node at (0.7, 4.5) (Alice2) {};
	\node[operation] at (1.2, 4.6+\shiftP) (Alice3) {\footnotesize${a_2}$};
	\node at (0.5, 2.8+\shiftP) (Alice4) {$a_1$};
	\node[operation] at (1.2, 2.3+\shiftQ) (Alice5) {\footnotesize$\overline{x}$};
	\node at (0.5, 0.5+\shiftQ) (Alice6) {$a_2$};
	\node at (0.7, -0.95+0.2) (Alice7) {};
	\node at (0.7, -1.7) (Alice8) {$a_1\!\oplus\!x\!\wedge\!(a_1\!\oplus\!a_2)$\!\!\!\!\!};
	\node at (4., 5.3) (Bob1) {$y$};
	\node at (3.3, 4.5) (Bob2) {};
	\node[operation] at (4-1.2, 4.6+\shiftP) (Bob3) {\footnotesize$\overline{\overline{y}\wedge\overline{b_2}}$};
	\node at (4-0.5, 2.8+\shiftP) (Bob4) {$b_1$};
	\node[operation] at (4-1.2, 2.3+\shiftQ) (Bob5) {\footnotesize$y$};
	\node at (4-0.5, 0.5+\shiftQ) (Bob6) {$b_2$};
	\node at (4-0.7, -0.95+0.2) (Bob7) {};
	\node at (4-0.7, -1.7) (Bob8) {$y\oplus b_1\oplus {b_2}$};
	
	\draw[my-arrow] (Alice1) .. controls +(right:2em) and +(up:1em).. (Alice2);
	\draw[my-arrow] (Alice3.south)  .. controls  +(down:1em) and +(up:0.8em) ..  (P.160);
	\draw[my-arrow] (P.200)  .. controls  +(down:1em) and +(right:0.8em) ..  (Alice4.east);
	\draw[my-arrow] (Alice5.south)  .. controls  +(south:1em) and +(up:0.8em) ..  (Q.160);
	\draw[my-arrow] (Q.200)  .. controls  +(down:1em) and +(right:0.8em) ..  (Alice6.east);
	\draw[my-arrow] (Alice7) to (Alice8);
	
	\draw[my-arrow] (Bob1) .. controls +(left:2em) and +(up:1em).. (Bob2);
	\draw[my-arrow] (Bob3.south)  .. controls  +(down:1em) and +(up:0.8em) .. (P.20);
	\draw[my-arrow] (P.340)  .. controls  +(down:1em) and +(left:0.8em) ..  (Bob4.west);
	\draw[my-arrow] (Bob5.south)  .. controls  +(down:1em) and +(up:0.8em).. (Q.20);
	\draw[my-arrow] (Q.340)  .. controls  +(down:1em) and +(left:0.8em) ..  (Bob6.west);
	\draw[my-arrow] (Bob7) to (Bob8);

\end{tikzpicture}
}
	\centering
	\caption{Examples of collapsing triangles, together with wirings that can be used in \refprop[Proposition]{prop:collapsing-if-good-multiplication-table} to show a collapse of communication complexity. The definition of the boxes $\P_{\text{L}}$ and $\P_{\text{NL}}$ can be found in \eqrefprop[Equation]{eq: extremal points of NS}.
	}
	\label{fig:new-collapsing-triangles}
\end{figure}
\vspace{-0.2cm}

\begin{proof}
	The proof follows directly from \refprop[Proposition]{prop:collapsing-if-good-multiplication-table} applied to what follows:
	\[
		\begin{array}{|c|c|c|c|c|}
			\hline
			\text{Triangle} & \P & \Q & \Rbox & \text{Wiring $\W\in\WW\subseteq \R^{32}$}\\
			\hline
			\hline
			\T & \PR & \P_0 & \P_1 & \text{\scriptsize$\Wbs = [0, 0, 1, 1, 0, 0, 1, 1, 0, 0, 0, 1, 0, 0, 0, 1, 0, 1, 1, 0, 0, 1, 1, 0, 0, 1, 1, 0, 0, 1, 1, 0]$}
			\\
			\hline
			\T_{1} & \PR & \P_{\text{L}}^{0111} & \P_0 & \text{\scriptsize$\W_{1} = [1, 0, 0, 1, 0, 0, 1, 0, 1, 1, 0, 0, 0, 0, 1, 1, 1, 0, 0, 1, 1, 0, 0, 1, 1, 0, 0, 1, 0, 1, 1, 0]$}
			\\
			\hline
			\T_{2} & \PR & \P_0 & \P_{\text{L}}^{1101} & \text{\scriptsize$\W_{2} = [0, 0, 1, 0, 0, 1, 1, 0, 0, 0, 1, 1, 1, 1, 0, 0, 0,1, 1, 0, 0, 0, 1, 1, 0, 1, 1, 0, 0, 1, 1, 0]$}
			\\
			\hline
			\T_{3} & \PR & \P_{\text{L}}^{0010} & \P_{\text{L}}^{1011} & \text{\scriptsize$\W_{3} = [0, 1, 1, 1, 1, 0, 1, 0, 0, 0, 1, 1, 1, 1, 0, 0, 0,  0, 1, 1, 0, 1, 1, 0, 0, 0, 1, 1, 0, 0, 1, 1]$}
			\\
			\hline
			\T_{4} & \PR & \P_{\text{L}}^{1000} & \P_{\text{L}}^{1110} & \text{\scriptsize$\W_{4} = [0    , 1    , 0    , 1    , 0    , 1    , 1    , 1    , 1    , 1    , 0    , 0    , 0, 0, 1    , 1    , 0    , 0    , 1    , 1    , 0    , 0    , 1    , 1    , 0    , 1    , 1    , 0    , 1    , 0    , 0    , 1    ]$}
			\\
			\hline
		\end{array}
	\]
	where the notation $\W\in\R^{32}$ comes form \eqrefprop[Equation]{eq:wiring-as-a-vector}.
	See a representation of these wirings in \refprop[Figure]{fig:new-collapsing-triangles}, bottom row.
\end{proof}

The wirings of \refprop[Figure]{fig:new-collapsing-triangles} are arbitrary examples of collapsing wirings that were obtained using \refprop[Algorithm]{alg:algo-B}; many more wirings can be found in other triangles of $\NS$ using the same algorithm, which is accessible via our GitHub page~\cite{GitHub-algebra-of-boxes}. {Notice that these wirings are all different from the ones used in the proof of~\cite{BMRC19}.}
Now, an interesting problem would be to understand better the structure of the set $\WW$ so that, given a triangle in $\NS$, we know how to construct a collapsing wiring $\W$ without using a search algorithm.

\bib

\section*{Conclusion}   \label{conclusion}

Our new algebraic perspective on nonlocal boxes allowed us to discover a surprising structure of what we called the orbit of a box, with some strong alignment and parallelism properties (see \refprop[Figure]{figure: orbit of a box}).
{As a consequence, this deeper understanding of the algebraic structure of nonlocal boxes enabled us to recover in \refprop[Section]{section: Analytical Proofs of the New Collapsing Regions} some collapsing regions of~$\NS$ that were recently also found in~\cite{BMRC19, EWC22PRL}}---for instance boxes in the triangle joining $\PR, \P_0,\P_1$ (except boxes in the segment $\P_0, \P_1$).
The importance of our results is emphasized by considering the many known impossibility results~\cite{BG15, Mori16, SWH20}.
Note also that according to our present intuition of Nature~\cite{vD99, BBLMTU06, BS09, BG15}, a direct consequence is that the collapsing boxes we presented are unlikely to exist in Nature.

{We made advances towards answering the open question of determining which nonlocal boxes do indeed collapse communication complexity}, but there is still a gap to be filled: for instance, as plotted in light blue in \refprop[Figure]{fig:numerical-new-collapsing-regions}, there are still regions of $\NS$ for which it is unknown whether there is a collapse of communication complexity.

Further work includes the study of the ``square root" of a box: here we introduced and studied the properties of the product $\P\boxtimes\P$, but one might be interested in finding all the boxes $\Q$ such that $\Q\boxtimes_{\W}\Q=\P$ for some fixed wiring $\W\in\WW$. If one knows that $\P$ is collapsing, then the wiring $\W$ yields a collapsing protocol, thus all the square roots $\Q$ are also collapsing.

Another interesting problem is the one mentioned at the end of \refprop[Section]{section: Analytical Proofs of the New Collapsing Regions}. Given a triangle of boxes in $\NS$, we already provided in \refprop[Section]{sec:numerical-approach} and in our GitHub page~\cite{GitHub-algebra-of-boxes} an algorithm that intends to find a wiring $\W$ that makes this triangle collapse communication complexity. Now, it would be interesting to address the question of how to construct such a collapsing wiring $\W$ without using a search algorithm, simply by knowing the target triangle of $\NS$. This can then help to understand better the structure of the set $\WW$ and importantly to better understand protocols for correlation distillation.

{In \refprop[Subsection]{subsection:analytical-results}, we give examples of collapsing sets of dimension $d=2$ (triangles). An open question from~\cite{BMRC19} consists in finding higher-dimensional sets of boxes that are all distillable to the $\PR$ box, which therefore collapses CC. For dimension $d=3$, the authors provide explicit examples of distillable sets, but it is still unknown whether these sets are collapsing. For dimensions $d\geq4$, they prove that distillable quantum voids are impossible due to the presence of isotropic boxes.
}

Another interesting problem would be to generalize and study the notion of ``product of boxes" with more than two boxes. Indeed, our study is limited to wirings connecting two boxes, but it is possible to consider more general wirings, connecting $n$ boxes, which are known to be strictly more powerful: \pierre{for instance, some genuine depth-$3$ wirings have significance in the context of trivial communication complexity~\cite{EWC22PRL}}. As a consequence, it may be that the multi-product of boxes $\W(\P_1, \dots, \P_n)$ gives rise to similar structures of orbits as the one we found in \refprop[Figure]{figure: orbit of a box}, which would be useful in the study of $n$-box distillation protocols.

In this work, we chose to study the principle of communication complexity. Although this principle alone cannot rule out the quantum set~\cite{NGHA15}, a clever idea would be to combine it with other principles, such as \emph{nonlocal computation}~\cite{LPSW07}, \emph{information causality}~\cite{PPKSWZ09, JGM23}, \emph{macroscopic locality}~\cite{NW09}, \emph{local orthogonality}~\cite{FSABCLA13}, \emph{nonlocality swapping}~\cite{SBP09}, \emph{many-box locality}~\cite{ZCBGS17}, in working towards a comprehensive information-based description of Quantum Mechanics.

\bib

\vspace{-0.25cm}
\section*{Acknowledgements}
\vspace{-0.2cm}
{\small
We thank Mirjam Weilenmann for the many helpful comments during the finalization process of the manuscript.
We also thank Giorgos Eftaxias for comments on a preliminary version of this work.
We acknowledge the support of the Natural Sciences and Engineering Research Council of Canada (NSERC), [funding reference number ALLRP 580876 - 22].
This work was also supported by the MITACS grant FR113029, and by the ANR projects \href{https://esquisses.math.cnrs.fr/}{ESQuisses}, grant number ANR-20-CE47-0014-01.
P.B.~was supported by the Institute for Quantum Technologies in Occitanie.
A.B.~acknowledges the support of the Air Force Office of Scientific Research under award number FA9550-20-1-0375.
R.C.~recognizes support from the ANR-3IA Artificial and Natural Intelligence Toulouse Institute.
I.N.~was supported by \href{https://www.math.univ-toulouse.fr/~gcebron/STARS.php}{STARS}, grant number ANR-20-CE40-0008, and by the PHC program \emph{Star} (Applications of random matrix theory and abstract harmonic analysis to quantum information theory).
C.P.~was supported by the ANR
projects Q-COAST ANR- 19-CE48-0003, “Quantum Trajectories” ANR-20-CE40-0024-01, and
“Investissements d’Avenir” ANR-11-LABX-0040 of the French National Research Agency.
}

{
	\printbibliography
}

\small
\appendix

	\section{Drawing of Some Orbits}
	\label{sec:drawing-of-some-orbits}
	
We present below some examples of orbits to illustrate \refprop[Subsection]{subsec:some-other-orbits}, using different wirings, each time in two different slices of $\NS$. Each orbit is drawn with depth going until $k=12$.
	The game $\G$ is defined by the winning rule $a=0$ and $b=y$.\\
	
\newcommand{\aaa}{\small}
\newcommand{\hh}{5.3cm}
\newcommand{\h}{4.55cm}
\newcommand{\wiringsize}{0.65}

	\noindent(a)~For $\W\oplus$ (see definition in \refprop[Subsection]{subsec:typical-examples-of-wirings}):\\
\begin{center}
	\includegraphics[height=\hh]{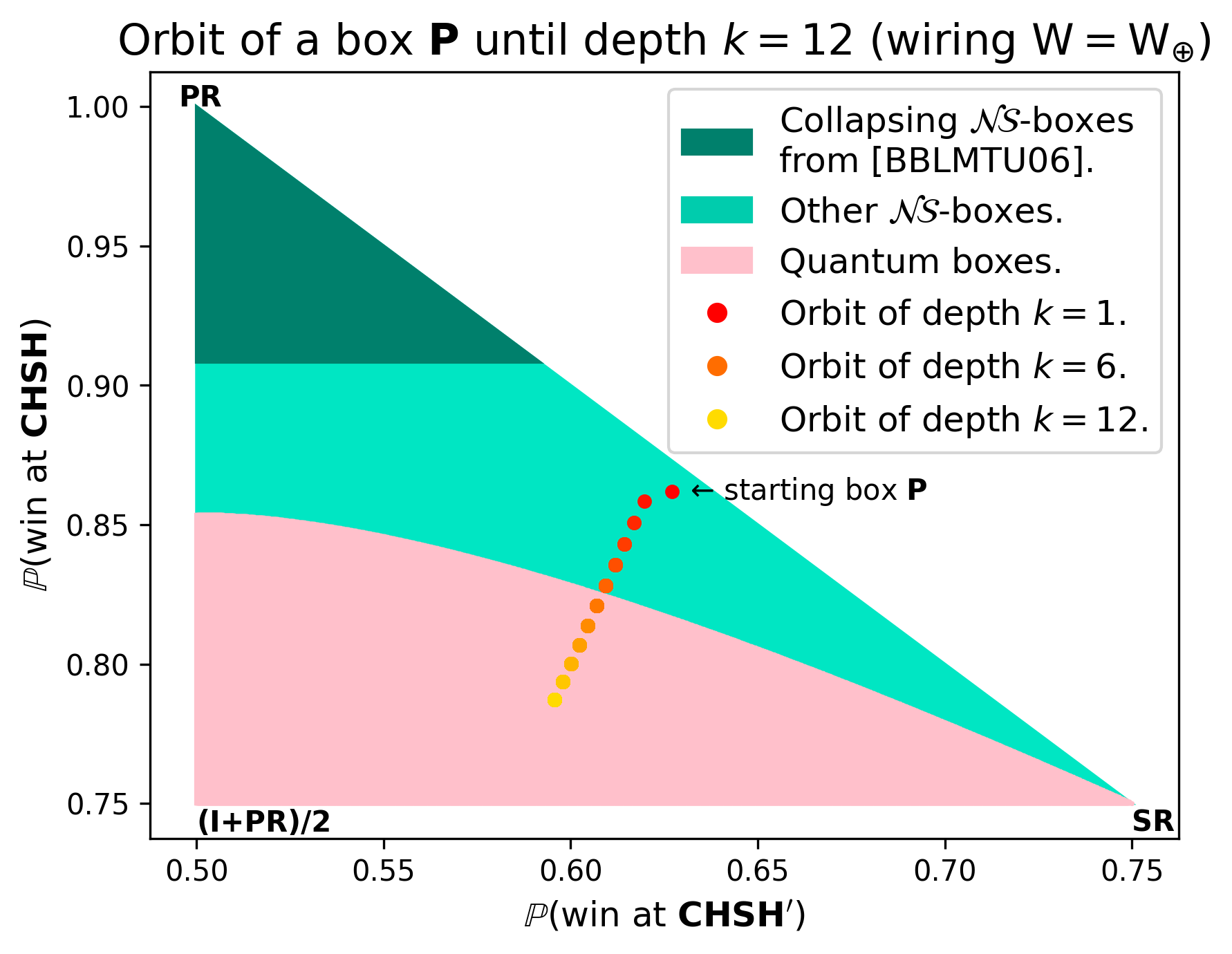}
	\includegraphics[height=\hh]{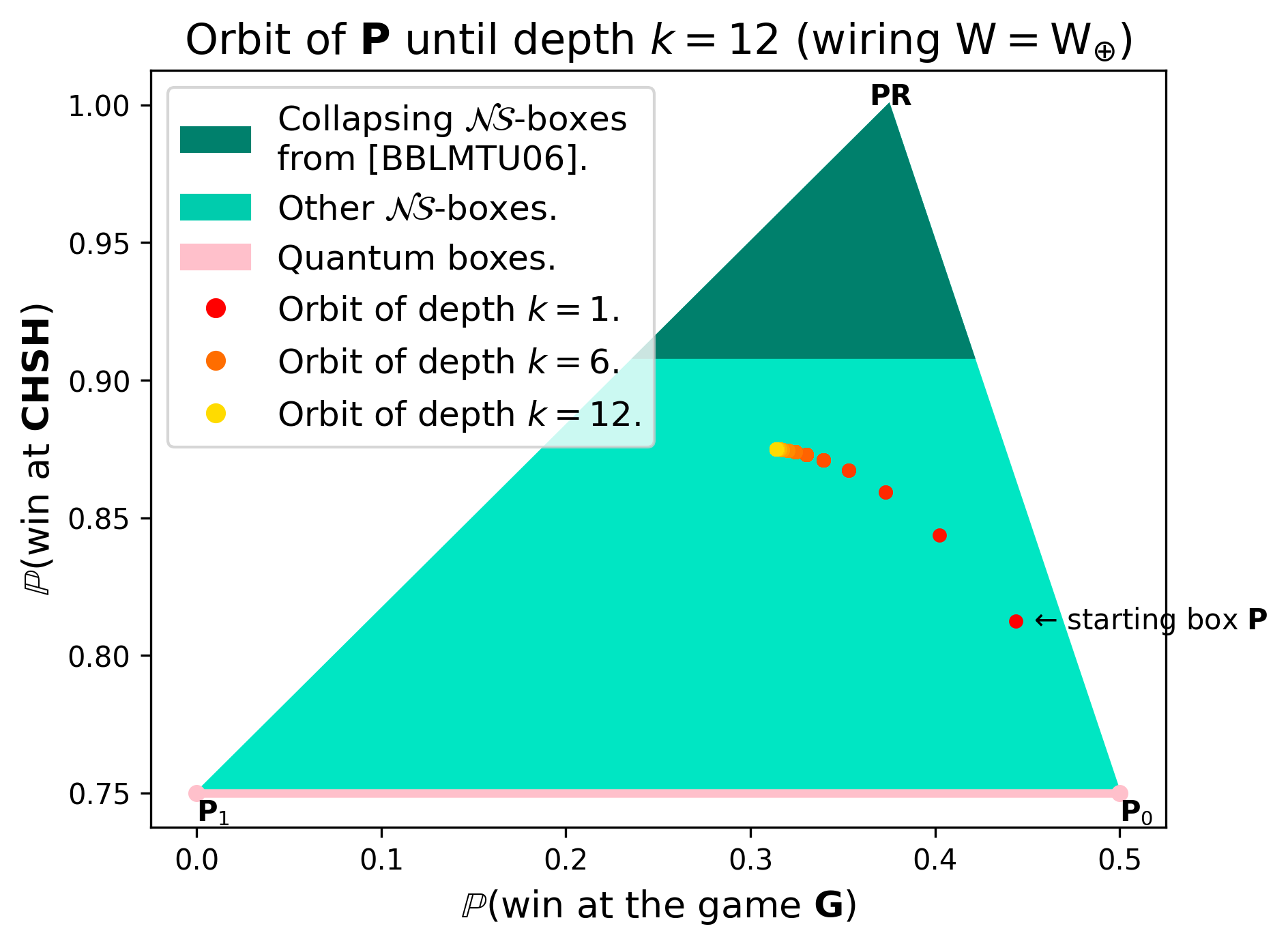}
\end{center}

	\noindent(b)~For {\aaa$\W_{(\text{b})}:=[0.,0.,1.,1.,0.,0.,1.,1.,0.,0.,0.,1.,0.,0.,0.,1.,1.,0.,0.,1.,1.,0.,0.,1.,1.,0.,0.,1.,1.,0.,0.,1.]$}:\\
\begin{center}
	
{
\newcommand{\shiftP}{-0.65}
\newcommand{\shiftQ}{-0.8}
\small
\begin{tikzpicture}[scale=\wiringsize, every node/.style={scale=\wiringsize},
small-NLB/.style={rectangle, draw=my-green, fill=my-green!35, line width=1.2pt, minimum width=60, minimum height=15, anchor=center},
big-NLB/.style={rectangle, draw=my-green, fill=my-green!9, line width=1.2pt, minimum width=60, minimum height=15},
operation/.style={rectangle, draw=black, line width=0.5pt},
my-arrow/.style={->, very thick, >=stealth, line width=0.8pt},
my-arrow-violet/.style={->, line width=0.5em, color=quantumviolet!8},
my-line/.style={very thick},
]

	\draw[big-NLB] (0., -1+0.2) rectangle (4., 4.5);
	\node at (2, 5) {\large$\W_{(\text{b})}$};
		
	\node[small-NLB] at (2, 3.5+\shiftP) (P) {\normalsize$\P$};
	\node[small-NLB] at (2, 1.2+\shiftQ) (Q) {\normalsize$\Q$};
	\node at (0., 5.3) (Alice1) {$x$};
	\node at (0.7, 4.5) (Alice2) {};
	\node[operation] at (1.2, 4.6+\shiftP) (Alice3) {\footnotesize$x$};
	\node at (0.5, 2.8+\shiftP) (Alice4) {$a_1$};
	\node[operation] at (1.2, 2.3+\shiftQ) (Alice5) {\footnotesize$x\wedge a_1$};
	\node at (0.5, 0.5+\shiftQ) (Alice6) {$a_2$};
	\node at (0.7, -0.95+0.2) (Alice7) {};
	\node at (0.7, -1.7) (Alice8) {$\overline{a_1}\oplus a_2$};
	\node at (4., 5.3) (Bob1) {$y$};
	\node at (3.3, 4.5) (Bob2) {};
	\node[operation] at (4-1.2, 4.6+\shiftP) (Bob3) {\footnotesize$y$};
	\node at (4-0.5, 2.8+\shiftP) (Bob4) {$b_1$};
	\node[operation] at (4-1.2, 2.3+\shiftQ) (Bob5) {\footnotesize$y\wedge b_1$};
	\node at (4-0.5, 0.5+\shiftQ) (Bob6) {$b_2$};
	\node at (4-0.7, -0.95+0.2) (Bob7) {};
	\node at (4-0.7, -1.7) (Bob8) {$\overline{b_1}\oplus{b_2}$};
	
	\draw[my-arrow] (Alice1) .. controls +(right:2em) and +(up:1em).. (Alice2);
	\draw[my-arrow] (Alice3.south)  .. controls  +(down:1em) and +(up:0.8em) ..  (P.160);
	\draw[my-arrow] (P.200)  .. controls  +(down:1em) and +(right:0.8em) ..  (Alice4.east);
	\draw[my-arrow] (Alice5.south)  .. controls  +(south:1em) and +(up:0.8em) ..  (Q.160);
	\draw[my-arrow] (Q.200)  .. controls  +(down:1em) and +(right:0.8em) ..  (Alice6.east);
	\draw[my-arrow] (Alice7) to (Alice8);
	
	\draw[my-arrow] (Bob1) .. controls +(left:2em) and +(up:1em).. (Bob2);
	\draw[my-arrow] (Bob3.south)  .. controls  +(down:1em) and +(up:0.8em) .. (P.20);
	\draw[my-arrow] (P.340)  .. controls  +(down:1em) and +(left:0.8em) ..  (Bob4.west);
	\draw[my-arrow] (Bob5.south)  .. controls  +(down:1em) and +(up:0.8em).. (Q.20);
	\draw[my-arrow] (Q.340)  .. controls  +(down:1em) and +(left:0.8em) ..  (Bob6.west);
	\draw[my-arrow] (Bob7) to (Bob8);

\end{tikzpicture}
}

	\hspace{-0.3cm}
	\includegraphics[height=\h]{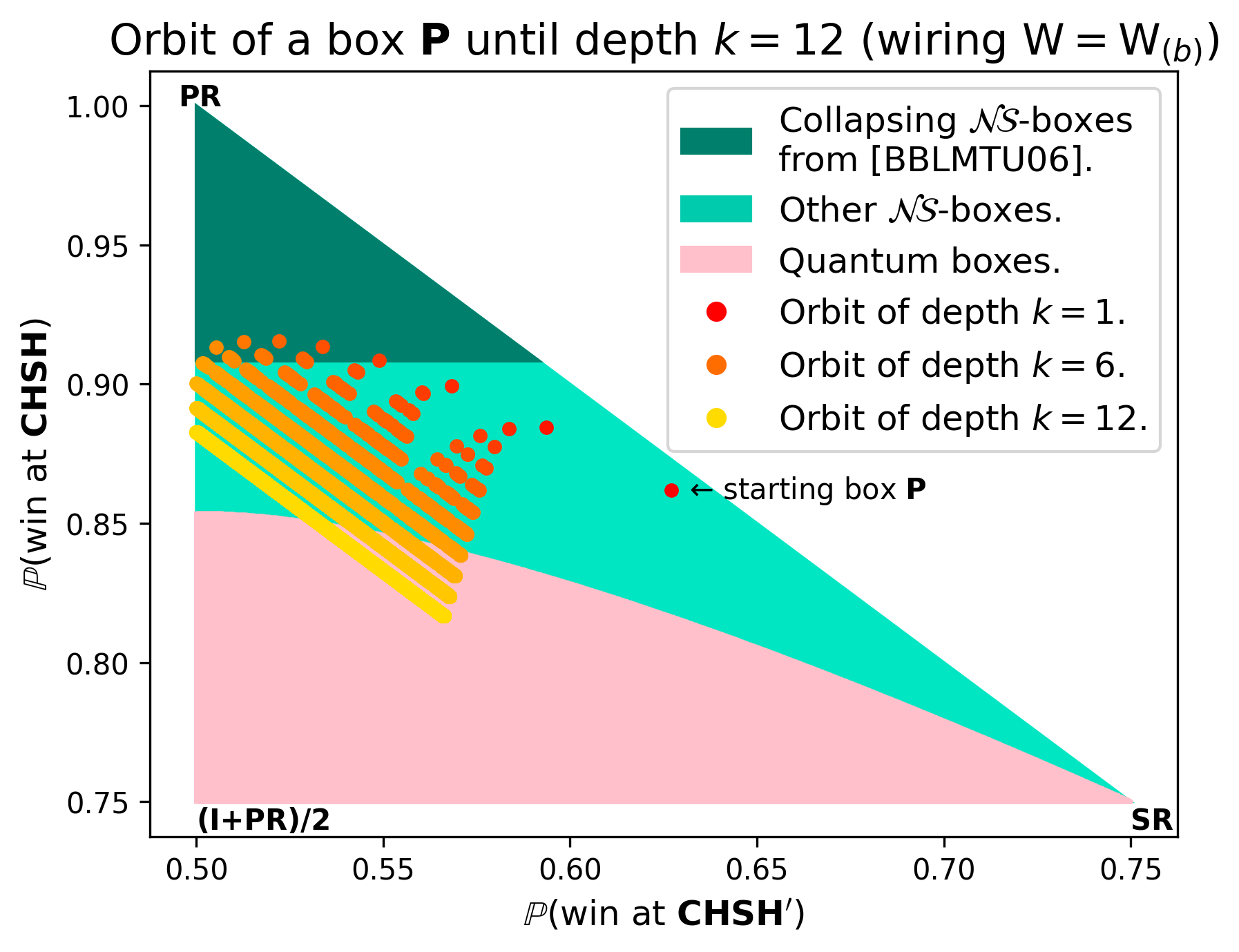}
	\hspace{-0.3cm}
	\includegraphics[height=\h]{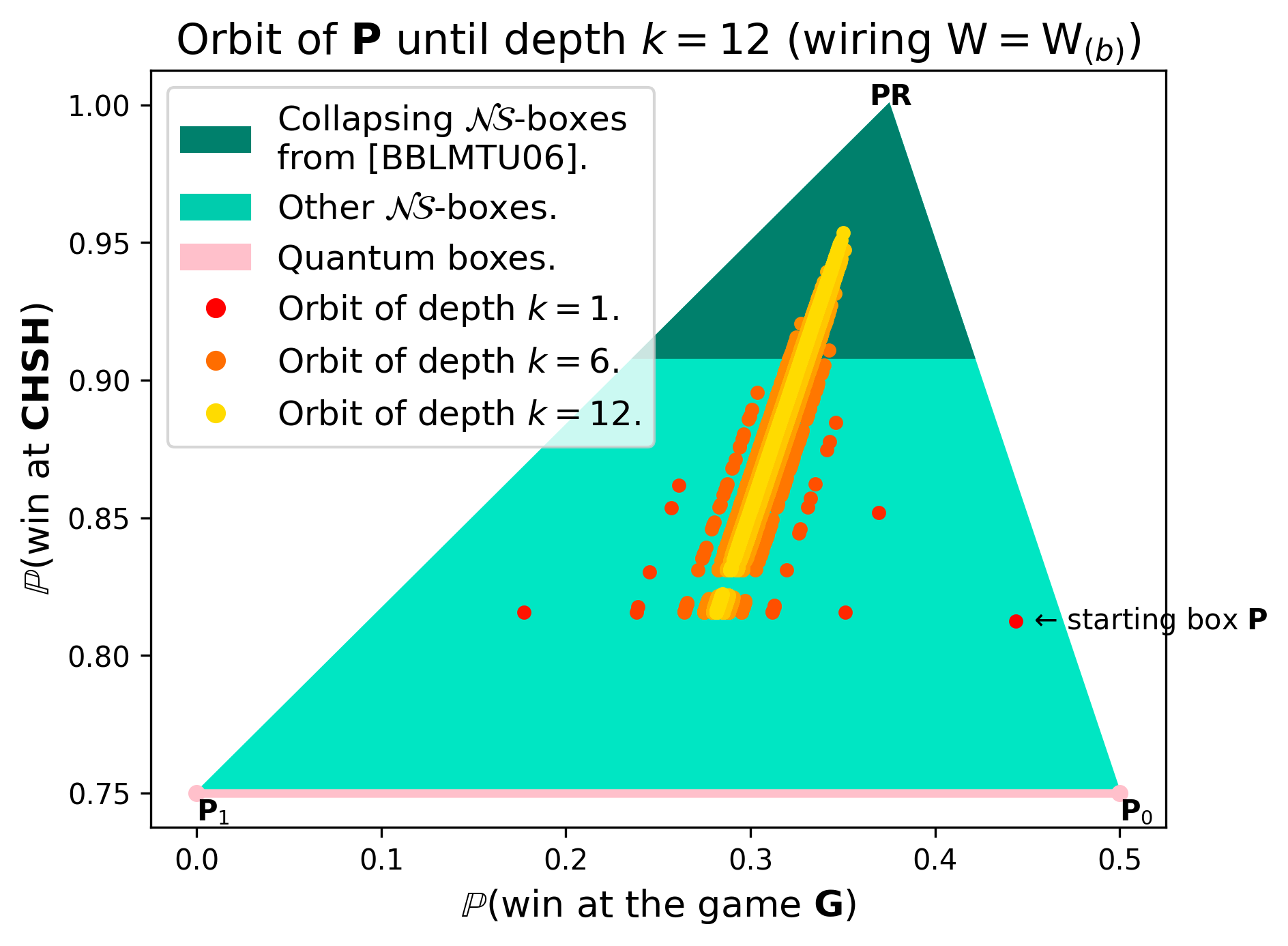}
\end{center}

	\noindent(c)~For {\aaa$\W_{(\text{c})}:=[0.,0.,1.,1.,0.,0.,1.,1.,0.,0.,1.,0.,0.,0.,1.,0.,1.,0.,0.,1.,1.,0.,0.,1.,1.,0.,0.,1.,1.,0.,0.,1.]$}:\\
\begin{center}
	
{
\newcommand{\shiftP}{-0.65}
\newcommand{\shiftQ}{-0.8}
\small
\begin{tikzpicture}[scale=\wiringsize, every node/.style={scale=\wiringsize},
small-NLB/.style={rectangle, draw=my-green, fill=my-green!35, line width=1.2pt, minimum width=60, minimum height=15, anchor=center},
big-NLB/.style={rectangle, draw=my-green, fill=my-green!9, line width=1.2pt, minimum width=60, minimum height=15},
operation/.style={rectangle, draw=black, line width=0.5pt},
my-arrow/.style={->, very thick, >=stealth, line width=0.8pt},
my-arrow-violet/.style={->, line width=0.5em, color=quantumviolet!8},
my-line/.style={very thick},
]

	\draw[big-NLB] (0., -1+0.2) rectangle (4., 4.5);
	\node at (2, 5) {\large$\W_{(\text{c})}$};
		
	\node[small-NLB] at (2, 3.5+\shiftP) (P) {\normalsize$\P$};
	\node[small-NLB] at (2, 1.2+\shiftQ) (Q) {\normalsize$\Q$};
	\node at (0., 5.3) (Alice1) {$x$};
	\node at (0.7, 4.5) (Alice2) {};
	\node[operation] at (1.2, 4.6+\shiftP) (Alice3) {\footnotesize$x$};
	\node at (0.5, 2.8+\shiftP) (Alice4) {$a_1$};
	\node[operation] at (1.2, 2.3+\shiftQ) (Alice5) {\footnotesize$x\wedge \overline{a_1}$};
	\node at (0.5, 0.5+\shiftQ) (Alice6) {$a_2$};
	\node at (0.7, -0.95+0.2) (Alice7) {};
	\node at (0.7, -1.7) (Alice8) {$\overline{a_1}\oplus a_2$};
	\node at (4., 5.3) (Bob1) {$y$};
	\node at (3.3, 4.5) (Bob2) {};
	\node[operation] at (4-1.2, 4.6+\shiftP) (Bob3) {\footnotesize$y$};
	\node at (4-0.5, 2.8+\shiftP) (Bob4) {$b_1$};
	\node[operation] at (4-1.2, 2.3+\shiftQ) (Bob5) {\footnotesize$y\wedge \overline{b_1}$};
	\node at (4-0.5, 0.5+\shiftQ) (Bob6) {$b_2$};
	\node at (4-0.7, -0.95+0.2) (Bob7) {};
	\node at (4-0.7, -1.7) (Bob8) {$\overline{b_1}\oplus{b_2}$};
	
	\draw[my-arrow] (Alice1) .. controls +(right:2em) and +(up:1em).. (Alice2);
	\draw[my-arrow] (Alice3.south)  .. controls  +(down:1em) and +(up:0.8em) ..  (P.160);
	\draw[my-arrow] (P.200)  .. controls  +(down:1em) and +(right:0.8em) ..  (Alice4.east);
	\draw[my-arrow] (Alice5.south)  .. controls  +(south:1em) and +(up:0.8em) ..  (Q.160);
	\draw[my-arrow] (Q.200)  .. controls  +(down:1em) and +(right:0.8em) ..  (Alice6.east);
	\draw[my-arrow] (Alice7) to (Alice8);
	
	\draw[my-arrow] (Bob1) .. controls +(left:2em) and +(up:1em).. (Bob2);
	\draw[my-arrow] (Bob3.south)  .. controls  +(down:1em) and +(up:0.8em) .. (P.20);
	\draw[my-arrow] (P.340)  .. controls  +(down:1em) and +(left:0.8em) ..  (Bob4.west);
	\draw[my-arrow] (Bob5.south)  .. controls  +(down:1em) and +(up:0.8em).. (Q.20);
	\draw[my-arrow] (Q.340)  .. controls  +(down:1em) and +(left:0.8em) ..  (Bob6.west);
	\draw[my-arrow] (Bob7) to (Bob8);

\end{tikzpicture}
}

	\hspace{-0.3cm}
	\includegraphics[height=\h]{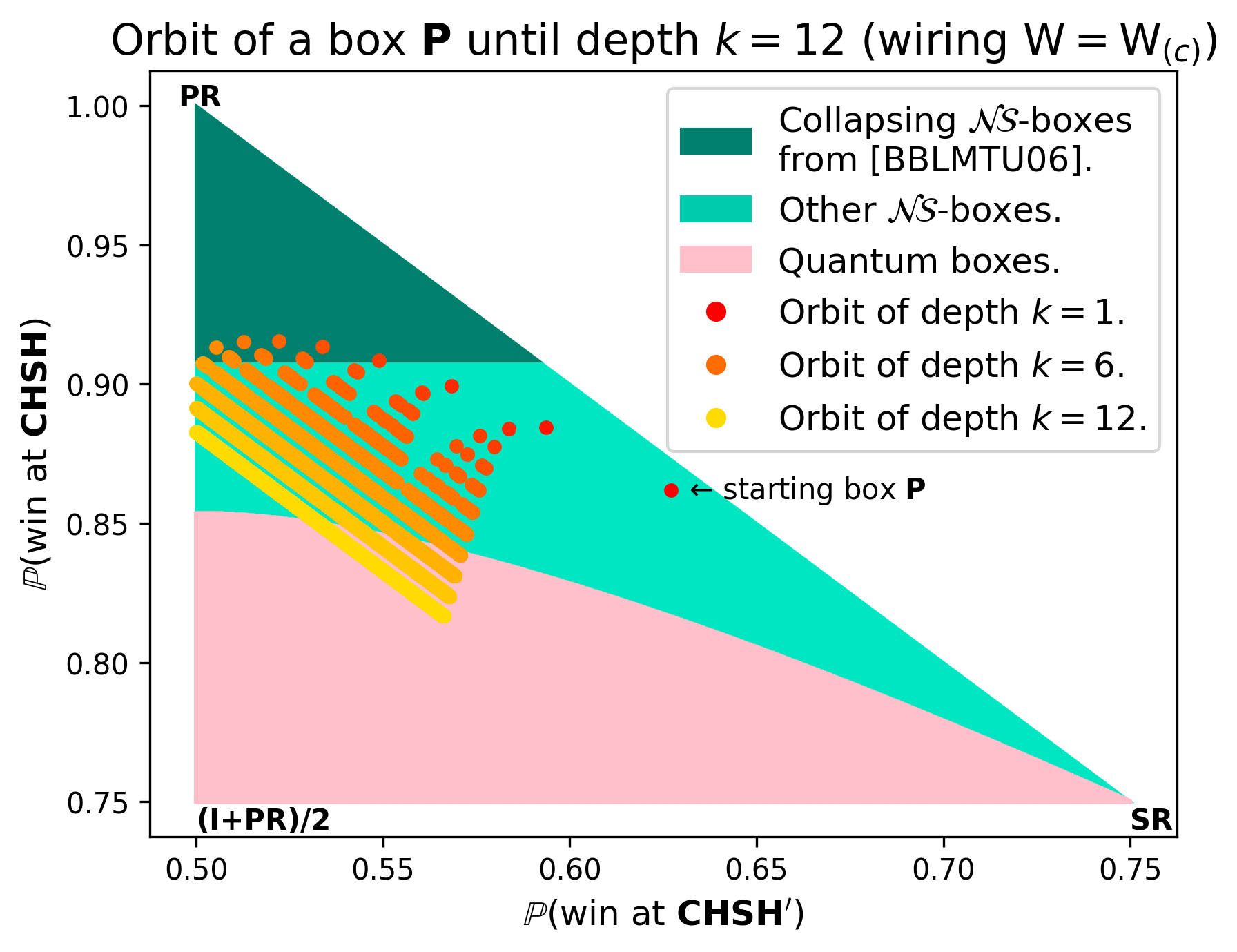}
	\hspace{-0.3cm}
	\includegraphics[height=\h]{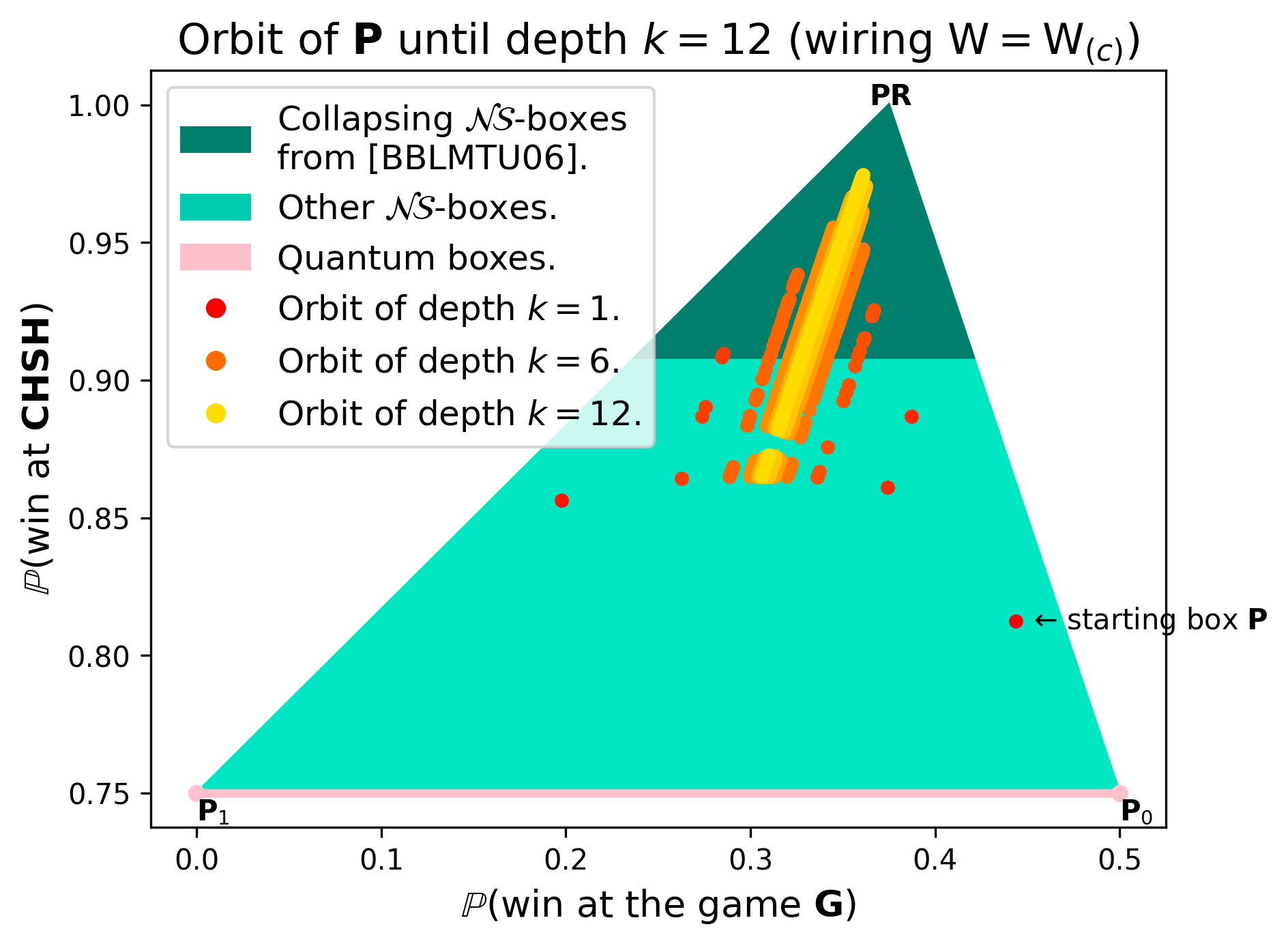}
\end{center}

	\noindent(d)~For {\aaa$\W_{(\text{d})}:=[0.,0.,1.,1.,0.,0.,1.,1.,0.,0.,1.,0.,0.,0.,1.,0.,0.,1.,1.,0.,0.,1.,1.,0.,0.,1.,1.,0.,0.,1.,1.,0.]$}:\\
\begin{center}
	
{
\newcommand{\shiftP}{-0.65}
\newcommand{\shiftQ}{-0.8}
\small
\begin{tikzpicture}[scale=\wiringsize, every node/.style={scale=\wiringsize},
small-NLB/.style={rectangle, draw=my-green, fill=my-green!35, line width=1.2pt, minimum width=60, minimum height=15, anchor=center},
big-NLB/.style={rectangle, draw=my-green, fill=my-green!9, line width=1.2pt, minimum width=60, minimum height=15},
operation/.style={rectangle, draw=black, line width=0.5pt},
my-arrow/.style={->, very thick, >=stealth, line width=0.8pt},
my-arrow-violet/.style={->, line width=0.5em, color=quantumviolet!8},
my-line/.style={very thick},
]

	\draw[big-NLB] (0., -1+0.2) rectangle (4., 4.5);
	\node at (2, 5) {\large$\W_{(\text{d})}$};
		
	\node[small-NLB] at (2, 3.5+\shiftP) (P) {\normalsize$\P$};
	\node[small-NLB] at (2, 1.2+\shiftQ) (Q) {\normalsize$\Q$};
	\node at (0., 5.3) (Alice1) {$x$};
	\node at (0.7, 4.5) (Alice2) {};
	\node[operation] at (1.2, 4.6+\shiftP) (Alice3) {\footnotesize$x\wedge\overline{a_2}$};
	\node at (0.5, 2.8+\shiftP) (Alice4) {$a_1$};
	\node[operation] at (1.2, 2.3+\shiftQ) (Alice5) {\footnotesize$x$};
	\node at (0.5, 0.5+\shiftQ) (Alice6) {$a_2$};
	\node at (0.7, -0.95+0.2) (Alice7) {};
	\node at (0.7, -1.7) (Alice8) {$x\oplus a_1\oplus \overline{a_2}$};
	\node at (4., 5.3) (Bob1) {$y$};
	\node at (3.3, 4.5) (Bob2) {};
	\node[operation] at (4-1.2, 4.6+\shiftP) (Bob3) {\footnotesize$y\oplus\overline{b_2}$};
	\node at (4-0.5, 2.8+\shiftP) (Bob4) {$b_1$};
	\node[operation] at (4-1.2, 2.3+\shiftQ) (Bob5) {\footnotesize$\overline{y}$};
	\node at (4-0.5, 0.5+\shiftQ) (Bob6) {$b_2$};
	\node at (4-0.7, -0.95+0.2) (Bob7) {};
	\node at (4-0.7, -1.7) (Bob8) {${b_1}\oplus\overline{b_2}$};
	
	\draw[my-arrow] (Alice1) .. controls +(right:2em) and +(up:1em).. (Alice2);
	\draw[my-arrow] (Alice3.south)  .. controls  +(down:1em) and +(up:0.8em) ..  (P.160);
	\draw[my-arrow] (P.200)  .. controls  +(down:1em) and +(right:0.8em) ..  (Alice4.east);
	\draw[my-arrow] (Alice5.south)  .. controls  +(south:1em) and +(up:0.8em) ..  (Q.160);
	\draw[my-arrow] (Q.200)  .. controls  +(down:1em) and +(right:0.8em) ..  (Alice6.east);
	\draw[my-arrow] (Alice7) to (Alice8);
	
	\draw[my-arrow] (Bob1) .. controls +(left:2em) and +(up:1em).. (Bob2);
	\draw[my-arrow] (Bob3.south)  .. controls  +(down:1em) and +(up:0.8em) .. (P.20);
	\draw[my-arrow] (P.340)  .. controls  +(down:1em) and +(left:0.8em) ..  (Bob4.west);
	\draw[my-arrow] (Bob5.south)  .. controls  +(down:1em) and +(up:0.8em).. (Q.20);
	\draw[my-arrow] (Q.340)  .. controls  +(down:1em) and +(left:0.8em) ..  (Bob6.west);
	\draw[my-arrow] (Bob7) to (Bob8);

\end{tikzpicture}
}

	\hspace{-0.3cm}
	\includegraphics[height=\h]{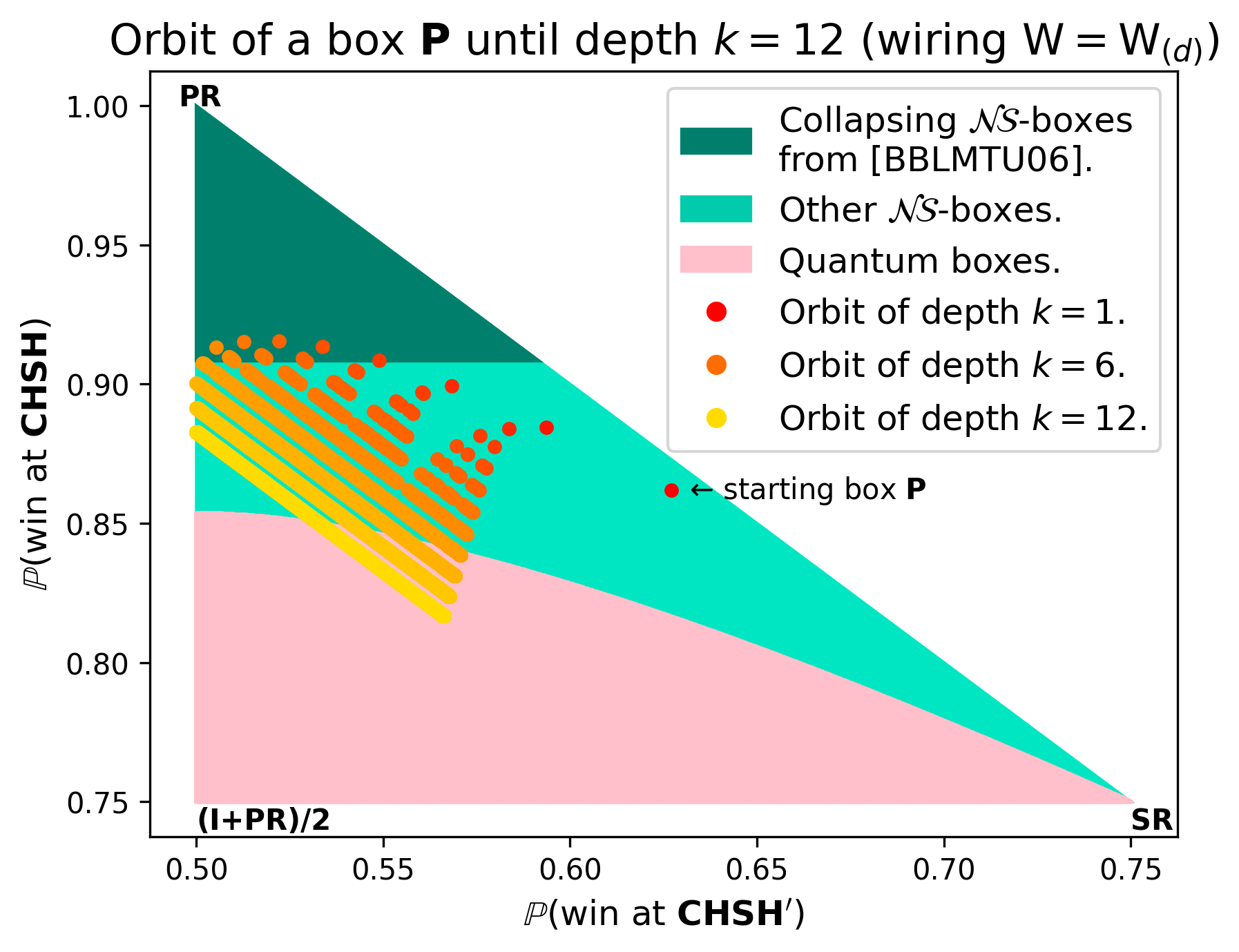}
	\hspace{-0.3cm}
	\includegraphics[height=\h]{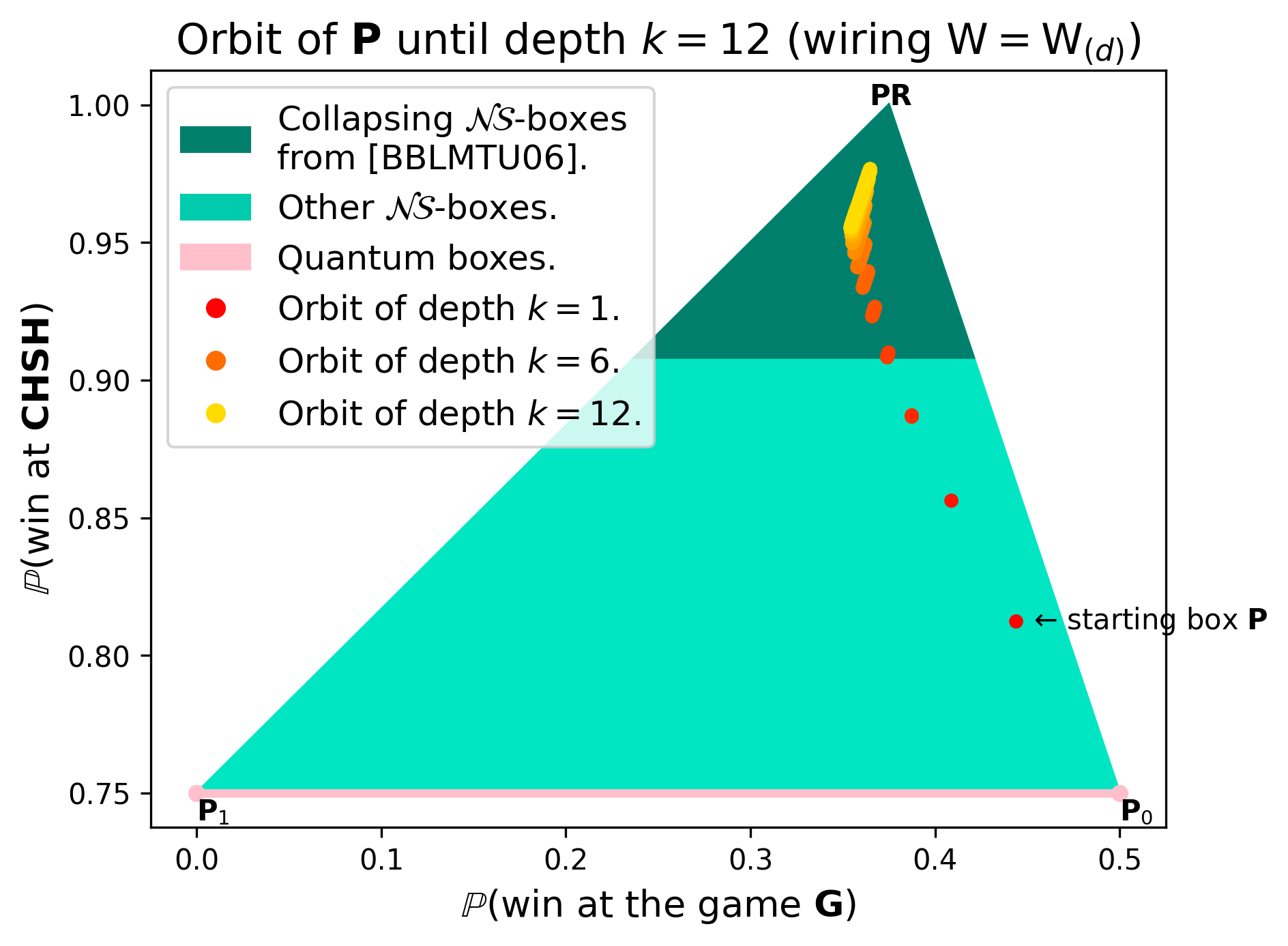}
\end{center}

\bib

\small

	\section{An Alternative Proof of \refprop[Theorem]{thm: the triangle PR-P0-P1 is collapsing}, using Pairwise Multiplication}
	\label{appendix: proof of the theorem using Pairwise Multiplication}

\begin{lemma}[Convergence sufficient conditions] \label{lem: Convergence of the first sequence}
	Let $\K\subseteq \R^2$ be a compact convex set, and let $F$, $f_\beta$, $g_\alpha$ be functions such that:
	\[
		F:\left\{
		\begin{array}{ccl}
			\K & \longrightarrow & \K\\
			(\alpha, \beta) & \longmapsto & \big( f_\beta(\alpha), g_\alpha(\beta) \big)\,.
		\end{array}
		\right.
	\]
	where $f_\beta: I_\beta \to I_\beta$ and $g_\alpha:J_\alpha \to J_\alpha$, where $I_\beta := \K\cap\big(\R\times \{\beta\}\big)$ and $J_\alpha:=\K\cap\big(\{\alpha\}\times \R\big)$ are viewed as intervals of $\R$. Suppose that:
	\begin{enumerate}[label=\normalfont(\roman*)]
		\item The function $F$ is continuous;
		\item The $f_\beta$'s are strictly increasing and uniformly bounded in the ``left part" of the interior of $\K$:
		\begin{equation}  \label{eq: growth and boundedness of the f_beta s}
		\exists\mu_*,\, 
				\forall (\alpha,\beta)\in\Int(\K),
				\,
				 \alpha < \mu_* 
				 \Longrightarrow 
				\alpha < f_\beta(\alpha) < \mu_*\,;
	\end{equation}
		\item For all boundary element $(\alpha,\beta)\in\partial \K$ such that $\alpha<\mu_*$, either the resulting inequalities of \eqref{eq: growth and boundedness of the f_beta s} are satisfied, or $\alpha=f_\beta(\alpha)$ and the image $F(\alpha, \beta)$ lies in the interior $\Int(\K)$ of $\K$.
	\end{enumerate}
	Then the sequence defined by $(\alpha_0, \beta_0)\in \K$, $\alpha_0<\mu_*$, and $(\alpha_{k+1}, \beta_{k+1}) = F(\alpha_k, \beta_k)$ ($k\geq0$), satisfies:
	\[
		\alpha_k
		\quad{}_{\overrightarrow{k\to+\infty}}\quad 
		\mu_*\,.
	\]
\end{lemma}

\begin{proof}
	A proof by induction based on (ii) and (iii) shows that the sequence $(\alpha_k)_k$ is increasing and bounded from above by $\mu_*$, so it converges to some $\alpha_\infty\leq \mu_*$. Let us show that $\alpha_\infty = \mu_*$.
	As $\K$ is compact, Bolzano–Weierstrass' Theorem tells us that the sequence $(\beta_k)_k$ admits a converging subsequence $\big(\beta_{\varphi(k)}\big)_k$, whose limit is $\beta_\infty$. Then, by continuity of $F$, we have:
	\[
		\big( \alpha_{\varphi(k)+1}, \beta_{\varphi(k)+1} \big)
		\,=\,
		F\big( \alpha_{\varphi(k)}, \beta_{\varphi(k)} \big)
		\quad
		{}_{\overrightarrow{k\to+\infty}}
		\quad
		F\big( \alpha_\infty, \beta_\infty \big)
		\,=\,
		\big( f_{\beta_\infty}(\alpha_\infty), g_{\alpha_\infty}(\beta_\infty) \big)\,.
	\]
	Therefore the sequence $\big( \alpha_{\varphi(k)+1} \big)_k$ tends to $f_{\beta_\infty}(\alpha_\infty)$, but we know from before that it also tends to $\alpha_\infty$, so the limits coincide:
	\[
		f_{\beta_\infty}(\alpha_\infty)
		\,=\,
		\alpha_\infty\,.
	\]
	In other words, we showed that $\alpha_\infty$ is a fixed point of $f_{\beta_\infty}$.
	Now, by contradiction if $\alpha_\infty$ were different from $\mu_*$, \ie if $\alpha_\infty<\mu_*$, then (ii) and (iii) 
	would imply either the contradiction $\alpha_\infty < f_{\beta_\infty}(\alpha_\infty)$, or the inlcusion $\big(f_{\beta_\infty}(\alpha_\infty), g_{\alpha_\infty}(\beta_\infty) \big)\in\Int(\K)$, but then (ii) would give $f_{\beta_\infty}(\alpha_\infty) < f_{\beta_\infty} \circ f_{\beta_\infty}(\alpha_\infty)$, which is  again the contradiction $\alpha_\infty < f_{\beta_\infty}(\alpha_\infty)$.
	Hence $\alpha_\infty = \mu_*$ and we obtain the wanted result.
\end{proof}

\begin{theorem}[Alternative proof of Theorem~\ref{thm: the triangle PR-P0-P1 is collapsing}]
	The set $T:=\Conv\{\PR, \P_0, \P_1\}\backslash\Conv\{\P_0, \P_1\}$ is collapsing.
\end{theorem}

\begin{proof}
	Consider boxes of the form $\P_{\alpha, \beta} := \alpha\PR + \beta\P_0 + (1-\alpha-\beta)\P_1$, where
	$\alpha, \beta\geq0$ and $\alpha\neq0$ and $\alpha+\beta\leq1$. Denote $\Delta$ the set of all such $\alpha$'s and $\beta$'s, so that $T = \{\P_{\alpha, \beta} : (\alpha, \beta)\in \Delta\}$.
	The triangle $\Delta$ is not compact but it can be written as a union of compact convex sets:
	$$
	\Delta
	\,=\,
	\bigcup_{n\geq1} \K_n\,,
	$$
	where $\K_n := \Delta \cap \{\alpha\geq1/n\}$.
	Denote $\boxtimes$ the box product induced by the wiring $\Wbs$ inspired from~\cite{BS09}.
	Fix $(\alpha_0, \beta_0)\in \Delta$, and fix an integer $n\geq1$ large enough so that $(\alpha_0, \beta_0)\in \K_n$.
	We want to build a protocol that collapses communication complexity using the starting box $\P_{\alpha_0, \beta_0}$ as a resource. To do so, we will define a sequence of boxes $(\P_{\alpha_k, \beta_k})_k$ that are eventually collapsing for $k$ large enough.
	First, see that the closure $\overline T$ of the triangle $T$ is stable under $\boxtimes$, because $\overline T$ equals $\NS \cap \Aff\{\PR, \P_0, \P_1\}$ (see \refprop[Proposition]{prop:PR-P0-P1-is-a-face-of-NS})
	and the two intersected sets are stable under $\boxtimes$~\footnote{Indeed, (i)~the set $\NS$ is stable under $\boxtimes$ because it is closed under wirings~\cite{ABLPSV09}, (ii)~the affine plane $\Aff\{\PR, \P_0, \P_1\}$ is stable under under $\boxtimes$ as a consequence of the multiplication table in \refprop[Figure]{fig: multiplication table} and by bilinearity of $\boxtimes$.
	}.
	Moreover, by bilinearity of $\boxtimes$ and using the multiplication table in \refprop[Figure]{fig: multiplication table}, we obtain:
	\begin{equation}  \label{eq: sequence of boxes}
		\P_{\alpha, \beta} \boxtimes \P_{\alpha, \beta} = \P_{\tilde\alpha, \tilde\beta}\,,
	\end{equation}
	with $\tilde\alpha := 2\alpha - \alpha^2 - \alpha\beta$ and $\tilde\beta := 1-2\alpha + \alpha^2 +\frac52 \alpha\beta - 2\beta + 2\beta^2$. 
	We then consider the function $F:\K_n\subseteq\R^2\to \R^2$ such that $F(\alpha, \beta) := (\tilde \alpha, \tilde\beta)$ and we define the sequence $(\alpha_{k+1}, \beta_{k+1}) := F(\alpha_k, \beta_k)$ for $k\geq0$. Let us show that the conditions of \refprop[Lemma]{lem: Convergence of the first sequence} are satisfied.
	Due to the stability of $\overline T$ under $\boxtimes$, we have the inclusion $F(\K_n)\subseteq \overline{\Delta}$, and as
	\begin{equation}   \label{eq: f beta increasing}
		\tilde\alpha 
		\,=\,
		2\alpha - \alpha(\alpha+\beta)
		\geq 
		\alpha \geq 1/n\,,
	\end{equation}
	we obtain the inclusion $F(\K_n)\subseteq\K_n$, where we used $\alpha+\beta\leq1$. 
	(i) The function $F$ is polynomial so continuous.  
	(ii) Condition \eqref{eq: growth and boundedness of the f_beta s} is satisfied for $\mu_*:=1$:
	on the one hand $f_\beta$ is strictly increasing due to \eqref{eq: f beta increasing} applied with $\alpha + \beta < 1$ in the interior $\Int(\K_n)$;
	 on the other hand,
	 by strict growth of $f_\beta$ we have $f_\beta(\alpha) = \tilde\alpha < f_{\tilde\beta}\big(\tilde\alpha\big)$, which is $\leq 1$ because $F(\K_n)\subseteq \K_n$. Hence we indeed have $\alpha < f_\beta(\alpha) < 1$ in the interior of $\K_n$.
	 (iii) Let $(\alpha, \beta)\in\partial\K_n=\{\beta = 0\} \cup \{\alpha=1/n\}\cup\{\alpha+\beta=1\}$ such that $\alpha<1$. If $\beta = 0$ or $\alpha = 1/n$, then inequalities of Condition \eqref{eq: growth and boundedness of the f_beta s} hold. Otherwise we have $\alpha + \beta = 1$ and in that case:
	 \[
	 \left\{
	 \begin{array}{ccl}
	 	\tilde\alpha & = & \alpha \in \big(\frac1n, 1\big)\,, \\
	 	\tilde\beta & = &\frac12(1-\alpha) (\, 1 + \beta) > 0\,, \\
		\tilde\alpha + \tilde\beta &=& 1 - \frac12(1-\alpha) ( 1-\beta ) < 1\,,
	\end{array}
	\right.
	 \]
	 which implies that $(\tilde\alpha, \tilde\beta)\in \Int(\K_n)$.
	 Finally, we may assume that $\alpha_0<1$ (otherwise $\alpha_0=1$ and then $\P_{\alpha_0, \beta_0} = \PR$, which is collapsing) and all the conditions of \refprop[Lemma]{lem: Convergence of the first sequence} are satisfied. 
	The lemma tells that the sequence $(\alpha_k)_k$ tends to $\mu_*=1$, so $(\P_{\alpha_k, \beta_k})_k\subseteq\R^{16}$ tends to $\PR$. 
	 But we know from~\cite{BBLMTU06} that there is a non-empty open set around $\PR$ that is collapsing. So there exists a finite $k$ for which the box $\P_{\alpha_{k}, \beta_{k}}$ is collapsing and we obtain the wanted collapsing protocol.	
\end{proof}

\bib

\section{Some Multiplication Tables}
\label{ap:the-multiplication-tables}

Find below the multiplication tables of $\boxtimes_\W$ for different wirings $\W$. Each cell displays the result of $\P\boxtimes_\W\Q$, where $\P$ lies in the first column and $\Q$ lies in the first line. 
	$\P_{10}:= \PL^{0100}$.
	$\P_{01}:= \PL^{0001}$.
	$\Q_1 := -\frac{1}{8}\big(\P_0 + \P_1\big)+ \frac{1}{4} \,\PR + \I $.
	$\Q_2 := \frac{1}{4}\big(\PR + \PNL^{011}\big) +\frac12 \, \I$.
	$\Q_3 := \frac14\,\PR + \frac12\,\I + \frac38\P_{0} - \frac18\P_{1}$.
	$\Q_4 := \frac{3}{8}\,\P_0 - \frac{1}{8}\,\P_1 + \frac{3}{4}\,\I$.
	$\Q_5 := \frac14\,\PR + \frac18(\P_0 +\P_1) + \frac12\,\P_{10}$. 
	$\Q_6 := \frac{3}{16} ( \P_0 + \P_1 ) + \frac{1}{16}\,\P_{01} + \frac{9}{16}\,\P_{10}$.
	Find an algorithm to compute a multiplication table in our GitHub page~\cite{GitHub-algebra-of-boxes}.
	Recall the definitions of $\PL$ and $\PNL$ in \eqrefprop[Equation]{eq: extremal points of NS}.
	

\newlength{\firetruck}
\setlength{\firetruck}{0.5cm}
\newlength{\apple}
\setlength{\apple}{0.15cm}
\setlength{\length}{0.2cm}

\begin{center}
{
	 \footnotesize
	
	\newcommand{\mycommand}{\normalsize}
	\newcommand{\mycommandd}{}
	
	\renewcommand{\arraystretch}{1.5}
	
	\begin{tabular}{| c ||c|c|c|c|c|c|c|c|}
		\hline
		 $\Wtriv$ & \hspace{\length}{\mycommand$\PR$}\hspace{\length} & \hspace{\length}{\mycommand$\P_0$}\hspace{\length} & \hspace{\length}{\mycommand$\P_1$}\hspace{\length}  & \hspace{\length}{\mycommand$\I$}\hspace{\length} \\
		\hline\hline
		{\mycommand$\PR$} &  $\PL^{1010}$ & $\PL^{1010}$ & $\PL^{1010}$ & $\PL^{1010}$  \\
		\hline
		{\mycommand$\P_0$} & $\PL^{1010}$  & $\PL^{1010}$  & $\PL^{1010}$ & $\PL^{1010}$  \\ \hline
		{\mycommand$\P_1$}  & $\PL^{1010}$ &  $\PL^{1010}$ & $\PL^{1010}$ & $\PL^{1010}$  \\ \hline
		{\mycommand$\I$}  & $\PL^{1010}$ &  $\PL^{1010}$ & $\PL^{1010}$ & $\PL^{1010}$  \\ \hline
	\end{tabular}
	\hspace{\firetruck}
	\begin{tabular}{| c ||c|c|c|c|c|c|c|c|}
		\hline
		 $\Woplus$~\cite{FWW09} & \hspace{\length}{\mycommand$\PR$}\hspace{\length} & \hspace{\length}{\mycommand$\P_0$}\hspace{\length} & \hspace{\length}{\mycommand$\P_1$}\hspace{\length}  & \hspace{\length}{\mycommand$\I$}\hspace{\length} \\
		\hline\hline
		{\mycommand$\PR$} &  $\frac{\P_0 + \P_1}{2}$ & $\PR$ & $\PR$ & $\I$  \\
		\hline
		{\mycommand$\P_0$} & $\PR$  & $\P_0$  & $\P_1$ & $\I$  \\ \hline
		{\mycommand$\P_1$}  & $\PR$ &  $\P_1$ & $\P_0$ & $\I$  \\ \hline
		{\mycommand$\I$}  & $\I$ &  $\I$ & $\I$ & $\I$  \\ \hline
	\end{tabular}

	\vspace{\apple}

	\hspace{-0.42cm}
	\begin{tabular}{| c ||c|c|c|c|c|c|c|c|}
		\hline
		$\Wbs$~\cite{BS09} & \hspace{\length}{\mycommand$\PR$}\hspace{\length} & \hspace{\length}{\mycommand$\P_0$}\hspace{\length} & \hspace{\length}{\mycommand$\P_1$}\hspace{\length} & \hspace{\length}{\mycommand$\I$}\hspace{\length} \\
		\hline\hline
		{\mycommand$\PR$} &  $\PR$ & $\PR$ & $\PR$ & $\I$  \\ \hline
		{\mycommand$\P_0$} & $\frac{\P_0 + \P_1}{2}$  & $\P_0$  & $\P_1$ & $\I$  \\
		\hline
		{\mycommand$\P_1$}  & $\PR$ &  $\P_1$ & $\P_0$ & $\I$  \\ \hline
		{\mycommand$\I$}  & $\Q_1$ &  $\I$ & $\I$ & $\I$  \\ \hline
	\end{tabular}	
	\hspace{\firetruck}
	\begin{tabular}{| c ||c|c|c|c|c|c|c|c|}
		\hline
		$\Wdist$~\cite{ABLPSV09} & \hspace{\length}{\mycommand$\PR$}\hspace{\length} & \hspace{\length}{\mycommand$\P_0$}\hspace{\length} & \hspace{\length}{\mycommand$\P_1$}\hspace{\length}  & \hspace{\length}{\mycommand$\I$}\hspace{\length} \\
		\hline\hline
		{\mycommand$\PR$} &  $\PR$ & $\PR$ & $\PR$ & $\I$  \\
		\hline
		{\mycommand$\P_0$} & $\frac{\P_0 + \P_1}{2}$  & $\P_1$  & $\P_0$ & $\I$  \\ \hline
		{\mycommand$\P_1$}  & $\PR$ &  $\P_0$ & $\P_1$ & $\I$  \\ \hline
		{\mycommand$\I$}  & $\Q_2$ &  $\I$ & $\I$ & $\I$  \\ \hline
	\end{tabular}
	
	\vspace{\apple}
	
	\hspace{0.64cm}
	\begin{tabular}{| c ||c|c|c|c|c|c|c|c|}
		\hline
		$\Wand$~\cite{ABLPSV09} & \hspace{\length}{\mycommand$\PR$}\hspace{\length} & \hspace{\length}{\mycommand$\P_0$}\hspace{\length} & \hspace{\length}{\mycommand$\P_1$}\hspace{\length}  & \hspace{\length}{\mycommand$\I$}\hspace{\length} \\
		\hline\hline
		{\mycommand$\PR$} &  $\frac{\P_0 + \P_1}{2}$ & $\P_0$ & $\PR$ & $\Q_3$  \\
		\hline
		{\mycommand$\P_0$} & $\P_0$  & $\P_0$  & $\P_0$ & $\P_0$  \\ \hline
		{\mycommand$\P_1$}  & $\PR$ &  $\P_0$ & $\P_1$ & $\I$  \\ \hline
		{\mycommand$\I$}  & $\Q_3 $ &  $\P_0$ & $\I$ & $\Q_4$  \\ \hline
	\end{tabular}
	\hspace{\firetruck}
	\begin{tabular}{| c ||c|c|c|c|c|c|c|c|}
		\hline
		$\Worand$~\cite{NSSRRB22PRL} & \hspace{\length}{\mycommand$\PR$}\hspace{\length} & \hspace{\length}{\mycommand$\P_0$}\hspace{\length} & \hspace{\length}{\mycommand$\P_1$}\hspace{\length}  & \hspace{\length}{\mycommand$\I$}\hspace{\length} \\
		\hline\hline
		{\mycommand$\PR$} &  \mycommandd$\frac{\PR + \P_{10}}{2}$ & \mycommandd$\frac{\P_0 + \P_{10}}{2}$ & \mycommandd$\frac{\P_1 + \P_{10}}{2}$ & $\Q_5$  \\
		\hline
		{\mycommand$\P_0$} & \mycommandd$\frac{\P_0 + \P_{10}}{2}$  & $\P_0$  & $\P_{10}$ & \mycommandd$\frac{\P_0 + \P_{10}}{2}$  \\ \hline
		{\mycommand$\P_1$}  & \mycommandd$\frac{\P_1 + \P_{10}}{2}$ &  $\P_{10}$ & $\P_1$ & \mycommandd$\frac{\P_1 + \P_{10}}{2}$  \\ \hline
		{\mycommand$\I$}  & $\Q_5$ &  \mycommandd$\frac{\P_0 + \P_{10}}{2}$ &\mycommandd $\frac{\P_1 + \P_{10}}{2}$ & $\Q_6$  \\ \hline
	\end{tabular}
	
	\renewcommand{\arraystretch}{1}

}
\end{center}

\bib

\end{document}